\documentclass[fleqn,usenatbib]{mnras}

\usepackage{newtxtext,newtxmath}
\usepackage[T1]{fontenc}

\DeclareRobustCommand{\VAN}[3]{#2}
\let\VANthebibliography\thebibliography
\def\thebibliography{\DeclareRobustCommand{\VAN}[3]{##3}\VANthebibliography}

\usepackage{graphicx}	% Including figure files
\usepackage{amsmath}	% Advanced maths commands

\usepackage{bm}  % bold math
\usepackage{booktabs}  % table
\usepackage[shortlabels]{enumitem}
\usepackage{balance}
\usepackage[normalem]{ulem} % normalem to prevent change of emph format

\usepackage[table,xcdraw,svgnames]{xcolor}
\hypersetup{
  pdfencoding=auto, bookmarksnumbered,
}

\usepackage{listings}  % code highlighting
\usepackage{xspace}
\newcommand{\opdf}{\textsc{opdf}\xspace}
\newcommand{\empdf}{\textsc{emPDF}\xspace}
\newcommand{\gaia}{\textit{Gaia}\xspace}

\newcommand{\refsec}[1]{\S\ref{#1}}
\newcommand{\reffig}[1]{Fig.~\ref{#1}}
\newcommand{\reftab}[1]{Table~\ref{#1}}
\newcommand{\refeqn}[1]{Eq.~(\ref{#1})}
\newcommand{\refeqnalt}[1]{Eq.~\ref{#1}}

\newcommand{\dif}{\mathrm{d}}
\newcommand{\abs}[1]{\left\lvert #1 \right\rvert}          % for absolute value
\newcommand{\avg}[1]{\left\langle #1 \right\rangle}        % for average
\renewcommand{\mid}{\,|\,}

\newcommand{\rvir}{R_\mathrm{vir}}

\newcommand{\Mh}{M_\mathrm{h}}
\newcommand{\Rh}{R_\mathrm{h}}

\newcommand{\msun}{M_\odot}
\newcommand{\kpc}{\mathrm{kpc}}
\newcommand{\rs}{r_\mathrm{s}}
\newcommand{\vs}{v_\mathrm{s}}
\newcommand{\rhos}{\rho_\mathrm{s}}

\newcommand{\robsmax}{r_{\mathrm{obs,max}}}
\newcommand{\robsmin}{r_{\mathrm{obs,min}}}
\newcommand{\avec}{\bm{\alpha}}

\graphicspath{{fig1/}}

\title[Empirical distribution function]{\empdf: Inferring the Milky Way mass with data-driven distribution function in phase space}

\author[Li et al.]{%
Zhaozhou Li,$^{1, 2, 3}$\thanks{\href{mailto:zhaozhou.li@mail.huji.ac.il}{zhaozhou.li@mail.huji.ac.il}; \href{mailto:lizz.astro@gmail.com}{lizz.astro@gmail.com}}\thanks{Marie Skłodowska-Curie Fellow}
Jiaxin Han,$^{1, 3, 4}$ %\thanks{jiaxin.han@sjtu.edu.cn}
Wenting Wang,$^{1, 3, 4}$
Yong-Zhong Qian,$^{5}$
Qingyang Li,$^{1, 3, 4}$\newauthor
Yipeng Jing,$^{1, 3, 4, 6}$
Ting S. Li$^{7}$
\\
$^{1}$ Department of Astronomy, Shanghai Jiao Tong University, Shanghai 200240, China\\
$^{2}$ Centre for Astrophysics and Planetary Science, Racah Institute of Physics, The Hebrew University, Jerusalem, 91904, Israel\\
$^{3}$ Shanghai Key Laboratory for Particle Physics and Cosmology, Shanghai 200240, China\\
$^{4}$ Key Laboratory for Particle Astrophysics and Cosmology (MOE), Shanghai 200240, China\\
$^{5}$ School of Physics and Astronomy, University of Minnesota, Minneapolis, MN 55455, USA\\
$^{6}$ Tsung-Dao Lee Institute, Shanghai Jiao Tong University, Shanghai, 200240, People’s Republic of China\\
$^{7}$ Department of Astronomy and Astrophysics, University of Toronto, 50 St. George Street, Toronto ON, M5S 3H4, Canada\\
}

\pubyear{2024}

\begin{document}
\label{firstpage}
\pagerange{\pageref{firstpage}--\pageref{lastpage}}
\maketitle

\begin{abstract} 
We introduce the \empdf (Empirical Distribution Function), a novel dynamical modeling method that infers the gravitational potential from kinematic tracers with optimal statistical efficiency under the minimal assumption of steady state. 
\empdf determines the best-fit potential by maximizing the similarity between instantaneous kinematics and the time-averaged phase-space distribution function (DF), which is empirically constructed from observation upon the theoretical foundation of oPDF (Han et al. 2016).
This approach eliminates the need for presumed functional forms of DFs or orbit libraries required by conventional DF- or orbit-based methods.
\empdf stands out for its flexibility, efficiency, and capability in handling observational effects, making it preferable to the popular Jeans equation or other minimal assumption methods, especially for the Milky Way (MW) outer halo where tracers often have limited sample size and poor data quality.
We apply \empdf to infer the MW mass profile using \gaia DR3 data of satellite galaxies and globular clusters, obtaining 
enclosed masses of $M({<}r){=}26{\pm}8, 46{\pm}8, 90{\pm}13$, and $149{\pm}40\times10^{10}\msun$
at $r{=}30, 50, 100$, and 200 kpc, respectively.
These are consistent with the updated constraints from simulation-informed DF fitting (Li et al. 2020).
While the simulation-informed DF offers superior precision owing to the additional information extracted from simulations, \empdf is independent of such supplementary knowledge and applicable to general tracer populations. 
\empdf is currently implemented for tracers with complete 6D kinematics within spherical potentials, but it can potentially be extended to address more general problems.
\end{abstract}

\begin{keywords}
Galaxy: halo -- Galaxy: structure -- Galaxy: kinematics and dynamics -- galaxies: dwarf -- dark matter
\end{keywords}

%%%%%%%%%%%%%%%%%%%%%%%%%%%%%%%%%%%%%%%%%%%%%%%%%%

%%%%%%%%%%%%%%%%% BODY OF PAPER %%%%%%%%%%%%%%%%%%

\section{Introduction}
\label{sec:intro}

\defcitealias{Binney2008a}{BT08}

Gravity is the predominant force in shaping cosmic structures.
A fundamental task in galactic dynamics is dynamical modeling,
determining the underlying mass distribution (equivalently gravitational potential $\Phi$) from the kinematics of a set of tracers assumed to be in equilibrium.
The mass distribution is typically characterized by some parameters to be estimated.
The \gaia mission (\citeyear{GaiaCollaboration2016}) has recently made it possible to acquire nearly complete 6D kinematic data (positions ${\bm{r}}$ and velocities ${\bm{v}}$) for tracer populations of the Milky Way (MW). This significant advancement not only enables more accurate measurements of the MW's mass distribution (see \citealt{Wang2019b} for a comprehensive review), but also necessitates developing new methodologies to fully exploit the data.

The 6D phase-space distribution function (DF) provides a complete description of steady-state systems \citep[\citetalias{Binney2008a} hereafter]{Binney2008a}.
Ideally, if we have a model for the expected DF of tracers in arbitrary potential, $f({\bm{r}}, {\bm{v}} \mid \Phi)$, the best-fit potential can be derived with optimal statistical efficiency. 
Most dynamical modeling methods rely on DFs in some form, although some methods use only part of the information contained in the full DF. For example, the \emph{Jeans equation} uses moments of the DF (tracer density and velocity dispersion profiles), thereby losing information from higher-order moments \citep[see][]{Lokas2003, Evslin2017, Read2020}. This limitation also applies to mass estimators based on the Jeans equation (aka. virial estimators, e.g., \citealt{Bahcall1981, Walker2009a, Wolf2010a, Watkins2010a}). Similarly, the \emph{escape velocity} method only considers a small fraction of stars in the tail of the velocity distribution \citep{Caldwell1981, Leonard1990}. Therefore, using the full DF is expected to provide tighter constraints. Moreover, observational errors and incompleteness can be naturally incorporated into DF models via forward modeling, which is nontrivial for many other methods.

Yet finding the appropriate DF model for a given tracer population is a challenging task. 
Various approaches exist to construct DF models analytically, including the \citet{Eddington1916} inversion of density profiles with assumed velocity anisotropy \citep[e.g.,][]{Osipkov1979, Merritt1985, Cuddeford1991, Gerhard1991, Wojtak2008} and analytical functions of orbital integrals (e.g., energy and angular momentum:  \citealt{Michie1963, Kent1982, Evans1997,Gieles2015}; actions: \citealt{Binney2011, Ting2013, Binney2014, Posti2015, Trick2016, Vasiliev2019, Binney2023}). 
Several DF models have been employed to infer the MW mass \citep[e.g.,][]{Wilkinson1999, Eadie2018, Deason2020b, Wang2022c, Slizewski2022, Shen2022}.
A major concern of such presumed analytical DFs is that they may not accurately represent specific tracer populations, potentially leading to substantially biased results \citep{Wang2015b,Han2016a}.

A more realistic approach uses template DFs derived from numerical simulations. 
\citet{Li2017} developed a method to rescale the orbital distribution of satellite galaxies in simulations to different halo masses, which was applied to estimate the MW halo mass by \citet{Callingham2018}. However, this method is biased and requires additional calibration because it uses the likelihood of orbits instead of direct observables $({\bm{r}}, {\bm{v}})$. \citet{Li2019} resolved this issue by further deriving the full phase-space DF arising from the satellite orbits under the steady-state assumption~\citep{Han2016b}, with which \citet{Li2020a} obtained
one of the most reliable mass constraints on the MW outer halo. The MW virial mass is inferred with a statistical uncertainty of 20\% and systematics $\lesssim 10\%$. 
While this technique offers optimal statistical efficiency (see \refsec{sec:benchmark}), it is limited to tracers that are well-understood in simulations, such as satellites. 

Concerns about model dependence in the above methods have prompted the development of more flexible DF models with large degrees of freedom to adequately describe realistic tracers. 
The \emph{orbit superposition} method describes the DF using a library of discrete orbits \citep{Schwarzschild1979, vandenBosch2008a,McMillan2012, Zhu2018b, Vasiliev2020} or a series of basis functions (e.g., \citetalias{Binney2008a}, eq. 4.116, \citealt{Bovy2010, Magorrian2014, Magorrian2019}) with adjustable weights.
Recently, machine learning techniques such as normalizing flows have been introduced to construct highly flexible models \citep{Green2020, Green2022, An2021, Naik2022, Buckley2023,Lim2023}.
In general, such flexible models with numerous free parameters are computationally demanding and often rely on ad-hoc regularization to mitigate parameter degeneracy.

Another appealing approach is to use the observed tracers themselves as the orbit library, thus eliminating the need for assumptions on the functional form of the DF 
and avoiding ad-hoc construction and regularization of orbit libraries. 
The main idea is that a steady-state system does not evolve over time when moving the tracers along their orbits under the correct potential.
Therefore, the similarity between the instantaneous observation and time-averaged statistics in a trial potential represents its goodness of fit. 
As steady-state is the only essential assumption here, such methods can be dubbed as \emph{minimal assumption} methods.
An example is the orbital roulette method (\citealt{Beloborodov2004}, also cf. \citealt{Price-Whelan2021}), which compares the phase angle distribution of tracers with a uniform distribution (as expected for time average).
A natural step forward is to compare the full 6D DF with the maximal available information.

Taking the time average is equivalent to dispersing each particle according to the travel time distribution along its orbit, aka. the orbital Probability Density Function (oPDF, \citealp{Han2016b}), $\mathrm{d}P(\lambda\mid\mathrm{orbit})\propto \mathrm{d}t$, where $\lambda$ denotes position in orbit. 
\citet{Han2016b} showed that the oPDF principle is the microscopic equivalence to the macroscopic condition of steady-state, the Jeans theorem, or the so-called ``random phase principle''~\citep[where the phase angle is a proxy for the travel time]{Beloborodov2004}. 
Summing up the oPDFs sampled along each tracer's orbit then provides an empirical realization of the time-averaged 6D DF (see also \refsec{sec:empdf}). 
However, the discreteness in the distribution of tracer orbits prevents direct comparison between this DF and observation.
To circumvent this difficulty, the \opdf method%
\footnote{We use oPDF to refer to the conditional distribution, $\mathrm{d}P(\lambda\mid\mathrm{orbit})\propto \mathrm{d}t$, while \opdf refers to the radial likelihood method based on the oPDF principle in \citet{Han2016b}.}
by \citet{Han2016b} marginalizes orbital distribution and considers the likelihood between the observed radial distribution of tracers and its time average.

In this work, we develop a new method, \empdf (empirical DF).
It is a natural extension of the \opdf method, but constructs a continuous empirical time-averaged DF nonparametrically from discrete tracers under the trial potential,
thus allowing direct comparison with observations in 6D kinematics.
The key difference from \opdf is that it replaces the discrete orbit distribution in \opdf with a smooth distribution using kernel density estimation (KDE).
The likelihood between the empirical DF and the observation sample then enables Bayesian inference for the true potential well.
As shown in this paper, the new \empdf method is intuitive, efficient, flexible, and simple to implement.

\empdf shares some spirits with the orbit superposition methods, especially \citealt{Magorrian2014} (see comments in Appendix \ref{sec:other_de} and \citealt{Han2016b}, \S6.3), but is much easier to implement and compute.
Unlike the orbit superposition methods, \empdf does not involve free parameters such as adjustable weights for the orbital library or any ad-hoc regulating procedures.

The high efficiency of \empdf enables its application to very small tracer samples with reasonable precision. 
As a DF method, it also facilitates the handling of the observational selection function.
For the first demonstration, we apply this method to infer the MW mass using globular clusters and satellite galaxies as tracers (\refsec{sec:application}). 

The layout of this paper is as follows. 
We present the \empdf method in \refsec{sec:model}.
Its validity and performance are evaluated using mock samples and compared to other methods in \refsec{sec:validate}.
We apply \empdf to estimate the MW mass profile in \refsec{sec:application}.
Possible future extensions are discussed in \refsec{sec:discuss}, followed by conclusions in \refsec{sec:conclusion}.

Throughout this paper, the abbreviation ``DF'' exclusively represents the
6D phase-space \emph{{{distribution function}}}, $f({\bm{r}}, {\bm{v}})$. 
For the probability distribution of other quantities ($X$), we simply use ``{{{distribution}}}'', denoted as $p(X)\equiv\dif P/\dif X$.
Specifically, the \emph{orbital distribution} refers to the distribution of orbits in terms of orbital integrals, e.g., $p(E,L)\equiv \dif^2 P/\dif E\dif L$ and should not be confused with oPDF.
We denote ${\bm{w}} \equiv ({\bm{r}}, {\bm{v}})$ for the 6D phase space coordinates.
The halo mass $\Mh$ refers to the \emph{total} mass (including  baryons) enclosed by the radius $\Rh$, within which the average density is 200 times the critical density of the present universe, $\rho_{\rm crit}={3H_0^2}/({8\pi G})$, 
with $H_0=70\,\mathrm{km\,s^{-1}Mpc^{-1}}$ and $G$ the gravitational constant.
We use a more specific notation $M_\mathrm{200c}$ when referring to $\Mh$ of the MW for clarity.

\section{Dynamical modeling with empirical DF}
\label{sec:model}

We aim to infer the potential $\Phi (r)$, which is specified by a set of
free parameters ${\bm{\Theta}}_{\Phi}$, from
the 6D kinematics $\{{\bm{r}}_i,
{\bm{v}}_i \}_{i = 1}^N$ of a tracer sample observed within a radial
range $r_{\min} < r < r_{\max}$.

\begin{figure}
  \centering
  \includegraphics[width=1\columnwidth]{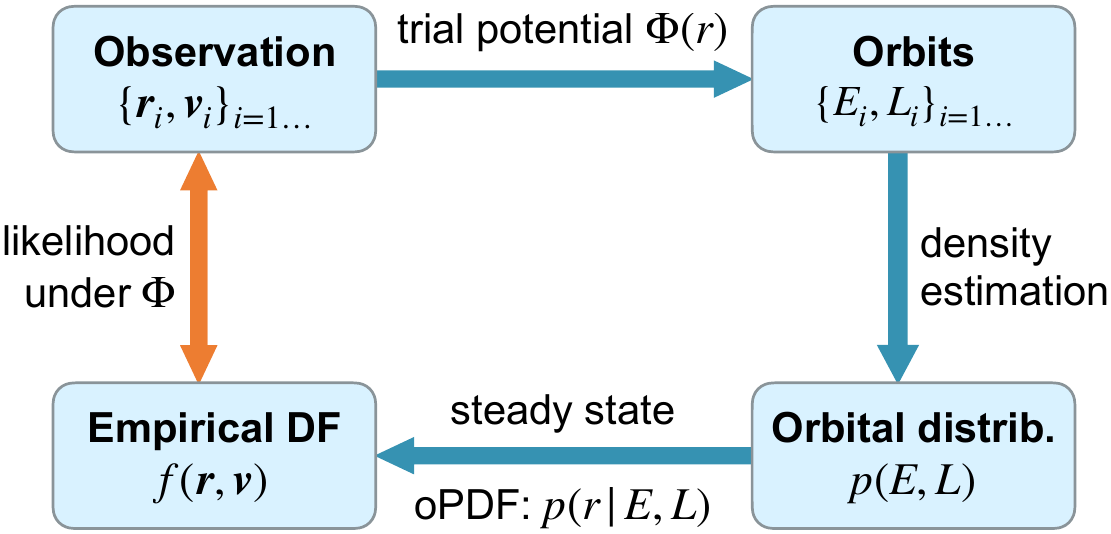}
  \vspace{-1em}
  \caption{
  Flowchart of the \empdf method.
  The best-fit parameters of the potential can be inferred by maximizing the likelihood (minimizing KL divergence) between the observed instantaneous kinematics, $\{{\bm{r}}_i, {\bm{v}}_i \}$, and the time-averaged DF, $f(\bm{r}, \bm{v})$, constructed empirically from observation under the trial potential. Other orbital integrals such as actions can be used instead of $(E,L)$ to extend this method for non-spherical systems.
  }
  \label{fig:flowchart}
\end{figure}

\begin{figure*}
  \includegraphics[width=0.95\linewidth]{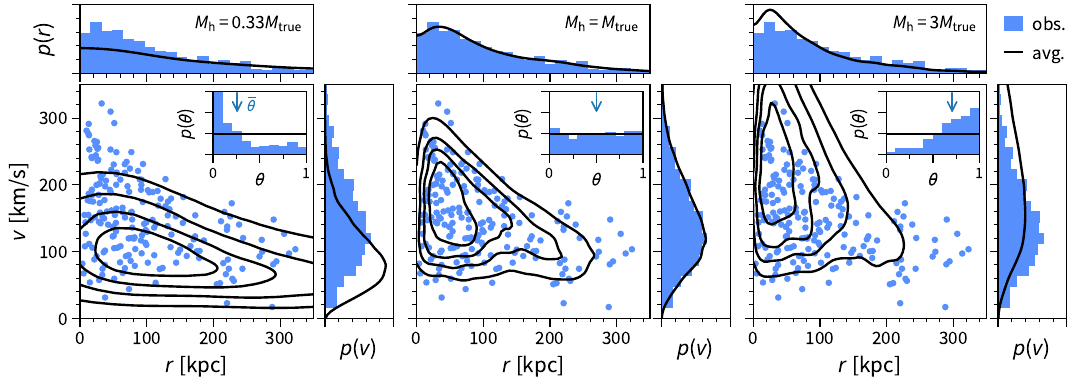}
  \vspace{-0.5em}
  \caption{Conceptual illustration of the empirical time-averaged distribution.
  The instantaneous kinematics of mock tracers (blue, same in three panels) are compared to their expected
  time-averaged distribution (black curves) in different trial potentials (panels).
  The true potential is adopted in the middle panel, and a smaller (larger) trial halo mass in the left (right) panel.
  The projected kinematics, $p(r, v)$, $p(r)$, $p(v)$, and $p({\theta})$, are displayed in the main, upper, right, and inset subpanels, respectively.
  Here, $r$ and $v$ are the Galactocentric distance and velocity, and $\theta \in [0, 1]$ is the normalized radial phase angle ($\theta=0$ and 1 for peri- and apocenter).
  In the main subpanels ($v$ vs $r$), the blue dots represent tracers (only 200 of 1000 are shown for clarity),
  while the black contours show the 20, 40, 60, 80th percentiles of the smoothed time-averaged distribution.
  In each inset subpanel, the arrow indicates the mean $\theta$.
  As expected, the time-averaged distribution matches best with the instantaneous observation across all dimensions under the correct potential.
  The proposed \empdf method uses the likelihood of the complete 6D DF, $f(\bm{r},\bm{v})$, to infer the best-fit parameters of the potential.
  }
  \label{fig:illustrate}
\end{figure*}

Same as \opdf, the \empdf method is based on the minimal common
assumptions that the tracers are in \emph{{{steady state}}} and the
potential is \emph{{{static}}}. While the general idea of \empdf can be applied to
less symmetrical potentials (see \refsec{sec:discuss}), 
this paper focuses on \emph{{{spherical potentials}}},
where the tracers therein can nevertheless follow arbitrary symmetry or velocity anisotropy.

The main idea of \empdf is that the underlying DF of tracers does not change over time under the steady-state assumption. 
If we evolve the tracers along their orbits in the \emph{{{correct}}} potential, the kinematics at an arbitrary epoch or the time-averaged distribution is expected to be statistically the same as the observed instantaneous kinematics. 
Therefore, we can infer the potential through the following procedures (see flowchart in \reffig{fig:flowchart}):
\setlist{nosep}
\begin{itemize}[leftmargin=*]
  \item Construct the time-averaged DF of the tracers empirically in given trial potential (\refsec{sec:empdf});
  \item Find (the parameters of) the best-fit potential that maximizes the similarity between this empirical DF and the observation (\refsec{sec:infer}).
\end{itemize}
Taking advantage of the complete DF, observational selection (i.e., incompleteness, \refsec{sec:selectfunc}) and measurement errors (\refsec{sec:obs_err}) can be incorporated when necessary.

To illustrate this idea, 
we generate 1000 mock tracers in equilibrium for an NFW halo \citep{Navarro1996} of $\Mh=10^{12}\msun$ with a fixed concentration $c=10$ (see \refsec{sec:mock}).
We then evolve these tracers in different trial potentials and compute their time-averaged statistics. % over 10 virial time using Agama \citep{Vasiliev2019}.
As expected, the time-averaged distribution in the {correct potential} matches best with the instantaneous observation across all dimensions shown in \reffig{fig:illustrate}.
In contrast, when placed in a potential shallower (deeper) than the true one, the tracers appear to crowd around their pericenter (apocenter) with a non-uniform phase angle distribution and tend to expand (contract) in space over time. 

The time-averaged DF can be compared to observations in different dimensions,
such as the mean phase angle, $\avg{\theta}$,
the phase angle distribution \citep[orbital roulette]{Beloborodov2004}, $p (\theta)$, 
the spatial distribution \citep[\opdf]{Han2016b}, $p (r)$,
and the 6D DF (\empdf), $f(\bm{r},\bm{v})$.
We refer the readers to \citet{Han2016b} for a summary and public implementation\footnote{\url{https://github.com/Kambrian/oPDF}} of these minimal assumption methods except for \empdf.
The proposed \empdf is expected to have optimal efficiency of its kind by fully exploiting the complete 6D DF.

\subsection{Constructing the empirical DF}
\label{sec:empdf}

In a spherical potential $\Phi (r)$, an orbit is characterized by specific energy,
$E = \Phi (r) + \frac{1}{2} (v_r^2 + v_t^2) $, and angular momentum, $\bm{L} = \bm{r}\times\bm{v}$, where $v_r$ and $v_t$ are the radial and tangential velocities
respectively. When the tracer distribution is also spherically symmetric, only $L=\abs{\bm{L}}$ matters, and the phase space coordinate of a particle can be fully specified by $(r,E,L)$. One can also use other orbital integrals instead, such as actions (\refsec{sec:action}). The resultant DF $f ({\bm{r}}, {\bm{v}})$ does not depend on the particular choice.
Here we adopt $(E, L)$ for the ease of computing.

The normalized phase-space DF, $f ({\bm{r}}, {\bm{v}}) \equiv \dif^6 P / \dif^3 {\bm{r}}\dif^3 {\bm{v}}$, 
can be decomposed into two parts, the \emph{{{orbital distribution}}},
$p (E, L) \equiv \dif^2 P / \dif E \dif L$, and the \emph{conditional radial distribution} along each orbit, $p (r\mid E, L)$, as
\begin{gather}
  f ({\bm{r}}, {\bm{v}}) = \frac{| v_r |}{8 \pi^2 L} p (r\mid E, L) p (E, L),
  \label{eq:f1}
\end{gather}
where the first term on the right-hand side is the Jacobian due to change of
variables (see Appendix A of \citealt{Li2019}).

\citet{Han2016b} proposed to model the orbital distribution empirically as $p (E, L) = \frac{1}{N} \sum_{i = 1}^N \delta (E - E_i, L - L_i)$, 
where $(E_i, L_i)$ represent the $i$-th tracer under the trial potential, and $\delta$ is the Dirac delta function. 
However, this distribution cannot be directly applied in the likelihood due to its discreteness \citep[see][Fig. 2]{Han2016b}. 
In \empdf, we propose to estimate the underlying smooth distribution using \emph{{{kernel density estimation}}} (KDE, \citealt{Silverman1986}),
\begin{gather}
  p (E, L) = \frac{\sum_{i = 1}^N \omega_i \mathcal{K} (E - E_i, L - L_i)}{\sum_{i=1}^N
  \omega_i},
  \label{eq:nel}
\end{gather}
where $\omega_i$ is the weight of each tracer, which is 1 by default (but see \refsec{sec:selectfunc} when involving selection function), and $\mathcal{K}$ is the smoothing kernel
with kernel size automatically determined by \citet{Scott1979}'s rule (see technical details in Appendix \ref{sec:kde_odf}).%
\footnote{%
  \refeqn{eq:nel} is intended for conceptual understanding and slightly different from our actual implementation,
  where we instead perform KDE smoothing for $p(E, \varepsilon^2)$
  and then convert it to $p (E, L)$. Here $\varepsilon \in [0, 1]$ denotes the orbital circularity (see Appendix \ref{sec:kde_odf} for rationale).
}
The orbital distribution of $(E,L)$ is assumed to be moderately smooth yet may contain clumpy substructures,
which will be resolved naturally with sufficient samples.

Under the steady-state assumption, the probability of a particle appearing at a position $r$ along a given orbit is proportional to the time spent around (oPDF, \citealp{Han2016b}), $p (r\mid E,
L) d r \propto d t = d r / | v_r |$. Normalizing over a radial orbital period%
\footnote{Here $2\int_{r_{{\mathrm{per}}}}^{r_{{\mathrm{apo}}}} p (r|E, L)\dif r = 1$, because $v_r$ in \refeqn{eq:f1} and (\ref{eq:opdf}) can be either positive or negative at $r$, effectively contributing a factor of 2.}
yields
\begin{gather}
  p (r\mid E, L) = \frac{1}{| v_r | T_r},
  \label{eq:opdf}
\end{gather}
where $T_r (E, L) = 2 \int_{r_{{\mathrm{per}}}}^{r_{{\mathrm{apo}}}} dr / | v_r |$ is the radial period of an orbit moving between pericenter
and apocenter $[r_{{\mathrm{per}}}, r_{{\mathrm{apo}}}]$.
Starting from oPDF, another useful quantity is the radial phase angle defined through $\mathrm{d}\theta\propto\mathrm{d}t/T_r$, with $\theta$ following a uniform distribution.
As we limit the tracers to $r \in [r_{\min}, r_{\max}]$, one should interpret $r_{{\mathrm{per}}}$ as 
$\max\{r_{{\mathrm{per}}}, r_{\min}\}$ and $r_{{\mathrm{apo}}}$ as $\min \{r_{{\mathrm{apo}}}, r_{\max}\}$ 
\emph{throughout} this paper (see \citealt{Han2016b,Li2019}).%
  \footnote{Though not obvious, this treatment is statistically equivalent to the incompleteness correction in Equations (\ref{eq:s2}--\ref{eq:lik_selfunc}) in \refsec{sec:selectfunc}, as if truncating the parent sample of $r\in[0, \infty]$ to the range $[r_{\min}, r_{\max}]$.
}

Combining Equations (\ref{eq:f1}--\ref{eq:opdf}), the empirical DF of the tracer sample in a {trial} potential $\Phi$ writes
\begin{gather}
f_\Phi ({\bm{r}}, {\bm{v}}) = \frac{p (E, L)}{8 \pi^2 L T_r (E, L)},
  \label{eq:f2}
\end{gather}
which represents the expected time-averaged distribution of the tracers.
Intuitively, \refeqn{eq:f2} redistributes each tracer along its orbit according to the steady-state configuration (oPDF), 
effectively averaging the tracer distribution over time.
The DF is denoted as $f_\Phi$ with a subscript $\Phi$, indicating its dependence on the potential (via $E$ and $T_r$).
By construction, we have $\int_{r \in [r_{\min}, r_{\max}]} f_\Phi ({\bm{r}}, \bm{v}) \dif^3 {\bm{r}}\dif^3 {\bm{v}}=1$.
This DF is fully specified by $(E, L)$ without
additional dependence on ${\bm{r}}$ or
${\bm{v}}$, as per the Jeans theorem (\citetalias{Binney2008a}, sec. 4.2).

We emphasize that the empirical DF is completely \emph{data-driven}.
There are no free parameters and no assumptions about the specific form of the DF, density profile, velocity anisotropy, or separability of DF.
As \citet{Han2016b} commented,
any dynamical modeling of steady-state tracers must utilize the $p (r \mid E, L)$ information in some manner, though the way to handle $p(E,L)$ can vary.
There exists a continuous spectrum from fixed forms (analytical, simulation-informed) with strong prior information at risk of bias
to more flexible models (Schwarzschild, Magorrian) that typically have more free parameters and ad-hoc regulation.
A particularity of our \empdf is its position at the most flexible end allowing arbitrary DFs but without free parameters, aligning with the original \opdf method.
Moreover, it allows unbound orbits as long as tracers in such orbits satisfy the steady-state assumption according to \refeqn{eq:opdf} (e.g., in an open system with constant inflows and outflows; \citealt{Han2016b}).

\subsection{Inferring the potential}
\label{sec:infer}

Once the expected time-averaged DF is obtained empirically, we can compute the likelihood
between the DF and the observed instantaneous kinematics, as the goodness of the trial potential.
Maximizing the likelihood is equivalent to minimizing the
statistical distance, Kullback-Leibler divergence (aka. relative entropy, see Appendix \ref{sec:lik_KLdiv}),
between the sample and the DF.
As shown in \refsec{sec:action}, this maximum likelihood also corresponds to the minimum entropy and maximum clustering of the \emph{actions} of the sample.

Under a trial potential specified by parameters ${\bm{\Theta}}_{\Phi}$, the
probability of observing a tracer with ${\bm{w}_i} \equiv ({\bm{r}_i}, {\bm{v}_i})$ at an arbitrary epoch is
\begin{gather}
  p ({\bm{w}_i}\mid {\bm{\Theta}}_{\Phi}) = f_\Phi ({\bm{w}_i}),
  \label{eq:lik_single}
\end{gather}
% Note that we have $\int_{r \in [r_{\min}, r_{\max}]} p ({\bm{w}}|{\bm{\Theta}}_{\Phi}) \dif^6{\bm{w}}=1$ by construction.
The posterior probability of parameters for a tracer sample is then
\begin{gather}
  p ({\bm{\Theta}}_{\Phi} \mid  \{{\bm{w}}_i \}) 
  \propto p ({\bm{\Theta}}_{\Phi})  {\textstyle\prod\nolimits}_{i = 1}^N p ({\bm{w}}_i \mid  \Phi),
  \label{eq:posterior}
\end{gather}
where $p ({\bm{\Theta}}_{\Phi})$ on the right-hand side represents the prior
knowledge about the parameters (constant for flat priors). One can further obtain the Bayesian evidence by
integrating the right-hand side over the parameter space, which is useful for
model comparison (e.g., between the standard NFW potential and its contracted/cored
alternatives).

The best-fit potential is then given by maximizing the posterior.
We use the Bayesian optimization method
(\texttt{scikit-optimize}, \citealt{skopt})%
\footnote{\url{https://github.com/scikit-optimize/scikit-optimize}}
to search the best-fit parameters. 
To obtain the uncertainty and joint distribution of parameters, the common practice is to use grid scanning (efficient for one or two free parameters) or Monte Carlo sampling 
\citep[e.g.,][]{Foreman-Mackey2013,Speagle2019}, which are however computationally expensive (recall that computing $T_r$ and selection function in \empdf involve integrations).
More efficient modern techniques include the Bayesian active learning {\citep[e.g.,][]{Kandasamy2015,Acerbi2018,ElGammal2022}},
% \footnote{\url{https://github.com/jonaselgammal/GPry}}, 
which can reduce the computational cost by two orders of magnitude or more.

Tests with mock samples indicate that the formal error of parameters reported by our implementation
slightly underestimates the actual uncertainty, likely due to systematics in the KDE technique.
The plain KDE with a constant kernel always overestimates the density around low-density points, especially with smaller kernel sizes.
It thus tends to overestimate the likelihood near the true potential, where the kernel size is often smaller.
This inconsistency disappears when using the likelihood between the mock tracers and an empirical DF constructed from a large independent sample drawn from the same parent tracer distribution (which is known for mock samples).
This suggests that the underestimated formal error is technical rather than physical.
A similar bias has been reported for density estimation based on the $k$-th nearest neighbor method \citep[e.g.,][]{Silva2024}.
A more rigorous formalism should account for such KDE uncertainties.
We leave possible technical refinements to future work but adopt an empirical correction in this paper with plain KDE. Practically, we find that using $\ln \mathcal{L}=0.6 \ln p ({\bm{\Theta}}_{\Phi} \mid  \{{\bm{w}}_i \})$ instead of the original $\ln p ({\bm{\Theta}}_{\Phi} \mid  \{{\bm{w}}_i \})$ during parameter inference manually increases the formal uncertainty and provides a more accurate estimate of the confidence region.

\subsection{Incorporating selection function}
\label{sec:selectfunc}

Real data are often incomplete due to observational limitations, such as insufficient survey depth and sky coverage.
The selection function, $S ({\bm{w}}) \in [0, 1]$, describes the completeness among all tracers at given phase-space locations. 
For example, $S (r) = 0.5$ indicates that a tracer can only be observed/selected with half probability at $r$.
The selection function often depends on luminosity and possibly other intrinsic properties such as color. 
It thus can vary for individual tracers, $S(\bm{w}\mid\avec_i)$, where $\avec_i$ denotes the property vector of the $i$-th tracer that affects the selection function.

While an ensemble angular selection does not affect dynamical modeling in spherical potentials, the radial incompleteness breaks the steady state and potentially introduces severe bias. 
Fortunately, it is straightforward to correct the selection function in DF methods \citep[e.g,][]{Han2016b,Li2019,Vasiliev2019}.

The selection function affects our method in two aspects.

1. Each tracer is a biased representative of its orbit $(E_i, L_i)$. 
To account for the undetected tracers and construct the underlying orbital distribution of the complete sample, 
in \refeqn{eq:nel} we reweigh each observed tracer with
\begin{gather}
\omega_i = \dfrac{1}{\int p ({\bm{w}}'\mid E_i, L_i) S ({\bm{w}}'\mid\avec_i) \dif^6{\bm{w}}^{\prime}} \geqslant 1.
\label{eq:w1}
\end{gather}
If not specified otherwise, the integration over the phase space is performed 
for the radial range $[r_{\min}, r_{\max}]$ and arbitrary velocities.%
\footnote{If $S\neq0$ is guaranteed in the space of concern, a simpler alternative is $\omega_i = 1/S({\bm{w}_i})$,
which is {statistically} equivalent to \refeqn{eq:w1} for the whole sample.} 

2. A tracer observed through its selection function follows a distribution different from the underlying DF. 
The likelihood then should be computed using
\begin{gather}
p ({\bm{w}_i} \mid {\bm{\Theta}}_{\Phi},\avec_i) = 
    \dfrac{f_\Phi ({\bm{w}_i}) S ({\bm{w}_i\mid\avec_i})}
          {\int_{} f_\Phi ({\bm{w}}') S ({\bm{w}}'\mid\avec_i) \dif^6{\bm{w}}^{\prime}} 
\label{eq:p1}
\end{gather}
instead of \refeqn{eq:lik_single}.%
\footnote{\refeqn{eq:p1} assumed that the underlying dynamics (DF) is independent of the intrinsic properties $\avec$, $f_\Phi(w\mid\avec)=f_\Phi(w)$, which is a fairly general assumption for many tracer populations.
When necessary, it is possible to extend \empdf to the joint distribution of $p(\bm{w},\avec)$.}

In many cases, the selection function undergoes a sharp transition from $S = 1$
to $0$ around a certain distance (e.g., the detectability of dwarf galaxies in flux-limited surveys, 
see \refsec{sec:obs_select}), 
which can be approximated by a simple radial selection,
\begin{gather}
  S (r \mid \avec_i) = \left\{ \begin{array}{ll}
    1, & \textrm{} r \in [\robsmin, \robsmax],\\
    0, & \mathrm{otherwise},
  \end{array} \right.
\label{eq:s2}
\end{gather}
where $[\robsmin, \robsmax]$ is the observable radial range of a given tracer.
Note that this observable range depends on $\avec_i$.
The weight then becomes
\begin{gather}
  \omega_i =
  \frac{1}
    {\int_{\robsmin}^{\robsmax} 
        p (r\mid E_i, L_i) \dif r} =
  \left.
  \frac{T_r [r_{{\mathrm{per}}}, r_{{\mathrm{apo}}}]}
    {T_r [\robsmin, \robsmax]}
  \right|_{E_i, L_i},
\label{eq:w2}
\end{gather}
where $T_r [a, b] \equiv 2 \int_{\max\{a,r_\mathrm{per}\}}^{\min\{b,r_\mathrm{apo}\}} \dif r / | v_r |$ is the time spent within the radius range for given orbit.
The likelihood of \refeqn{eq:lik_single} is rewritten as
\begin{gather}
  p ({\bm{w}}_i \mid {\bm{\Theta}}_{\Phi}, \avec_i) =
\cfrac{f_\Phi ({\bm{w}}_i)}{\int_{r_{\mathrm{obs},
\min}}^{r_{\mathrm{obs}, \max}} \varrho (r') {4 \pi r'}^2 \dif r'},
\label{eq:lik_selfunc}
\end{gather}
where $\varrho (r) = \int f_\Phi(E,L) 2\pi v^2\dif v\dif \cos\vartheta$ is the tracer number density profile of the underlying sample with $\cos\vartheta\equiv {\hat{\bm{r}} \cdot \hat{\bm{v}}} \in [-1, 1]$ and $L=\allowbreak r v \sin \vartheta$.

In practice, when selection function is involved,
we suggest only using tracers
more luminous than certain absolute magnitude threshold, 
thus excluding tracers with $r_{\mathrm{obs}, \max}$ comparable to
$r_{\min}$ of the sample  (see \refsec{sec:obs_data} for example).
This is because such tracers will be assigned very high weights
by \refeqn{eq:w2}, which may lead to nonrobust estimates
dictated by a few high-weight tracers.

\subsection{Treating observational errors}
\label{sec:obs_err}

Another important observational effect is the measurement error.
Intriguingly, \empdf is robust against moderate observational errors even without any special treatment.
As shown in \refsec{sec:mock_err} and \reffig{fig:benchmark_obserr},
for hundreds of tracer with typical \gaia DR3 observational errors for satellite galaxies or globular clusters, the systematic bias in mass estimates introduced by these errors is much smaller than the statistical uncertainty and can thus be safely ignored.
Therefore, it is not necessary to treat the observational errors explicitly for these tracers when inferring the MW mass.

However, more careful consideration might be necessary for tracers like halo stars due to their larger measurement errors and smaller statistical uncertainty given their large number. 
Overlooking observational errors may lead to artificial deviation from the steady state and biases in mass estimates.
Similar to the selection function in \refsec{sec:selectfunc}, observational errors affect our method in two aspects: (a) construction of empirical DF and (b) likelihood under this DF.

(a) It is possible to derive the underlying ``true'' DF, $f_\Phi ({\bm{w}})$,
by deconvolving the observational errors through a novel technique
based on {iterative reweighted importance sampling} (IRIS)
when performing the KDE in \refeqn{eq:nel} (see Appendix \ref{sec:error_corr} for details).

(b) The likelihood of an observed tracer ${\bm{w}_i}$ (\refeqnalt{eq:lik_single})
can be obtained by marginalizing over the possible true ${\bm{w}}_{\mathrm{tr}}$ under the observational error $p_{\mathrm{err}} ({\bm{w}_i}\mid {\bm{w}}_{\mathrm{tr}})$,
\begin{gather}
  \textstyle p_{\mathrm{ob}} ({\bm{w}_i}\mid {\bm{\Theta}}_{\Phi}) = 
  \int p_{\mathrm{err}} ({\bm{w}_i}\mid {\bm{w}}_{\mathrm{tr}}) f_\Phi ({\bm{w}})
   \dif^6{\bm{w}_\mathrm{tr}}.
\label{eq:mc_int_err}
\end{gather}
This can be computed efficiently using the Monte Carlo integration \citep[see e.g.,][\S3.2]{Li2020a}.
We note that $f_\Phi$ in above equation should be replaced with \refeqn{eq:lik_selfunc} when selection function is also involved.

A natural question might be why bother with deconvolving the observational scatter in step (a) if we will convolve it back in step (b).
While these two procedures partially cancel each other out, they do not do so entirely.
Nevertheless, this partial cancellation probably explains why this method is robust against moderate observational errors even without explicit correction.

\begin{figure*}
  \centering
  \includegraphics[width=0.95\columnwidth]{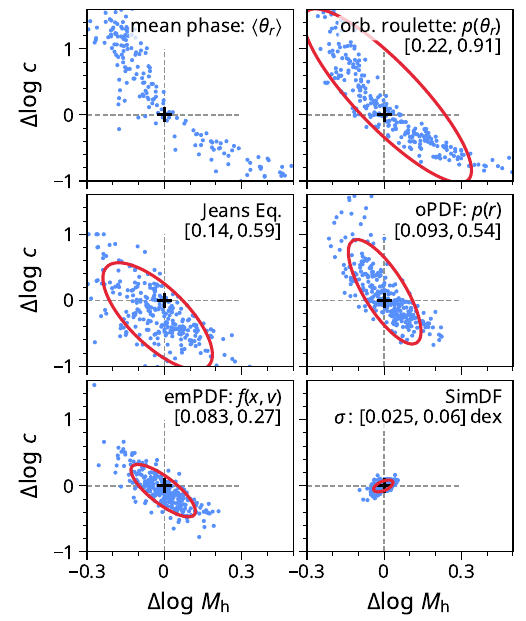}
  \hspace{1em}
  \includegraphics[width=0.95\columnwidth]{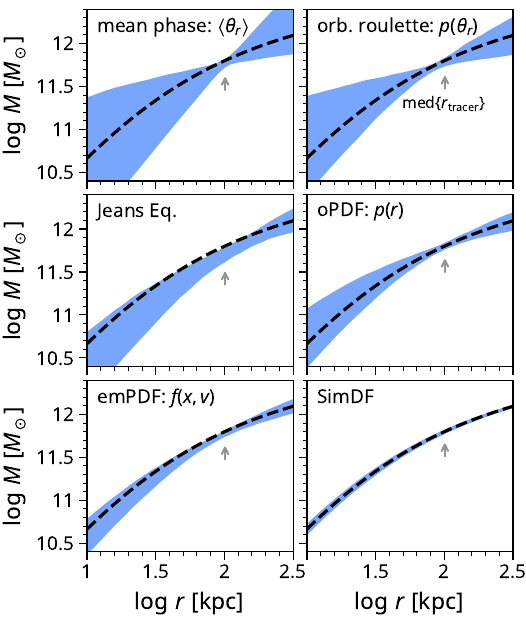}
  \vspace{-1em}
  \caption{
  Method comparison based on idealized mock samples.
  Results are shown for 300 mock halos, each with 160 tracers, using six methods separately, mean phase, orbit roulette \citep{Beloborodov2004}, spherical Jeans equation, \opdf \citep{Han2016b}, \empdf (this work), and simulation-informed DF \citep{Li2017, Li2019}.
  \textbf{Left panels}: Best-fit parameters (blue points) relative to their true values for individual mock halos. 
  Each panel includes a 1$\sigma$ ellipse, indicating the mean and covariance of the points (not shown for ``mean phase'' with infinite inherent uncertainty).
  The root-mean-square error of $\log \Mh$ and $\log c$ respective to their true values are provided in the upper-right corner of each panel.
  \textbf{Right panels}: 16th and 84th percentiles (blue shades) of the 300 best-fit mass profiles $M(<r)$.
  The black dashed curves show the true mass profile,
  and gray arrows indicate the median distance of tracers for reference.
  The simulation-informed DF shows the highest efficiency due to strong assumptions about tracer DFs but is limited to specific tracers like satellite galaxies.
  Among other methods for general tracers, \empdf demonstrates optimal efficiency by fully exploiting the DF.
  }
  \label{fig:benchmark}
\end{figure*}

\section{Method validation and comparison}
\label{sec:validate}

\subsection{Mock tracer sample}
\label{sec:mock}

We test our method using idealized mock samples and compare with several other methods. 
We consider halos that follow the spherical NFW density profile \citep{Navarro1996},
whose mass profile and potential are specified by the virial mass and concentration $(\Mh, c)$,
\begin{align}
\rho(r \mid  \Mh, c) &= \frac{\rhos}{(r/\rs)(1+r/\rs)^2},\\
\Phi(r \mid  \Mh, c) &= -4\pi G \rhos \rs^2 \frac{\ln(1+{r}/{\rs})}{r/\rs},
\end{align}
where the characteristic density and radius, $\rhos$ and $\rs$, are determined by solving $\Mh=M(<\!\Rh)=\frac{4\pi}{3}200\rho_\mathrm{crit}\Rh^3$ and $\Rh=c \rs$.
The goal is to infer $(\Mh,c)$ from mock tracers.

We generate tracers for an NFW halo with $M_\mathrm{h} = 10^{12} M_{\odot}$ and concentration $c = 10$.
For simplicity, we assume that the tracer spatial density follows the same profile as the host density (but with a squared-exponential truncation for $r\gtrsim500 \kpc$ for sampling efficiency) and the tracers have isotropic velocity anisotropy. Although this setup may not hold for MW tracers in reality, it does not affect the generality of our test. 
Following \citet{Li2023}, the tracers are randomly sampled 
according to the theoretical DF using the \citet{Eddington1916} formula and the \textsc{Agama} package \citep{Vasiliev2019}\footnote{\url{https://github.com/GalacticDynamics-Oxford/Agama}}.
The tracers are guaranteed to be in steady state by construction. 
To mimic the MW halo tracers, we limit the mock observation to tracers in a radial range of $[20, 300]\kpc$. We try different numbers of tracers $N_\mathrm{tracer}$ ranging from 10 to 2560 in a mock halo. For each $N_\mathrm{tracer}$, we generate samples for 300 mock halos independently.

In this section, we aim to test the statistical efficiency of the method with idealized mock samples and gain theoretical understanding. 
However, we note that the systematic errors due to the deviations from the model assumptions (steady state and spherical potential) are not involved in this test but can be important in practice.
To estimate such systematics in future work,
more realistic mocks can be taken from cosmological simulations \citep[see e.g.,][]{Han2016a,Wang2017d,Li2017,Li2019,Li2021d,Wang2022f}.

We apply the \empdf method to the mock tracers and find the best-fit parameters $(\Mh,c)$ of the halo profile using Bayesian optimization. The parameters are searched within the range of $11<\log \Mh/M_\odot<13$ and $-1<\log c<3$.

\subsection{Other methods for comparison}

We compare the performance of \empdf with five other methods,
including mean phase, orbit roulette \citep{Beloborodov2004}, \opdf \citep{Han2016b}, spherical Jeans equation, and simulation-informed DF \citep{Li2017, Li2019}.

The first three methods share similar spirit as \empdf, as mentioned in \refsec{sec:intro}.
We refer readers to \citet{Han2016b} for a summary and public implementation.
Specifically, \opdf is based on binned statistics of the radial distribution of tracers.
Through numerical experiments, we find that the optimal number of bins is approximately $m=\max \{\ln (N), 2\}$.
Binning with equal counts slightly outperforms binning with equal widths.
The result becomes insensitive to these details when $N \gtrsim 200$.

It is worth noting that the individual parameters are completely unconstrained by the mean phase method, except for their joint distribution. For each mock halo, there is not a unique best-fit point but rather a degeneracy curve in $(\Mh, c)$ space, as illustrated by \citet{Han2016b}. The parameter search algorithm just returns a random single value along this degeneracy curve. The degeneracy leaves the mass profile unconstrained, except around the median tracer radius.
% \footnote{Due to the prior range set for parameter searching, this degeneracy is not so apparent (but still clear) in the right panel of \reffig{fig:benchmark}.}

We also include the widely used spherical Jeans equation in the comparison.
There are different ways to use the Jeans equation, ranging from purely data-driven to more parametric forms \citep[also dubbed as ``backward'' and ``forward'' approaches,][]{Rehemtulla2022}.
Here we adopt the backward form used by \citet{Li2021d} without assuming parametric profiles for tracer density and anisotropy (see Appendix \ref{sec:SJE} for details). 
The Jeans equation demands a relatively large sample ($N \gtrsim 100$) to ensure reasonable statistics, because it relies on the gradients of the tracer number density and velocity dispersion across adjacent radial bins. 
% While the performance of the Jeans equation can depend on the specific implementation (e.g., how gradients are computed), it should be less sensitive to the details when $N$ is large.

It is interesting to compare with the simulation-informed DF (SimDF) method based on the scaling relation of satellite kinematics in cosmological simulations \citep{Li2017,Li2019},
which has shown optimal statistical efficiency.
This method assumes that the scaled orbital distribution, $p(E/\vs^2, L/\rs\vs)$, is universal across halos, as a natural generalization to the universal NFW density profile. 
Here $\vs=\rs\sqrt{4\pi G\rhos}$, with $r_s$ and $\rho_s$ being the characteristic radius and density of the NFW profile.
Extracting this distribution from simulations and combining it with the oPDF, $p(r\mid E,L)$, yields a steady-state template DF, $f_0$.
The expected DF for a halo of arbitrary $\Mh$ and $c$ is then given by the scaling relation
\begin{equation}
  f(\bm{r}, \bm{v} \mid \Mh, c)= r_\mathrm{s}^{-3}  v_\mathrm{s}^{-3} f_0 (\bm{r}/ r_\mathrm{s}, \bm{v}/ v_\mathrm{s}),
\end{equation}
where $(\rs, \vs)$ are the new characteristic scales determined by $(\Mh, c)$ (see an illustration in \reffig{fig:illustrate_scaling}).
In the original papers, the template DF $f_0$ was constructed nonparametrically by rescaling and stacking satellites in cosmological simulations, 
following a similar procedure as described in \refsec{sec:empdf}.
In this idealized test, we adopt the true halo DF as the template to mimic its performance.
Clearly, unlike the other methods applicable to general tracers, the simulation-informed DF is limited to very specific tracers that are known to follow the scaling relation (e.g., satellite galaxies).
% also consult CorreaMagnus \& Vasiliev 2022 for treatment of unbound systems.

We refer interested readers to the literature \citep{Han2016b,Han2016a,Wang2017d,Wang2018a,Li2019,Li2021d} for previous tests of these methods with idealized and realistic tracers.

\subsection{Test results}
\label{sec:benchmark}

\begin{figure}
  \centering
  \includegraphics[width=1\columnwidth]{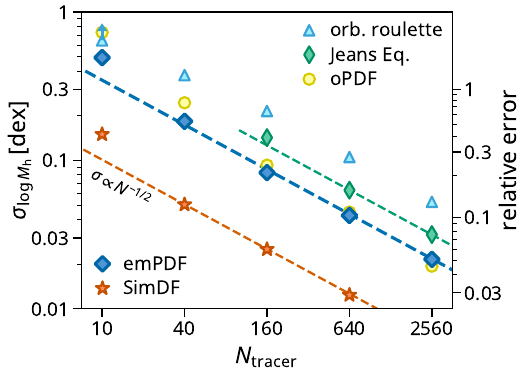}
  \vspace{-2em}
  \caption{
  Performance as function of tracer number.
  The symbols show the root-mean-square error of the best-fit $\log \Mh$ of 300 mock halos with varying number of tracers.
  The dashed lines display $\sigma\propto N^{-1/2}$ for reference.
  \empdf gives smaller uncertainty than orbit roulette and Jeans equation by a factor of 2.5 and 1.5, respectively.
  \opdf shows similar uncertainty on $\Mh$ as \empdf, but with a larger scatter 
  in the concentration and inner density profile as shown in \reffig{fig:benchmark}.
  The simulation-informed DF shows significantly smaller uncertainty owing to the additional model assumption.
  }
  \label{fig:ntracer}
\end{figure}

The left panel of \reffig{fig:benchmark} presents the best-fit parameters inferred from six different methods for 300 mock halos, each with $N_{\mathrm{tracer}}=160$ tracers. The right panel displays the corresponding best-fit profiles. In \reffig{fig:ntracer}, we illustrate the root-mean-square error, $\langle(x_\text{best-fit}-x_\text{true})^2\rangle^{1/2}$, of the best-fit $\log \Mh$ for 300 mock halos as a function of different $N_{\mathrm{tracer}}$.

As expected, \empdf provides the tightest constraints on both parameters among these methods (except for simulation-informed DF) due to its full exploitation of the DF. \empdf yields an uncertainty in $\log \Mh$ that is smaller by factors of 2.5 and 1.5 compared to the phase roulette and Jeans equation methods, respectively  (\reffig{fig:ntracer}). While \opdf shows similar uncertainty in $\Mh$ as \empdf, it exhibits a larger scatter in $c$, leading to a greater uncertainty in the inner halo mass profile (\reffig{fig:benchmark}). 
$\Mh$ alone does not fully describe the mass profile, because the essential constraint is for the entire profile or, equivalently, the joint distribution of $(\Mh, c)$.

As discussed in \citet{Han2016b}, the spatial distribution used by \opdf contains more information than the simple mean phase estimator or the phase angle distribution (orbit roulette). The difference also partly arises because the maximum likelihood estimator employed by \opdf is more efficient than the minimum distance estimator by orbital roulette. 
While \empdf shows better constraints, especially on $c$, than \opdf for $N_{\mathrm{tracer}}\lesssim 1000$, we find that their performance becomes similar for larger $N$. 
This suggests that \empdf does not gain much additional information from velocity components beyond the spatial distribution. 
This may not be surprising, considering that the position determines velocity in a given orbit and that the density profile largely determines the DF when the velocity anisotropy is known (implicitly provided here by the tracer sample).
At low $N_{\mathrm{tracer}}$, the better performance of \empdf largely benefits from smoothing compared to binned statistics in \opdf, which becomes less significant with large samples.
Nevertheless, besides its better performance for small samples, the complete DF provided by \empdf is useful as a complete description of tracer kinematics.

The Jeans equation demonstrates inferior efficiency compared to \empdf, as it uses only the first and second moments of the DF, thereby losing information in high-order moments.
In \reffig{fig:benchmark}, a mild bias is observed in the estimates from the Jeans equation, likely due to limited binned statistics given the relatively small sample size. 
We confirm that this bias diminishes with larger samples  (also cf. \citealt{Li2021d}, where a greater number of tracers was used).

Interestingly, except for the simulation-informed DF method, all methods show a similar ``sweet point'' around the median radius of the tracers, where the mass profile is best constrained with minimal uncertainty. This optimal constraint radius has been reported for various methods (see \citealt{Han2016b} for discussion) and has motivated the development of so-called Jeans estimators \citep[e.g.,][]{Walker2009a, Wolf2010a}. The uncertainty at the sweet point is similar for the four minimal assumption methods (mean phase, orbital roulette, \opdf, and \empdf), indicating a limit to the information that can be gained. This behavior explains the similar degeneracy in the inferred $(M, c)$ seen in the left panel of \reffig{fig:benchmark}. 
Nonetheless, the \empdf method provides significantly better constraints on the overall density profile.

The simulation-informed DF (simDF) method exhibits the highest efficiency, significantly surpassing the other five methods, owing to its strong assumptions about the specific form of the DF and potential, and the coupling between them through scaling relations. 
In this aspect, simDF resembles analytical DFs, both being characterized by a few parameters, although simDF is derived from simulations without an explicit analytical expression.
The additional coupling between DF and potential likely grants simDF even higher efficiency than analytical DF models that allow flexible combinations of DF and potential \citep[e.g.,][]{Posti2018,CorreaMagnus2022}.
However, simDF is limited to very specific tracers (satellites), as noted earlier, placing it in a distinct category.
In contrast, the \empdf method assumes only a functional form for the potential, leaving the DF ``free-form'', offering greater flexibility at lower efficiency.
Overall, the \empdf method offers theoretically optimal efficiency for \emph{general tracers} with minimal model assumptions.

\subsection{Tests with selection function and observational error}
\label{sec:mock_err}

\begin{figure}
  \centering
  \includegraphics[width=0.9\columnwidth]{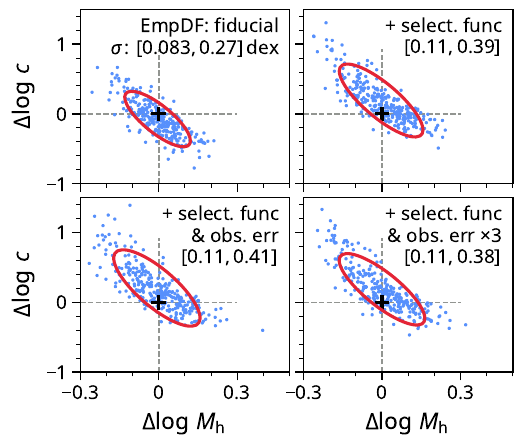}
  \caption{
  Similar to \reffig{fig:benchmark} left panels, but showing \empdf results for mock tracers with selection function and observational errors.
  \textbf{Upper left}: Fiducial samples without any observational effects, identical to the left panel of \reffig{fig:benchmark}.
  \textbf{Upper right}: Mocks with selection function.
  \textbf{Lower left}: Mocks with selection function and observational errors, mimicking the \gaia DR3 precision for MW satellites and globular clusters.
  \textbf{Lower right}: Same as previous, but with the observational errors tripled.
  We exert the selection function correction described in \refsec{sec:selectfunc} when applicable, but without any explicit treatment for observational error in this test.
  }
  \label{fig:benchmark_obserr}
\end{figure}

We test the validity of \empdf for tracers with observational effects. 
As detailed in Appendix \ref{sec:more_tests}, we generate mock data with a selection function and observational errors to mimic typical \gaia DR3 observations for MW dwarf satellites, where completeness and observational errors depend on the apparent magnitude of the tracers. 

We generate 300 mock halos, each with $N_\mathrm{tracer}=160$ tracers within $[20, 300]\kpc$. This sample size is higher than the actual number of available MW satellites, so any systematics will be more appreciable due to the reduced statistical error. We apply the selection function treatment described in \refsec{sec:selectfunc}, but do \emph{not} apply any explicit treatment for observational error. For reference, we also provide a fiducial mock sample with the same $N_\mathrm{tracer}$ and radial range, but without the observational effects. 

The results are shown in \reffig{fig:benchmark_obserr}, which generally justifies our correction for the selection function in \refsec{sec:selectfunc}.
Compared to the fiducial case, the uncertainty with incomplete data is slightly larger.
This increase is because of the reduced median distance of observed tracers, weakening constraints on the virial scale,
and the reduced effective tracer number due to unequal correction weights during the DF construction (see \refeqnalt{eq:w2}).
A slight bias appears in $c$, possibly also related to uneven correction weights. 
We leave possible technical improvements to future works.

Unmodeled observational errors may introduce systematic error into the inference in theory. However, in \reffig{fig:benchmark_obserr}, \empdf yields nearly identical results for mocks without and with observational errors, even when the observational errors are tripled. Clearly, this systematic error is much smaller than the statistical error for $N_\mathrm{tracer}=160$ (or smaller $N_\mathrm{tracer}$), and thus can be safely ignored without explicit treatment. The insensitivity to moderate observational error is noteworthy. As discussed in \refsec{sec:obs_err}, it is likely because both the instantaneous kinematics and the empirical DF are changed by observational errors in a similar way. Therefore, we do not apply any correction for observational errors in the next section, where satellite galaxies and globular clusters are used as MW tracers.

Further tests indicate that systematical errors due to unmodeled observational uncertainties become appreciable only for substantially larger sample size (reduced statistical error) or observational error (increased bias). In such cases, explicit modeling of observational error (see \refsec{sec:obs_err}) is necessary.

\section{Applying to the Milky Way}
\label{sec:application}

\begin{figure}
  \centering
  \includegraphics[width=0.9\columnwidth]{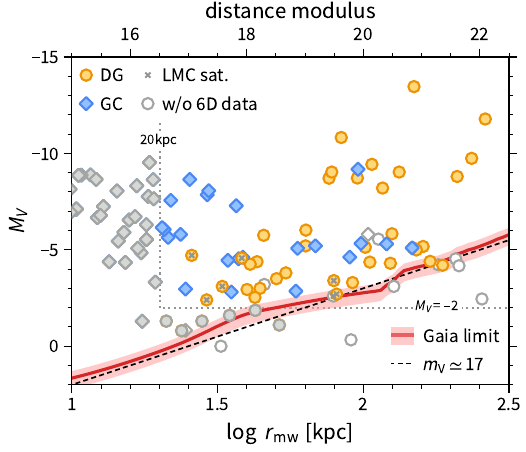}
  \caption{
  Galactocentric distance and absolute magnitude of MW tracers, including dwarf galaxies (DGs) and globular clusters (GCs).
  We select the tracers brighter than $M_V=-2$ within $[20, 300]\kpc$, as indicated by the dotted lines, tracers outside of which are shown in gray.
  Tracers without reasonable 6D kinematic measurements (open symbols) and possible LMC satellites (crosses) are excluded.
  The solid red curve shows the completeness distance $\robsmax(M_V)$ for \gaia limits, effectively distinguishing tracers with valid kinematic data from those without.
  For reference, the black dashed line indicates a constant apparent magnitude of $m_V=17$.
  Low-luminosity tracers are clearly incomplete at large distances.
  Thus, observed tracers represent a larger underlying sample.
  Correcting for selection function is crucial for unbiased dynamical modeling.
  }
  \label{fig:mw_tracers}
\end{figure}

\begin{figure*}
  \includegraphics[width=0.85\linewidth]{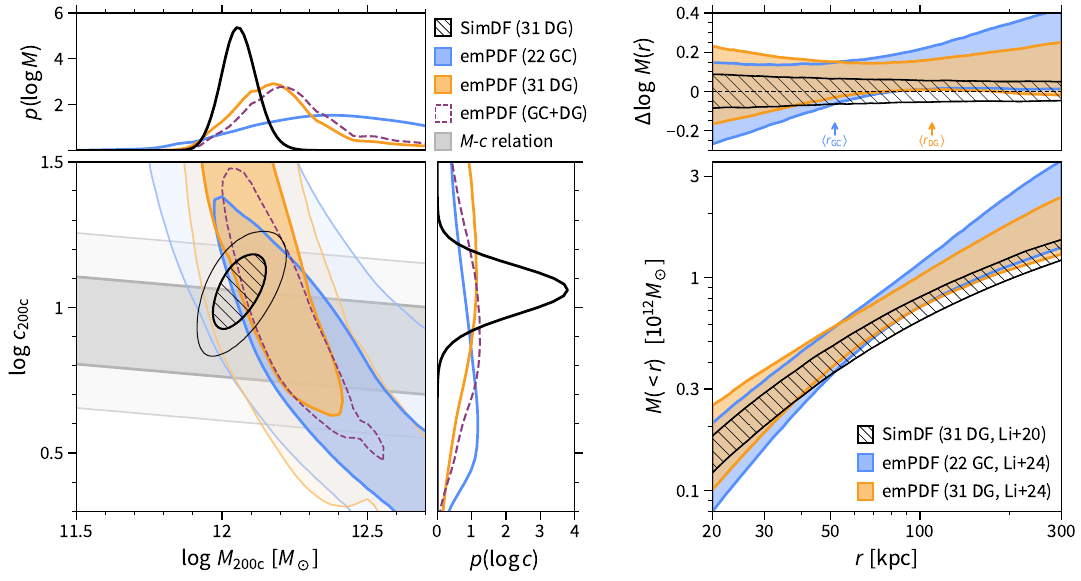}
  \caption{
  \textbf{Left}: MW halo mass and concentration inferred by \empdf using \gaia DR3 kinematics of DGs or/and GCs.
  The contours represent the $1\sigma$ (68.3\%) and $2\sigma$ (95.4\%) regions for different constraints (see legend),
  where flat priors for $\log M_\mathrm{200c}/\msun\in [11.5, 12.7]$ and $\log c_\mathrm{200c}\in [0.3, 1.5]$ are used.
  For clarity, only the $1\sigma$ contour is shown for combined constraints with DGs and GCs.
  An updated estimate using DGs and the simulation-informed DF \citep{Li2020a} is also shown for comparison.
  Marginalized distributions are displayed in the upper and right subpanels.
  The $M$-$c$ relation is shown in gray for reference but \emph{not} used.
  \textbf{Right}: 16th and 84th percentiles of best-fit MW profiles.
  The residuals relative to the best fit of the simulation-informed DF are shown in the upper subpanel.
  \empdf constraints with DGs and GCs exhibit slightly different degeneracy directions due to distinct spatial distributions, as indicated by their median radii (shown as arrows).
  The various constraints agree within $1\sigma$.
  }
  \label{fig:mw_mass}
\end{figure*}

We aim to measure the Milky Way (MW) mass profile using \empdf.
We assume the MW total mass distribution (including baryonic contribution) to be a spherical NFW profile, 
characterized by virial mass and concentration, $M_\mathrm{200c}$ and $c$.
Hydrodynamical cosmological simulations show that the total mass profile of MW-sized halos can be well approximated by the NFW profile for $r>0.05\rvir$ ($\sim\!10\kpc$ for MW) \citep[e.g., EAGLE,][]{Schaller2015}.
Though the inner halo is affected by the stellar disk, using a spherical potential for the outer halo is probably justifiable, especially recalling that the potential is always more spherical than the density distribution \citepalias{Binney2008a}. 
In the future, more flexible profiles can be considered \citep[e.g.,][]{Cautun2019,Dekel2017,Freundlich2020b} with additional stellar components.
Here we limit our analysis to the radial range $r\in[20,300]\kpc$. 

\subsection{Observation data}
\label{sec:obs_data}

In this work, we utilize the latest kinematic data of the MW tracers from \gaia DR3 
(\citeyear{GaiaCollaboration2022})
and complementary spectroscopic observations, including faint satellite dwarf galaxies (DGs, exclusive of the Magellanic Clouds, LMC and SMC) from \citet[with ``fixed'' background model]{Pace2022}%
\footnote{\url{https://zenodo.org/record/7158669}}, and globular clusters (GCs) from \citet{Vasiliev2021a}%
\footnote{\url{https://people.smp.uq.edu.au/HolgerBaumgardt/globular}}. The properties used in our analysis include the $V$-band magnitude, sky coordinates ($\alpha, \delta$), distances to the sun ($d_\odot$), line-of-sight velocities ($v_\mathrm{los}$), and proper motions ($\mu_\alpha^\ast,\mu_\delta$). 
Following \citet{Li2020a}, the kinematics are transformed from Heliocentric frame into Galactocentric frame \citep{Bland-Hawthorn2016a}.
We convert the apparent $V$-magnitude $m_V$ to absolute magnitude via $M_V = m_V - 5 \log (d_\odot/\kpc) - 10$.

Out of the 54 DGs (excluding the LMC and SMC) and 165 GCs in the parent sample, we have 53 DGs and 24 GCs within the radial range $r\in[20,300]\kpc$. We exclude tracers without line-of-sight velocity measurements or with large mean proper motion uncertainties $\widebar\epsilon_{\mu}=(\epsilon_{\mu_\alpha^\ast}+\epsilon_{\mu_\delta})/2>0.2\mathrm{mas/yr}$, which results in 43 DGs and 22 GCs.
We further exclude seven DGs likely associated with the LMC: Car II, Car III, Hor I, Hor II, Hyi I, Phx II, and Ret II \citep{Pardy2019,Pace2022,Battaglia2022,CorreaMagnus2022} to avoid potential bias.
We also exclude tracers fainter than $M_V=-2$ whose completeness is relatively less understood, leaving 31 DGs and 22 GCs in the final sample.%
\footnote{
Such faint satellites represent a largely unobserved parent sample, thus assigned high incompleteness correction weights in \empdf.
Any uncertainty in completeness correction will be amplified.
Moreover, extremely unequal weights reduce the effective tracer number, also leading to inferior inference.
The latter effect can possibly be mitigated by using weight-dependent smoothing kernel when constructing empirical DF in the future.}
The parent dataset and the subsample used in our analysis are shown in \reffig{fig:mw_tracers}.

\subsection{Selection function}
\label{sec:obs_select}

Clearly, faint tracers at large distances are largely missed in the lower-right corner in \reffig{fig:mw_tracers}.
To prevent the bias introduced by incompleteness (cf.\ discussion in \citealt{Li2020a}), we should either limit ourselves to a volume-limited subsample or consider the selection function as described in \refsec{sec:selectfunc}.
As the DGs and GCs were discovered by multiple sky surveys with different depths, the selection function is a complicated function of both distance and sky coordinates. Fortunately, requiring \gaia DR3 measurements further posed a uniform selection on the parent sample (assuming those surveys are deeper than \gaia), which simplifies the task.

We find a good approximate selection function for these tracers with \gaia DR3 kinematics following \citet{Li2020a}. Assuming that \gaia can measure the proper motion of a DG/GC reliably only when it contains at least $\sim\! 5$ member stars brighter than the limit $m_{V,\mathrm{star}} \!\sim\! 21$ \citep{GaiaCollaboration2022},
we can derive the maximum observable radius $\robsmax(M_V)$ as a function of the luminosity,
using synthetic member stars generated by the PARSEC stellar model library \citep{Bressan2012}\footnote{\url{http://stev.oapd.inaf.it/cgi-bin/cmd_3.7}.
We queried the synthetic samples in batch using a script written by Zhaozhou Li \citep[as part of][]{Li2020b}, \url{https://github.com/syrte/query_isochrone}.} for a typical stellar population with a \citet{Chabrier2001} mass function, an age of 12.5 Gyr, and a metallicity of $\mathrm{[Fe/H]} = -2.2$.
As shown by the solid red curve
in \reffig{fig:mw_tracers}, this $\robsmax(M_V)$
distinguishes the tracers with and without valid kinematic data very well,
suggesting a sharp transition in completeness,%
\footnote{In principle, we should use Heliocentric distance rather than the Galactocentric $r$. Nevertheless, the resulted difference in enclosed volume is small as $\robsmax$ is far beyond the solar location.} 
thus justifying the use of \refeqn{eq:s2}.
The shaded band around this curve corresponds to $N({m_{V,\mathrm{star}}<21})=5\pm 2$ for reference.
At the luminous end, the above curve is close to a total magnitude cut of $m_{V}=17$, which happens to be nearly identical to the detection limit of DGs in the SDSS survey \citep{Walsh2009}.

In addition to the selection on distance,
the spatial distribution of satellites is further truncated by the angular coverage of the sky surveys and partly blocked by foreground dust and stellar light of the disk.
However, as pointed out in \citealt{Li2020a}, this angular selection does not affect our analysis when assuming a spherical DF. 

We note that the above \gaia-based selection function is somewhat empirical.
We refer the interested readers to the general reviews of DG and GC populations \citep{Simon2019a,Gratton2019}
and dedicated discussions on completeness of DGs (\citealt{Koposov2008,Walsh2009,Jethwa2016,Nadler2019,Manwadkar2022})
and GCs (\citealt{Webb2020}).

Finally, we reiterate that a rigorous and straightforward treatment of the selection function is an important virtue of the DF methods, including \empdf.
This approach allows for full exploitation of the available observational data.
In contrast,
dynamical methods relying on a volume-limited complete sample must either remove abundant faint tracers or limit the analysis to a small volume.

\begin{table}
\caption{
  MW mass profile inferred from dwarf galaxies (DGs) and globular clusters (GCs) 
  with \empdf and simulated-based DF (SimDF).
}
\begin{center}
\def\arraystretch{1.5}
\setlength{\tabcolsep}{0pt}
% \begin{tabular}{>{\centering}p{5.2em}rrrr}
\begin{tabular*}{1\columnwidth}{r@{\extracolsep{\fill}}rrrr}
\toprule
 Method & \empdf & \empdf & \empdf & SimDF\\

\midrule
                   Tracer &       GC &       DG &           GC+DG &       DG \\[-1ex]
      $N_\mathrm{tracer}$ &       22 &       31 &              53 &       31 \\[-1ex]
         $r_\mathrm{med}$ &     36.0 &    101.6 &            68.5 &    101.6 \\[-1ex]
         $r_\mathrm{avg}$ &     51.7 &    109.7 &            85.6 &    109.7 \\

\midrule
    $M(\!<\!20\kpc)$ 
       & $0.12_{-0.04}^{+0.08}$ & $0.17_{-0.07}^{+0.08}$ & $0.15_{-0.05}^{+0.07}$ & $0.15_{-0.03}^{+0.03}$ \\ 
    25 & $0.17_{-0.05}^{+0.10}$ & $0.23_{-0.08}^{+0.09}$ & $0.20_{-0.06}^{+0.08}$ & $0.20_{-0.03}^{+0.04}$ \\ 
    30 & $0.22_{-0.06}^{+0.11}$ & $0.28_{-0.09}^{+0.09}$ & $0.26_{-0.07}^{+0.08}$ & $0.24_{-0.04}^{+0.05}$ \\ 
    40 & $0.32_{-0.07}^{+0.13}$ & $0.39_{-0.10}^{+0.09}$ & $0.36_{-0.08}^{+0.08}$ & $0.33_{-0.05}^{+0.06}$ \\ 
    50 & $0.43_{-0.08}^{+0.14}$ & $0.48_{-0.10}^{+0.10}$ & $0.46_{-0.07}^{+0.08}$ & $0.41_{-0.06}^{+0.07}$ \\ 
    65 & $0.58_{-0.09}^{+0.17}$ & $0.61_{-0.11}^{+0.10}$ & $0.60_{-0.07}^{+0.08}$ & $0.51_{-0.07}^{+0.08}$ \\ 
    80 & $0.74_{-0.14}^{+0.18}$ & $0.73_{-0.12}^{+0.12}$ & $0.74_{-0.09}^{+0.09}$ & $0.60_{-0.07}^{+0.09}$ \\ 
   100 & $0.95_{-0.22}^{+0.23}$ & $0.87_{-0.15}^{+0.15}$ & $0.90_{-0.13}^{+0.13}$ & $0.71_{-0.08}^{+0.10}$ \\ 
   125 & $1.19_{-0.33}^{+0.30}$ & $1.03_{-0.18}^{+0.20}$ & $1.07_{-0.18}^{+0.21}$ & $0.83_{-0.09}^{+0.11}$ \\ 
   160 & $1.48_{-0.48}^{+0.46}$ & $1.21_{-0.24}^{+0.29}$ & $1.29_{-0.25}^{+0.32}$ & $0.97_{-0.10}^{+0.12}$ \\ 
   200 & $1.78_{-0.64}^{+0.66}$ & $1.39_{-0.30}^{+0.41}$ & $1.49_{-0.34}^{+0.46}$ & $1.10_{-0.12}^{+0.14}$ \\ 
   250 & $2.10_{-0.83}^{+0.93}$ & $1.55_{-0.35}^{+0.57}$ & $1.70_{-0.40}^{+0.63}$ & $1.24_{-0.13}^{+0.16}$ \\ 
   320 & $2.46_{-1.03}^{+1.29}$ & $1.77_{-0.45}^{+0.73}$ & $1.95_{-0.51}^{+0.84}$ & $1.39_{-0.14}^{+0.17}$ \\ 
   400 & $2.81_{-1.24}^{+1.69}$ & $1.96_{-0.51}^{+0.92}$ & $2.18_{-0.62}^{+1.08}$ & $1.53_{-0.15}^{+0.19}$ \\ 

\midrule
    $M_\mathrm{200c}$ & $ 3.67_{-2.26}^{+0.73}$ & $ 1.68_{-0.62}^{+0.34}$ & $ 1.68_{-0.55}^{+0.69}$ & $ 1.14_{-0.14}^{+0.16}$ \\ 
    $R_\mathrm{200c}$ & $  318_{  -87}^{  +20}$ & $  245_{  -35}^{  +16}$ & $  245_{  -31}^{  +30}$ & $  215_{   -9}^{  +10}$ \\ 
    $c_\mathrm{200c}$ & $  3.0_{ -0.5}^{ +8.4}$ & $ 10.0_{ -3.1}^{+20.2}$ & $  9.1_{ -4.8}^{ +8.9}$ & $ 11.5_{ -1.9}^{ +2.3}$ \\ 

\midrule
   $\log M_\mathrm{200c}$ & $ 12.42 \pm  0.33$ & $ 12.16 \pm  0.16$ & $ 12.23 \pm  0.16$ & $ 12.06 \pm  0.06$ \\[-1ex]
   $\log c_\mathrm{200c}$ & $  0.71 \pm  0.47$ & $  1.12 \pm  0.37$ & $  0.98 \pm  0.29$ & $  1.06 \pm  0.08$ \\[-1ex]
       $\rho_\text{corr}$ & $ -0.89          $ & $ -0.80          $ & $ -0.86          $ & $  0.55          $ \\

\bottomrule
\end{tabular*}
\end{center}
\textit{Rows.} 
Type, number, median and mean Galactocentric distances $[\kpc]$ of tracers;
MW mass within a series of radii, $M(<r) [10^{12}\msun]$;
inferred virial mass, radius, and concentration, $M_\mathrm{200c} [10^{12}\msun]$, $R_\mathrm{200c} [\kpc]$, $c_\mathrm{200c}$;
2D Gaussian fitting to the joint distribution of $\log (M_\mathrm{200c}/\msun)$ and $\log c_\mathrm{200c}$, including mean $\pm$ standard deviation
and correlation coefficient $\rho_\mathrm{corr}$.

\textit{Notes.}
(a) The estimated masses are for the total mass, including baryons.
(b) The uncertainties are for 16\% and 84\% percentiles ($1\sigma$).
(c) We emphasize that the constrained mass profile is more fundamental than the fitting parameters.
Merely quoting the inferred halo mass can be meaningless when the $M$-$c$ degeneracy is strong.
\label{tab:result}
\end{table}

\begin{figure*}
  \centering
  \includegraphics[width=0.9\linewidth]{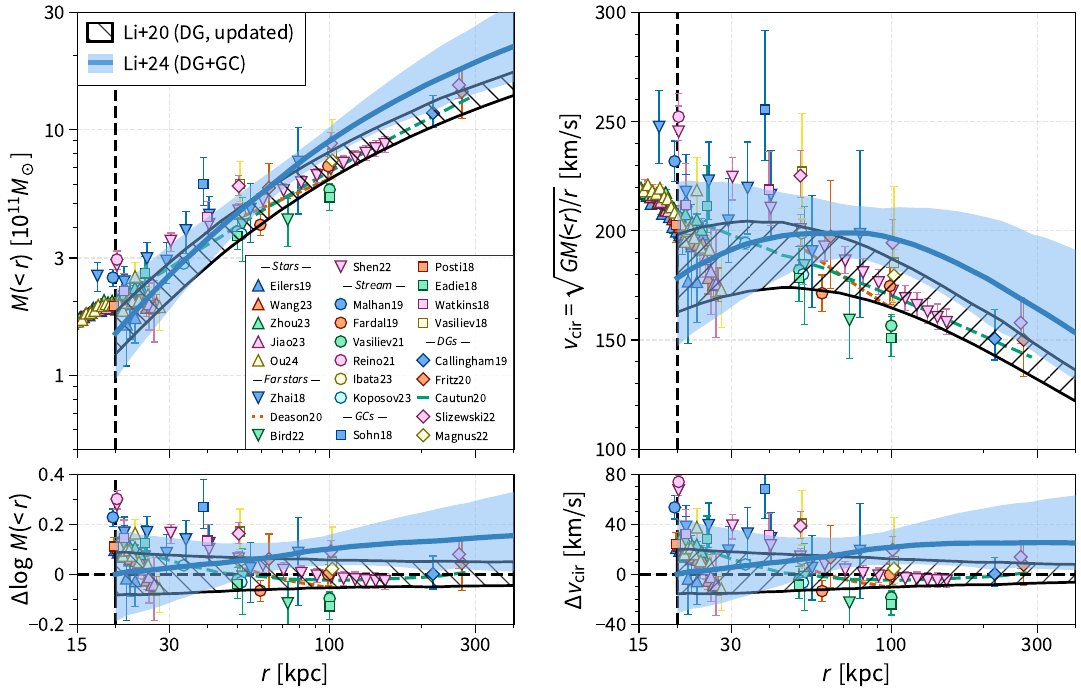}
  \caption{
  Comparison of the inferred MW mass profile (left) and rotation curve (right).
  Shown are our new estimates based on GC+DGs (blue curves with 1$\sigma$ error bands)
  and the updated estimate of \citet{Li2020a} based on DGs (striped $1\sigma$ error bands).  Other symbols or lines represent recent measurements since 2018 using various tracers,
  including stars within 30 kpc \citep{Eilers2019,Wang2023f,Zhou2023a,Jiao2023,Ou2024},
  more distant halo stars \citep{Zhai2018,Deason2020b,Bird2022,Shen2022},
  stellar streams \citep{Malhan2019,Fardal2019,Vasiliev2020a,Reino2020,Ibata2023,Koposov2023},
  GCs \citep{Sohn2018,Posti2018,Eadie2018,Watkins2018,Vasiliev2018},
  and DGs \citep{Callingham2018,Cautun2019,Fritz2020,Slizewski2022,CorreaMagnus2022}.
  Studies that used multiple tracer populations are categorized by the most distant tracers.
  Small offsets in $r$ are added to reduce overlap.
  Most recent estimates of the MW mass profile are consistent within $\pm0.15$ dex.
  }
  \label{fig:mw_compare}
\end{figure*}

\subsection{Constraining the MW potential}
\label{sec:mw_mass}

We apply the \empdf method to DGs and GCs separately.
For each tracer labeled $i$, we consider the selection function (\refsec{sec:obs_select}) with a maximum observable radius $\max\{\robsmax(M_{V,i}), r_i\}$ individually.
The correction for the selection function (described in \refsec{sec:selectfunc}) is used, but without explicit treatment for observational errors (justified in \refsec{sec:mock_err}).
The likelihood is estimated on a dense 2D grid of $(\log M_\mathrm{200c}, \log c)$
for $\log M_\mathrm{200c}/\msun \in [11.5, 12.7]$ and $\log c \in [0.3, 1.5]$ with
\emph{flat priors}.
This range of $c$ covers the 4$\sigma$ region of the $M$-$c$ relation for this mass range in hydrodynamical simulations \citep{Schaller2015}.
Flat priors are used to avoid relying on extra information.
The confidence regions and the marginalized distributions of parameters are estimated from the posterior on the grid.

The likelihood contours of the inferred parameters are shown in the left panel of \reffig{fig:mw_mass}.
The corresponding mass profiles are displayed in the right panel.
These parameters and mass profiles are also provided in \reftab{tab:result}.
Similar to \reffig{fig:benchmark}, a strong degeneracy between $M$ and $c$ is observed,
corresponding to the most constrained ``sweet point'' in the mass profile near the mean tracer radius.%
\footnote{In contrast to \reffig{fig:benchmark}, here
the best-constrained radius is not exactly at the median tracer radius due to selection function correction.
}
The inferences with DGs and GCs are mutually consistent.
DGs provide better constraints than GCs for $r>60\kpc$, owing to their larger sample size and more extended spatial distribution.
We also present joint constraints from DGs and GCs by multiplying their likelihoods,
which results in tighter overall constraints.

We emphasize that the constrained mass profile is more fundamental than the inferred virial mass.
The virial mass is merely one of the parameters and does not fully describe the mass profile alone.
Its value may have large uncertainty and even bias if the virial radius extends far beyond the tracer sample \citep[cf.][fig.\ 9]{Li2020a}.
The extrapolation becomes unreliable because it depends on an assumed functional form of the profile, which may not be suitable for the full radial range.
However, over the radial range covered by tracers, the mass profile remains well-constrained.
For example, one might think that MW virial mass by GCs is significantly overestimated with unrealistic concentration according to the left panel of \reffig{fig:mw_mass}, but upon examining the right panel, this profile is actually reasonably constrained.

For comparison, we also show an updated constraint by the simulation-informed DF using scaling relations of satellite kinematics in the EAGLE simulation \citep{Li2019,Li2020a} in \reffig{fig:mw_mass}.
The original estimates by \citet{Li2020a} were made for a similar DG sample with \gaia DR2 data \citep{GaiaCollaboration2018d}, but for $r\in[40, 300]\kpc$.
Using the same simulation-informed DF model and a flat prior,
we update the constraints using the updated DG sample with \gaia DR3 kinematics and the corresponding selection function presented in this paper.
The updated parameters are very close to those originally reported by \citet[$M_\mathrm{200c}=1.29_{-0.20}^{+0.24}\msun$ and $c=11.0_{-3.3}^{+4.8}$]{Li2020a},
but with reduced uncertainty of concentration due to additional tracers used.

While consistent with the results of \empdf within $1\sigma$,
the simulation-informed DF provides much tighter constraints.
This improvement benefits from additional knowledge learnt from simulation on orbital distribution and its scaling relation with potential, as discussed in \refsec{sec:benchmark}.
In contrast, the \empdf method constructs the orbital distribution from observed tracers.
This difference also leads to a different parameter degeneracy direction.
As shown in the right panel, the inferred profile by simulation-informed DF does not exhibit a ``sweet point'',
but shows a rather uniform uncertainty that is only slightly larger inward.
This demonstrates the advantage of the simulation-informed method,
despite at the cost of dependence on simulations.
Nevertheless, \empdf with the same DG sample shows similar uncertainties as simulation-informed DF around the median tracer radius.

% Tests in \citet{Li2020a} show that the MW mass inferred by simulation-informed DF is insensitive to the inclusion of LMC satellites.
% This insensitivity occurs because the DF is constructed from simulation independently from the tracers.
% In this sense, the simulation-informed DF method is more robust.

We also show the mass-concentration ($M$-$c$) relation in cosmological hydrodynamical simulations in \reffig{fig:mw_mass} for reference.
Shown in the median relation, $\log c=0.912 - 0.087 \log (M_\mathrm{200c}/10^{12} \msun)$ \citep[EAGLE simulation]{Schaller2015}, with a scatter of $0.15$dex \citep{Jing2000a}.
This relation is only slightly different from that in dark-matter-only simulations.
All the constraints above overlap with this relation.

\subsection{Comparison with previous work}

There is a considerable effort devoted to the MW mass estimates,
based on different tracer populations and methods.
We refer readers to \citealt{Wang2019b} for a comprehensive review (also \citealt{Bobylev2023,Hunt2025} for recent compilations).
Below we list some recent work (not exhaustive) in several categories.
\setlist{nosep}
\begin{itemize}[leftmargin=*]
  \item Stars of the inner halo with $30\kpc$ \citep{Eilers2019,Ablimit2020,Wang2023f,Zhou2023a,Jiao2023,Ou2024} and more distant halos stars \citep{Xue2008,Huang2016,Ablimit2017,Zhai2018,Deason2020b,Bird2022,Shen2022}. 
  Most studies use the Jeans equation, while some employ DF models \citep{Deason2020b,Shen2022}.
  \item Stellar streams \citep{Gibbons2014,Malhan2019,Fardal2019,Reino2020,Vasiliev2020a,Koposov2023,Ibata2023}.
  \item Globular clusters \citep{Sohn2018,Posti2018,Eadie2018,Watkins2018,Vasiliev2018,Wang2022c}
  and satellite galaxies \citep{Boylan-Kolchin2013,Patel2018,Callingham2018,Cautun2019,Li2020a,Fritz2020,Slizewski2022,CorreaMagnus2022,Kravtsov2024}.
  Jeans mass estimators are widely used due to the limited tracer sample size.
  Some studies utilize DF models as well \citep{Li2017,Li2020a,Eadie2018,Callingham2018,Wang2022c}.
  \item Escape velocity of stars \citep{Smith2007,Deason2019,Grand2019,Necib2022,Roche2024}.
  \item Dynamics of the Local Group
  \citep{Li2008,Boylan-Kolchin2013,Penarrubia2016,Penarrubia2017,Li2021b}.
\end{itemize}
In \reffig{fig:mw_compare},
we compare our new measurements with selected recent literature after 2018,
most of which utilize proper motion measurements by HST or \gaia.
We display the mass profile and its alternative expression in terms of the rotation curve $v_\mathrm{cir}=\sqrt{GM(<r)/r}$.
Most recent estimates of the MW mass profile are consistent within $\pm0.15$ dex.
The estimates by \empdf appear slightly higher than the other measurements, but this deviation is not very significant given the uncertainty band.

It is noted that the MW masses inferred by \empdf and simulation-informed DF show a trend lower than estimates by stars for $r\lesssim 20\,\kpc$ and likely more so inwards.
This discrepancy arises because a single NFW profile cannot adequately describe the additional stellar components within this radius, prompting us to limit our analysis to $r>20\,\kpc$.
Future work may address this issue by using a more flexible model with multiple components for the MW mass profile.

Many studies also provide estimates of the MW virial mass.
We reiterate that it is more meaningful to compare inferred density profiles in regions directly constrained by tracers.
Extrapolation is unreliable,
especially when inferring virial mass from tracers located in the very inner halo,
which is highly sensitive to the assumed functional form for mass profiles.
A workaround involves combining multiple tracer populations to
obtain better constraints on a full radial range
\citep[e.g.,][]{McMillan2017,Cautun2019,Karukes2019,Li2020a}.

\section{Discussion: extensions of the method}
\label{sec:discuss}

\citet{Han2016b} discussed various extensions to the \opdf method and its connections to other methods, including analytical DFs~\citep{Wang2015,Han2016a}, the Schwartszchild (\citeyear{Schwarzschild1979}) method, and the orbital roulette~\citep{Beloborodov2004}. Most of these discussions apply to \empdf as an upgrade to the \opdf method. We briefly discuss some of them below.

The \empdf method can be extended to axisymmetric or triaxial systems by using the actions $\bm{J}$ instead of $(E,L)$ to characterize the orbital distribution.
When such calculations are difficult, direct numerical integration of the orbits can be used to obtain the time-averaged 6D distribution in a way similar to the Schwarzschild method.
Following the phase-mark method in \citet{Han2016b}, it is possible to apply \empdf to separate radial bins to obtain local potential estimates and reconstruct the entire potential profile non-parametrically from these segmented estimates. 
When some dimensions of tracer kinematics are lacking in observation, the missing information may be recovered statistically based on the symmetry of the system \citep{Dehnen1998,Wegg2018} or treated as observations with extremely large errors (cf. \refsec{sec:obs_err}).

In what follows, we provide more extended discussions on \empdf expressed in action space and its connection to the minimum entropy principle.

\subsection{\empdf in action space and  minimum entropy}
\label{sec:action}

We used $p(E,L)$ to characterize the orbital distribution in this work.
One can use combinations of other orbital integrals as well, such as circular radius and orbital circularity $(r_\mathrm{cir},\epsilon)$, or actions $\bm{J}$.
A particular interest in actions is that $\bm{J}$ and their conjugated angles, $\bm{\theta}$, are the so-called canonical coordinates \citepalias[sec. 3.5]{Binney2008a}
such that
$f ({\bm{r}}, {\bm{v}}) = p(\bm{J},\bm{\theta})=(2\pi)^{-3}p(\bm{J})$
with unity Jacobian.
Similar to $p(E,L)$, the orbital distribution $p(\bm{J})$ can be constructed empirically from the tracer sample $\{\bm{J}_i\}$ using KDE.%
\footnote{In spherical symmetry, one can choose the actions $(J_r, L, L_z)$, where $L_z\equiv\bm{L}\cdot\bm{e}_z$ follows a uniform distribution in $[-L, L]$. The KDE procedure thus becomes simpler with $p(J_r, L, L_z)=\frac{1}{2L}p(J_r, L)$.
Also note that the transformation to action–angle coordinates is feasible only if the potential is integrable and not dominated by resonances or chaotic regions (see e.g., \citetalias{Binney2008a}, \citealt{Price-Whelan2021}). This condition is typically met for many simple potential models.}

The log-likelihood between the empirical DF and observation (\refeqnalt{eq:posterior}, without prior) is rewritten as
\begin{gather}
\textstyle \ln \mathcal{L}= \sum_{i = 1}^N \ln f ({\bm{w}}_i) = \sum_{i = 1}^N \ln p ({\bm{J}}_i) -3N\ln(2\pi)
\end{gather}
This is the \emph{same} likelihood, just in a different form.
On the other hand, the entropy of the action distribution of the sample is
\begin{gather}
H_{\bm{J}} = - {\textstyle\int} p ({\bm{J}}) \ln p ({\bm{J}}) \dif^3 {\bm{J}} \simeq -\frac{1}{N} \textstyle\sum_{i = 1}^N \ln p ({\bm{J}_i})
\end{gather}
Therefore, with $\ln \mathcal{L}=-NH+\mathrm{const.}$, maximizing the \empdf likelihood is equivalent to \emph{minimizing the entropy of actions}, $H_{\bm{J}}$,
or \emph{minimizing the entropy of the phase-mixed DF}, $H_{\bm{w}}\equiv H_{\bm{J}}+\ln(2\pi)^3$.
It is also equivalent to \emph{maximizing the clustering in action space or phase space} (recall that a flat distribution has maximum entropy).

The concept of \empdf theoretically justifies why we should minimize the entropy of actions to infer the potential, rather than the entropy of other coordinates such as $(E, L)$.
Essentially, \empdf seeks the minimum difference in terms of KL-divergence (aka. relative entropy) between the time-averaged distribution, $f(\bm{r}, \bm{v})$, and the instantaneous observed kinematics, $\{\bm{r}_i, \bm{v}_i\}$, which involves two distributions rather than a single distribution.
Though the KL-divergence has a similar form as entropy when these two distributions are both expressed in terms of actions, $f(\bm{J})=(2\pi)^{-3}p(\bm{J})$ and $\{\bm{J}_i\}$, they hold different meanings nevertheless.

\citet{Silva2024} recently proposed a dynamical modeling method based on minimum entropy in action space, or equivalently, minimum entropy of the phase-mixed distribution function (DF). They first recognize that the entropy of the phase-mixed DF for a tracer sample is minimized in the correct potential, given that the entropy of the DF always increases during phase-mixing in an incorrect potential. The minimum entropy method provides an alternative viewpoint of the same steady-state principle used in \empdf, bearing the same mathematical expressions as above but with a distinct physical motivation. 
However, unlike the \empdf\ method, which brings a natural likelihood function for Bayesian inference, the minimum entropy method does not provide a direct likelihood and must resort to Approximate Bayesian Computation (at least in its present form). Moreover, it is probably more straightforward to handle the observational selection function from the perspective of \empdf.

Despite the different contexts and physical motivations,
it is also intriguing to note the resemblance between \empdf and the clustering-based inference with stellar streams.
The latter searches for the best-fit Galactic potential by maximizing the clustering of stream member stars in action space (\citealt{Helmi1999,Sanderson2015,Buckley2019,Yang2020,Reino2020}, also cf.\ \citealt{Penarrubia2012,Brooks2024} in energy space).
% It is noteworthy that \empdf is indeed possibly applicable to stellar streams if the stream members follow the oPDF \refeqn{eq:opdf} locally in the concerned sky region.

% Loosely speaking, $p (r|E, L)$ is determined by dynamics and thus reflecting the underlying potential, while the $p(E,L)$ part is primarily set by the tracer population. 
% Adding extra assumptions in the construction of a DF is a trade-off between statistical uncertainty (formal error) and bias to the model dependence. The \empdf method is specifically designed to minimize extra assumptions to reduce the bias.

\section{Conclusions}
\label{sec:conclusion}

We propose the \empdf, a novel dynamical modeling method for general tracers with the optimal statistical efficiency under the minimal assumption of steady state. \empdf searches for the best-fit potential by maximizing the similarity between instantaneous kinematics and the time-averaged DF.
This DF is empirically constructed from observation upon the theoretical foundation of oPDF \citep{Han2016b}, thereby eliminating the reliance on presumed functional forms or orbit libraries inherent in conventional DF- or orbit-based methods. Standing out with its flexibility, efficiency, and adeptness in managing observational effects, \empdf warrants wide applications.

The key features of \empdf include 
\begin{itemize}[leftmargin=*]
  \item Minimal assumptions except for steady state. No assumptions are made on the specific form of the DF, and thus, no free parameters in the DF construction under a given trial potential.
  \item Flexibility. \empdf accommodates arbitrary DFs, including those with unbound orbits and substructures in orbital integrals, as long as the tracers are phase-mixed within the concerned volume.
  \item High precision. Tests demonstrate that \empdf outperforms other minimal assumption methods and the Jeans equation (\refsec{sec:validate}). The high efficiency also results in lower requirements for sample size and data quality.
  \item Effective treatment of observational effects. \empdf treats selection functions (\refsec{sec:selectfunc}) and observational errors (\refsec{sec:obs_err}) effectively. And it is robust against moderate observational errors even without explicit treatment (\refsec{sec:mock_err}).
  \item Extendable framework. Though currently implemented for complete kinematics in spherical potentials, the concept of \empdf can be applied to more general problems (\refsec{sec:discuss}).
  \item Useful by-product. \empdf provides non-binned estimates of various statistics for the tracers, such as the energy distribution, number density, velocity dispersion, and anisotropy $p(E)$, $\varrho(r), \sigma(r)$, and $\beta(r)$.
\end{itemize}

\empdf is a natural improvement upon the theoretical foundation of \opdf \citep{Han2016b} but exploiting the full DF.
It outperforms \opdf for small samples due to the smoothing of DF, whereas their performances become comparable when the sample size exceeds $\sim\!1000$.
This suggests that the information gained from the velocities beyond the spatial distribution used in \opdf is somehow limited.
Nevertheless, \empdf excels with small samples and provides a complete description of tracer kinematics with useful byproducts and an extendable framework.

\empdf can also be regarded as a lightweight \citet{Schwarzschild1979} method.
Instead of constructing an orbital library and solving for their weights under regulation, we obtain the orbital distribution directly and deterministically from the observed kinematics, with each tracer representing one orbit family.
Thus, \empdf has less computational complexity and parameter degeneracy compared to the Schwarzschild method.
\empdf is also less model-dependent than presumed analytical or simulation-informed DFs.
A particularity of our \empdf is that it has the maximum flexibility allowing arbitrary DFs but without free parameters except for the potential well.

Interestingly,
maximizing the likelihood of \empdf is mathematically equivalent to minimizing the entropy of the action distributions of the tracers
and the entropy of their phase-mixed DF (\refsec{sec:action}).
The minimum entropy principle has recently been independently proposed for dynamical modeling by \citet{Silva2024}, with distinct physical motivations,
providing an alternative viewpoint on the same steady-state principle.
On the other hand, the perspective of \empdf offers a natural likelihood, enabling more straightforward Bayesian inference with observational effects.

We applied \empdf to infer the MW mass profile using \gaia DR3 data for satellite galaxies and globular clusters separately, carefully modeling the sample selection function.
Combining these two datasets, we determined the enclosed masses to be
$M({<}r){=}26{\pm}8$, $46{\pm}8$, $90{\pm}13$, and $149{\pm}40\times10^{10}\msun$
at $r\!=\!30$, 50, 100, and 200 kpc, respectively (see \reffig{fig:mw_mass} and \reftab{tab:result}).
In parallel, we updated the estimates from \citet{Li2020a} using \gaia DR3 data and the DF based on scaling relations of satellite kinematics from the Eagle simulation \citep{Li2019}.
The halo mass and concentration were refined to
$M_\mathrm{200c,tot}{=}1.14{\pm}0.15\times 10^{12}\msun$ and $c_\mathrm{200c}{=}11.5{\pm}2.1$, 
yielding enclosed masses of
$M({<}r){=}24{\pm}4$, $41{\pm}6$, $71{\pm}9$, and $110{\pm}13\times10^{10}\msun$
at the same radii. 
While the two results are consistent, the simulation-informed DF achieves higher precision, representing one of the most accurate MW mass constraints to date, owing to the additional information extracted from simulations.
In general, \empdf offers greater flexibility for general tracers, whereas the simulation-informed DF is more efficient but limited to specific tracers, such as satellites.

The MW mass profiles inferred by \empdf are consistent with previous measurements in general (\reffig{fig:mw_compare}).
We reiterate that the fitted parameters should be regarded as describing the local profile.
In particular, the virial mass is not a complete or even a reliable description of the actual profile,
because extrapolation based on local estimates can be risky.
This may explain discrepancies in MW halo mass previously inferred from halo stars and satellite/GC tracers.
We should focus on comparison within the radial range covered by tracers.
Using multiple tracer populations with very different spatial distributions (e.g., halo stars and DGs) may help break the $M$-$c$ degeneracy \citep[e.g.,][]{Walker2011, Li2020a}.

\empdf can also be used in combination with other information about tracer populations, such as chemical abundances or stellar ages.
Dynamical modeling can be performed for different tracer populations separately \citep[mono-abundance populations,][]{Bovy2012, Bovy2013, Ting2013, Mackereth2020} or for the joint chemodynamical distribution \citep[extended distribution functions,][]{Sanders2015b, Das2016, Price-Whelan2021, Horta2023}.
This could provide stronger constraints and insights into the MW's chemodynamical structure in the future.
In addition, it is possible to use \empdf constructed from observed regions to partly infer the dynamical properties in the MW regions lacking direct observations \citep[cf.][]{Yang2022}.

However, since \empdf relies on the assumption of steady-state tracers and a static potential, it is subject to systematics when these assumptions are not met. 
Significant stochastic biases have been found in simulated tracers of various types~\citep{Han2016a,Wang2017d,Wang2018a,Han2019}, in both galactic and cluster halos~\citep{Li2021d,Li2022}.
This concern becomes more severe when using the stars as tracers.
Future exploration with more realistic mocks is worthwhile to gauge the level of such systematics. 
It is also possible to estimate the degree of deviation from equilibrium from observation \citep[cf.][]{Kipper2021}.
As for \empdf, the difference between the best-fit empirical DF and observed kinematics itself provides useful diagnosis for deviations from steady-state, warranting future applications and explorations.

In particular, the LMC is believed to be a major perturbator of the MW \citep[e.g.,][]{Petersen2020,Erkal2020,Kravtsov2024}, leading to deviations from equilibrium in the MW tracers. Although \citet{Li2020a} showed that the MW mass (including the LMC contribution) derived from simulation-informed DF is robust against the LMC effect, this may not necessarily be the case for \empdf.
\citet{CorreaMagnus2022} developed an orbit-rewinding method to compensate for LMC perturbations, where the corrected kinematics can be used with usual dynamical modeling methods that rely on equilibrium, including \empdf.

\section*{Acknowledgements}

ZZL thanks Ling Zhu, Adrian M. Price-Whelan, and Leandro Beraldo e Silva for helpful discussions,
and Eugene Vasiliev for public pedagogical materials of dynamical modeling. % at \url{http://eugvas.net/},
We thank the anonymous referee for constructive comments and suggestions.
This work was done on the \textsc{Gravity} supercomputer at the Department of Astronomy, SJTU,
and partially on the super cluster Moriah at HUJI.
This work was supported in part by National Key R\&D Program of China (2023YFA1607800, 2023YFA1607801, 2023YFA1605600, 2023YFA1605601), NSFC (12273021, 12133006), the science research grants from the China Manned Space Project (No.\ CMS-CSST-2021-A03), 111 project (No.\ B20019) and the U.S. Department of Energy under grant DE-FG02-87ER40328 (YZQ).
ZZL acknowledges the Marie Skłodowska-Curie Actions Fellowship under the European Union’s Horizon 2020 programme (101109759, “CuspCore”) and the Israel Science Foundation Grant ISF 861/20. ZZL was partially supported by grant NSF PHY-2309135 to the Kavli Institute for Theoretical Physics (KITP) during the preparation of this manuscript. JH acknowledges the sponsorship from the Yangyang Development Fund.

This research made use of the following software:
\texttt{Agama} \citep{Vasiliev2019}, 
  %\footnote{\url{https://github.com/GalacticDynamics-Oxford/Agama}},
\texttt{astropy} \citep{astropy}, 
\texttt{Cython} \citep{cython},
\texttt{gsl} \citep{gsl},
\texttt{Jupyter} \citep{jupyter},
  %\footnote{\url{https://jupyter.org}},
\texttt{KDEpy} \citep{kdepy},
  % https://github.com/tommyod/KDEpy
\texttt{Matplotlib} \citep{matplotlib},
  %\footnote{\url{https://matplotlib.org}},
\texttt{Numpy} \citep{numpy},
  %\footnote{\url{https://numpy.org}},
\texttt{ProPlot} \citep{proplot},
  %\footnote{\url{https://proplot.readthedocs.io}},
\texttt{scikit-optimize} \citep{skopt},
  % https://github.com/scikit-optimize/scikit-optimize
and \texttt{Scipy} \citep{scipy}.
  %\footnote{\url{https://www.scipy.org}}.
Some of the figures were created with the colorblind friendly scheme developed by \citet{Petroff2021}.

%%%%%%%%%%%%%%%%%%%%%%%%%%%%%%%%%%%%%%%%%%%%%%%%%%
\section*{Data Availability}

The data used and generated in this work are available upon reasonable request to the authors.
The numerical implementation of \empdf is pending public release and will be available at \url{https://github.com/syrte/empdf}.

% The inclusion of a Data Availability Statement is a requirement for articles published in MNRAS. Data Availability Statements provide a standardised format for readers to understand the availability of data underlying the research results described in the article. The statement may refer to original data generated in the course of the study or to third-party data analysed in the article. The statement should describe and provide means of access, where possible, by linking to the data or providing the required accession numbers for the relevant databases or DOIs.



\bibliographystyle{mnras}
\bibliography{mylib} % if your bibtex file is called example.bib

\begin{thebibliography}{}
\makeatletter
\relax
\def\mn@urlcharsother{\let\do\@makeother \do\$\do\&\do\#\do\^\do\_\do\%\do\~}
\def\mn@doi{\begingroup\mn@urlcharsother \@ifnextchar [ {\mn@doi@}
  {\mn@doi@[]}}
\def\mn@doi@[#1]#2{\def\@tempa{#1}\ifx\@tempa\@empty \href
  {http://dx.doi.org/#2} {doi:#2}\else \href {http://dx.doi.org/#2} {#1}\fi
  \endgroup}
\def\mn@eprint#1#2{\mn@eprint@#1:#2::\@nil}
\def\mn@eprint@arXiv#1{\href {http://arxiv.org/abs/#1} {{\tt arXiv:#1}}}
\def\mn@eprint@dblp#1{\href {http://dblp.uni-trier.de/rec/bibtex/#1.xml}
  {dblp:#1}}
\def\mn@eprint@#1:#2:#3:#4\@nil{\def\@tempa {#1}\def\@tempb {#2}\def\@tempc
  {#3}\ifx \@tempc \@empty \let \@tempc \@tempb \let \@tempb \@tempa \fi \ifx
  \@tempb \@empty \def\@tempb {arXiv}\fi \@ifundefined
  {mn@eprint@\@tempb}{\@tempb:\@tempc}{\expandafter \expandafter \csname
  mn@eprint@\@tempb\endcsname \expandafter{\@tempc}}}

\bibitem[\protect\citeauthoryear{{Ablimit} \& {Zhao}}{{Ablimit} \&
  {Zhao}}{2017}]{Ablimit2017}
{Ablimit} I.,  {Zhao} G.,  2017, \mn@doi [\apj] {10.3847/1538-4357/aa83b2},
  \href {https://ui.adsabs.harvard.edu/abs/2017ApJ...846...10A} {846, 10}

\bibitem[\protect\citeauthoryear{{Ablimit}, {Zhao}, {Flynn}  \&
  {Bird}}{{Ablimit} et~al.}{2020}]{Ablimit2020}
{Ablimit} I.,  {Zhao} G.,  {Flynn} C.,   {Bird} S.~A.,  2020, \mn@doi [\apjl]
  {10.3847/2041-8213/ab8d45}, \href
  {https://ui.adsabs.harvard.edu/abs/2020ApJ...895L..12A} {895, L12}

\bibitem[\protect\citeauthoryear{{Acerbi}}{{Acerbi}}{2018}]{Acerbi2018}
{Acerbi} L.,  2018, \mn@doi [arXiv e-prints] {10.48550/arXiv.1810.05558}, \href
  {https://ui.adsabs.harvard.edu/abs/2018arXiv181005558A} {p. arXiv:1810.05558}

\bibitem[\protect\citeauthoryear{{An}, {Naik}, {Evans}  \& {Burrage}}{{An}
  et~al.}{2021}]{An2021}
{An} J.,  {Naik} A.~P.,  {Evans} N.~W.,   {Burrage} C.,  2021, \mn@doi [\mnras]
  {10.1093/mnras/stab2049}, \href
  {https://ui.adsabs.harvard.edu/abs/2021MNRAS.506.5721A} {506, 5721}

\bibitem[\protect\citeauthoryear{Archambeau, Lee  \& Verleysen}{Archambeau
  et~al.}{2003}]{Archambeau2003a}
Archambeau C.,  Lee J.~A.,   Verleysen M.,  2003, in {{ESANN}} 2003, {{European
  Symposium}} on {{Artificial Neural Networks}}.

\bibitem[\protect\citeauthoryear{{Astropy Collaboration} et~al.,}{{Astropy
  Collaboration} et~al.}{2018}]{astropy}
{Astropy Collaboration} et~al., 2018, \mn@doi [\aj] {10.3847/1538-3881/aabc4f},
  \href {https://ui.adsabs.harvard.edu/abs/2018AJ....156..123A} {156, 123}

\bibitem[\protect\citeauthoryear{{Bahcall} \& {Tremaine}}{{Bahcall} \&
  {Tremaine}}{1981}]{Bahcall1981}
{Bahcall} J.~N.,  {Tremaine} S.,  1981, \mn@doi [\apj] {10.1086/158756}, \href
  {https://ui.adsabs.harvard.edu/abs/1981ApJ...244..805B} {244, 805}

\bibitem[\protect\citeauthoryear{{Battaglia}, {Taibi}, {Thomas}  \&
  {Fritz}}{{Battaglia} et~al.}{2022}]{Battaglia2022}
{Battaglia} G.,  {Taibi} S.,  {Thomas} G.~F.,   {Fritz} T.~K.,  2022, \mn@doi
  [\aap] {10.1051/0004-6361/202141528}, \href
  {https://ui.adsabs.harvard.edu/abs/2022A&A...657A..54B} {657, A54}

\bibitem[\protect\citeauthoryear{Behnel, Bradshaw, Citro, Dalcin, Seljebotn  \&
  Smith}{Behnel et~al.}{2011}]{cython}
Behnel S.,  Bradshaw R.,  Citro C.,  Dalcin L.,  Seljebotn D.~S.,   Smith K.,
  2011, Computing in Science \& Engineering, 13, 31

\bibitem[\protect\citeauthoryear{{Beloborodov} \& {Levin}}{{Beloborodov} \&
  {Levin}}{2004}]{Beloborodov2004}
{Beloborodov} A.~M.,  {Levin} Y.,  2004, \mn@doi [\apj] {10.1086/422908}, \href
  {https://ui.adsabs.harvard.edu/abs/2004ApJ...613..224B} {613, 224}

\bibitem[\protect\citeauthoryear{{Beraldo e Silva}, {Valluri}, {Vasiliev},
  {Hattori}, {de Siqueira Pedra}  \& {Daniel}}{{Beraldo e Silva}
  et~al.}{2024}]{Silva2024}
{Beraldo e Silva} L.,  {Valluri} M.,  {Vasiliev} E.,  {Hattori} K.,  {de
  Siqueira Pedra} W.,   {Daniel} K.~J.,  2024, \mn@doi [arXiv e-prints]
  {10.48550/arXiv.2407.07947}, \href
  {https://ui.adsabs.harvard.edu/abs/2024arXiv240707947B} {p. arXiv:2407.07947}

\bibitem[\protect\citeauthoryear{{Binney}}{{Binney}}{2014}]{Binney2014}
{Binney} J.,  2014, \mn@doi [\mnras] {10.1093/mnras/stu297}, \href
  {https://ui.adsabs.harvard.edu/abs/2014MNRAS.440..787B} {440, 787}

\bibitem[\protect\citeauthoryear{{Binney} \& {McMillan}}{{Binney} \&
  {McMillan}}{2011}]{Binney2011}
{Binney} J.,  {McMillan} P.,  2011, \mn@doi [\mnras]
  {10.1111/j.1365-2966.2011.18268.x}, \href
  {https://ui.adsabs.harvard.edu/abs/2011MNRAS.413.1889B} {413, 1889}

\bibitem[\protect\citeauthoryear{Binney \& Tremaine}{Binney \&
  Tremaine}{2008}]{Binney2008a}
Binney J.,  Tremaine S.,  2008, Galactic {{Dynamics}}: {{Second Edition}}, 2nd
  edn.
Princeton University Press, Princeton

\bibitem[\protect\citeauthoryear{{Binney} \& {Vasiliev}}{{Binney} \&
  {Vasiliev}}{2023}]{Binney2023}
{Binney} J.,  {Vasiliev} E.,  2023, \mn@doi [\mnras] {10.1093/mnras/stad094},
  \href {https://ui.adsabs.harvard.edu/abs/2023MNRAS.520.1832B} {520, 1832}

\bibitem[\protect\citeauthoryear{{Bird} et~al.,}{{Bird}
  et~al.}{2022}]{Bird2022}
{Bird} S.~A.,  et~al., 2022, \mn@doi [\mnras] {10.1093/mnras/stac2036}, \href
  {https://ui.adsabs.harvard.edu/abs/2022MNRAS.516..731B} {516, 731}

\bibitem[\protect\citeauthoryear{Bishop}{Bishop}{2006}]{Bishop2006}
Bishop C.~M.,  2006, Pattern {{Recognition}} and {{Machine Learning}}.
Springer, New York

\bibitem[\protect\citeauthoryear{{Bland-Hawthorn} \&
  {Gerhard}}{{Bland-Hawthorn} \& {Gerhard}}{2016}]{Bland-Hawthorn2016a}
{Bland-Hawthorn} J.,  {Gerhard} O.,  2016, \mn@doi [\araa]
  {10.1146/annurev-astro-081915-023441}, \href
  {https://ui.adsabs.harvard.edu/abs/2016ARA&A..54..529B} {54, 529}

\bibitem[\protect\citeauthoryear{{Bobylev} \& {Baykova}}{{Bobylev} \&
  {Baykova}}{2023}]{Bobylev2023}
{Bobylev} V.~V.,  {Baykova} A.~T.,  2023, \mn@doi [Astronomy Reports]
  {10.1134/S1063772923080024}, \href
  {https://ui.adsabs.harvard.edu/abs/2023ARep...67..812B} {67, 812}

\bibitem[\protect\citeauthoryear{{Bovy} \& {Rix}}{{Bovy} \&
  {Rix}}{2013}]{Bovy2013}
{Bovy} J.,  {Rix} H.-W.,  2013, \mn@doi [\apj] {10.1088/0004-637X/779/2/115},
  \href {https://ui.adsabs.harvard.edu/abs/2013ApJ...779..115B} {779, 115}

\bibitem[\protect\citeauthoryear{{Bovy}, {Murray}  \& {Hogg}}{{Bovy}
  et~al.}{2010}]{Bovy2010}
{Bovy} J.,  {Murray} I.,   {Hogg} D.~W.,  2010, \mn@doi [\apj]
  {10.1088/0004-637X/711/2/1157}, \href
  {https://ui.adsabs.harvard.edu/abs/2010ApJ...711.1157B} {711, 1157}

\bibitem[\protect\citeauthoryear{{Bovy}, {Hogg}  \& {Roweis}}{{Bovy}
  et~al.}{2011}]{Bovy2011}
{Bovy} J.,  {Hogg} D.~W.,   {Roweis} S.~T.,  2011, \mn@doi [Annals of Applied
  Statistics] {10.1214/10-AOAS439}, \href
  {https://ui.adsabs.harvard.edu/abs/2011AnApS...5.1657B} {5, 1657}

\bibitem[\protect\citeauthoryear{{Bovy}, {Rix}, {Liu}, {Hogg}, {Beers}  \&
  {Lee}}{{Bovy} et~al.}{2012}]{Bovy2012}
{Bovy} J.,  {Rix} H.-W.,  {Liu} C.,  {Hogg} D.~W.,  {Beers} T.~C.,   {Lee}
  Y.~S.,  2012, \mn@doi [\apj] {10.1088/0004-637X/753/2/148}, \href
  {https://ui.adsabs.harvard.edu/abs/2012ApJ...753..148B} {753, 148}

\bibitem[\protect\citeauthoryear{{Boylan-Kolchin}, {Bullock}, {Sohn}, {Besla}
  \& {van der Marel}}{{Boylan-Kolchin} et~al.}{2013}]{Boylan-Kolchin2013}
{Boylan-Kolchin} M.,  {Bullock} J.~S.,  {Sohn} S.~T.,  {Besla} G.,   {van der
  Marel} R.~P.,  2013, \mn@doi [\apj] {10.1088/0004-637X/768/2/140}, \href
  {https://ui.adsabs.harvard.edu/abs/2013ApJ...768..140B} {768, 140}

\bibitem[\protect\citeauthoryear{{Bressan}, {Marigo}, {Girardi}, {Salasnich},
  {Dal Cero}, {Rubele}  \& {Nanni}}{{Bressan} et~al.}{2012}]{Bressan2012}
{Bressan} A.,  {Marigo} P.,  {Girardi} L.,  {Salasnich} B.,  {Dal Cero} C.,
  {Rubele} S.,   {Nanni} A.,  2012, \mn@doi [\mnras]
  {10.1111/j.1365-2966.2012.21948.x}, \href
  {https://ui.adsabs.harvard.edu/abs/2012MNRAS.427..127B} {427, 127}

\bibitem[\protect\citeauthoryear{{Brooks}, {Sanders}, {Lilleengen}, {Petersen}
  \& {Pontzen}}{{Brooks} et~al.}{2024}]{Brooks2024}
{Brooks} R. A.~N.,  {Sanders} J.~L.,  {Lilleengen} S.,  {Petersen} M.~S.,
  {Pontzen} A.,  2024, \mn@doi [\mnras] {10.1093/mnras/stae1565}, \href
  {https://ui.adsabs.harvard.edu/abs/2024MNRAS.532.2657B} {532, 2657}

\bibitem[\protect\citeauthoryear{{Buckley}, {Hogg}  \&
  {Price-Whelan}}{{Buckley} et~al.}{2019}]{Buckley2019}
{Buckley} M.~R.,  {Hogg} D.~W.,   {Price-Whelan} A.~M.,  2019, \mn@doi [arXiv
  e-prints] {10.48550/arXiv.1907.00987}, \href
  {https://ui.adsabs.harvard.edu/abs/2019arXiv190700987B} {p. arXiv:1907.00987}

\bibitem[\protect\citeauthoryear{{Buckley}, {Lim}, {Putney}  \&
  {Shih}}{{Buckley} et~al.}{2023}]{Buckley2023}
{Buckley} M.~R.,  {Lim} S.~H.,  {Putney} E.,   {Shih} D.,  2023, \mn@doi
  [\mnras] {10.1093/mnras/stad843}, \href
  {https://ui.adsabs.harvard.edu/abs/2023MNRAS.521.5100B} {521, 5100}

\bibitem[\protect\citeauthoryear{{Caldwell} \& {Ostriker}}{{Caldwell} \&
  {Ostriker}}{1981}]{Caldwell1981}
{Caldwell} J.~A.~R.,  {Ostriker} J.~P.,  1981, \mn@doi [\apj] {10.1086/159441},
  \href {https://ui.adsabs.harvard.edu/abs/1981ApJ...251...61C} {251, 61}

\bibitem[\protect\citeauthoryear{{Callingham} et~al.,}{{Callingham}
  et~al.}{2019}]{Callingham2018}
{Callingham} T.~M.,  et~al., 2019, \mn@doi [\mnras] {10.1093/mnras/stz365},
  \href {https://ui.adsabs.harvard.edu/abs/2019MNRAS.484.5453C} {484, 5453}

\bibitem[\protect\citeauthoryear{{Cautun} et~al.,}{{Cautun}
  et~al.}{2020}]{Cautun2019}
{Cautun} M.,  et~al., 2020, \mn@doi [\mnras] {10.1093/mnras/staa1017}, \href
  {https://ui.adsabs.harvard.edu/abs/2020MNRAS.494.4291C} {494, 4291}

\bibitem[\protect\citeauthoryear{{Chabrier}}{{Chabrier}}{2001}]{Chabrier2001}
{Chabrier} G.,  2001, \mn@doi [\apj] {10.1086/321401}, \href
  {https://ui.adsabs.harvard.edu/abs/2001ApJ...554.1274C} {554, 1274}

\bibitem[\protect\citeauthoryear{{Correa Magnus} \& {Vasiliev}}{{Correa Magnus}
  \& {Vasiliev}}{2022}]{CorreaMagnus2022}
{Correa Magnus} L.,  {Vasiliev} E.,  2022, \mn@doi [\mnras]
  {10.1093/mnras/stab3726}, \href
  {https://ui.adsabs.harvard.edu/abs/2022MNRAS.511.2610C} {511, 2610}

\bibitem[\protect\citeauthoryear{{Cuddeford}}{{Cuddeford}}{1991}]{Cuddeford1991}
{Cuddeford} P.,  1991, \mn@doi [\mnras] {10.1093/mnras/253.3.414}, \href
  {https://ui.adsabs.harvard.edu/abs/1991MNRAS.253..414C} {253, 414}

\bibitem[\protect\citeauthoryear{{Das} \& {Binney}}{{Das} \&
  {Binney}}{2016}]{Das2016}
{Das} P.,  {Binney} J.,  2016, \mn@doi [\mnras] {10.1093/mnras/stw744}, \href
  {https://ui.adsabs.harvard.edu/abs/2016MNRAS.460.1725D} {460, 1725}

\bibitem[\protect\citeauthoryear{{Davis}}{{Davis}}{2021}]{proplot}
{Davis} L. L.~B.,  2021, {ProPlot}, \mn@doi{10.5281/zenodo.5602155}

\bibitem[\protect\citeauthoryear{{Deason}, {Fattahi}, {Belokurov}, {Evans},
  {Grand}, {Marinacci}  \& {Pakmor}}{{Deason} et~al.}{2019}]{Deason2019}
{Deason} A.~J.,  {Fattahi} A.,  {Belokurov} V.,  {Evans} N.~W.,  {Grand} R.
  J.~J.,  {Marinacci} F.,   {Pakmor} R.,  2019, \mn@doi [\mnras]
  {10.1093/mnras/stz623}, \href
  {https://ui.adsabs.harvard.edu/abs/2019MNRAS.485.3514D} {485, 3514}

\bibitem[\protect\citeauthoryear{{Deason} et~al.,}{{Deason}
  et~al.}{2021}]{Deason2020b}
{Deason} A.~J.,  et~al., 2021, \mn@doi [\mnras] {10.1093/mnras/staa3984}, \href
  {https://ui.adsabs.harvard.edu/abs/2021MNRAS.501.5964D} {501, 5964}

\bibitem[\protect\citeauthoryear{{Dehnen} \& {Binney}}{{Dehnen} \&
  {Binney}}{1998}]{Dehnen1998}
{Dehnen} W.,  {Binney} J.,  1998, \mn@doi [\mnras]
  {10.1046/j.1365-8711.1998.01282.x10.1111/j.1365-8711.1998.01282.x}, \href
  {https://ui.adsabs.harvard.edu/abs/1998MNRAS.294..429D} {294, 429}

\bibitem[\protect\citeauthoryear{{Dekel}, {Ishai}, {Dutton}  \&
  {Maccio}}{{Dekel} et~al.}{2017}]{Dekel2017}
{Dekel} A.,  {Ishai} G.,  {Dutton} A.~A.,   {Maccio} A.~V.,  2017, \mn@doi
  [\mnras] {10.1093/mnras/stx486}, \href
  {https://ui.adsabs.harvard.edu/abs/2017MNRAS.468.1005D} {468, 1005}

\bibitem[\protect\citeauthoryear{Delaigle \& Meister}{Delaigle \&
  Meister}{2008}]{Delaigle2008}
Delaigle A.,  Meister A.,  2008, \mn@doi [Bernoulli] {10.3150/08-BEJ121}, 14,
  562

\bibitem[\protect\citeauthoryear{{Dinh}, {Sohl-Dickstein}  \& {Bengio}}{{Dinh}
  et~al.}{2016}]{Dinh2016}
{Dinh} L.,  {Sohl-Dickstein} J.,   {Bengio} S.,  2016, \mn@doi [arXiv e-prints]
  {10.48550/arXiv.1605.08803}, \href
  {https://ui.adsabs.harvard.edu/abs/2016arXiv160508803D} {p. arXiv:1605.08803}

\bibitem[\protect\citeauthoryear{{Eadie}, {Keller}  \& {Harris}}{{Eadie}
  et~al.}{2018}]{Eadie2018}
{Eadie} G.,  {Keller} B.,   {Harris} W.~E.,  2018, \mn@doi [\apj]
  {10.3847/1538-4357/aadb95}, \href
  {https://ui.adsabs.harvard.edu/abs/2018ApJ...865...72E} {865, 72}

\bibitem[\protect\citeauthoryear{{Eddington}}{{Eddington}}{1916}]{Eddington1916}
{Eddington} A.~S.,  1916, \mn@doi [\mnras] {10.1093/mnras/76.7.572}, \href
  {https://ui.adsabs.harvard.edu/abs/1916MNRAS..76..572E} {76, 572}

\bibitem[\protect\citeauthoryear{{Eilers}, {Hogg}, {Rix}  \& {Ness}}{{Eilers}
  et~al.}{2019}]{Eilers2019}
{Eilers} A.-C.,  {Hogg} D.~W.,  {Rix} H.-W.,   {Ness} M.~K.,  2019, \mn@doi
  [\apj] {10.3847/1538-4357/aaf648}, \href
  {https://ui.adsabs.harvard.edu/abs/2019ApJ...871..120E} {871, 120}

\bibitem[\protect\citeauthoryear{{El Gammal}, {Sch{\"o}neberg}, {Torrado}  \&
  {Fidler}}{{El Gammal} et~al.}{2023}]{ElGammal2022}
{El Gammal} J.,  {Sch{\"o}neberg} N.,  {Torrado} J.,   {Fidler} C.,  2023,
  \mn@doi [\jcap] {10.1088/1475-7516/2023/10/021}, \href
  {https://ui.adsabs.harvard.edu/abs/2023JCAP...10..021E} {2023, 021}

\bibitem[\protect\citeauthoryear{{Erkal}, {Belokurov}  \& {Parkin}}{{Erkal}
  et~al.}{2020}]{Erkal2020}
{Erkal} D.,  {Belokurov} V.~A.,   {Parkin} D.~L.,  2020, \mn@doi [\mnras]
  {10.1093/mnras/staa2840}, \href
  {https://ui.adsabs.harvard.edu/abs/2020MNRAS.498.5574E} {498, 5574}

\bibitem[\protect\citeauthoryear{{Evans}, {Hafner}  \& {de Zeeuw}}{{Evans}
  et~al.}{1997}]{Evans1997}
{Evans} N.~W.,  {Hafner} R.~M.,   {de Zeeuw} P.~T.,  1997, \mn@doi [\mnras]
  {10.1093/mnras/286.2.315}, \href
  {https://ui.adsabs.harvard.edu/abs/1997MNRAS.286..315E} {286, 315}

\bibitem[\protect\citeauthoryear{{Evslin} \& {Del Popolo}}{{Evslin} \& {Del
  Popolo}}{2017}]{Evslin2017}
{Evslin} J.,  {Del Popolo} A.,  2017, \mn@doi [\apj]
  {10.3847/1538-4357/aa7205}, \href
  {https://ui.adsabs.harvard.edu/abs/2017ApJ...841...90E} {841, 90}

\bibitem[\protect\citeauthoryear{{Fardal}, {van der Marel}, {Law}, {Sohn},
  {Sesar}, {Hernitschek}  \& {Rix}}{{Fardal} et~al.}{2019}]{Fardal2019}
{Fardal} M.~A.,  {van der Marel} R.~P.,  {Law} D.~R.,  {Sohn} S.~T.,  {Sesar}
  B.,  {Hernitschek} N.,   {Rix} H.-W.,  2019, \mn@doi [\mnras]
  {10.1093/mnras/sty3428}, \href
  {https://ui.adsabs.harvard.edu/abs/2019MNRAS.483.4724F} {483, 4724}

\bibitem[\protect\citeauthoryear{{Foreman-Mackey}, {Hogg}, {Lang}  \&
  {Goodman}}{{Foreman-Mackey} et~al.}{2013}]{Foreman-Mackey2013}
{Foreman-Mackey} D.,  {Hogg} D.~W.,  {Lang} D.,   {Goodman} J.,  2013, \mn@doi
  [\pasp] {10.1086/670067}, \href
  {https://ui.adsabs.harvard.edu/abs/2013PASP..125..306F} {125, 306}

\bibitem[\protect\citeauthoryear{{Freundlich} et~al.,}{{Freundlich}
  et~al.}{2020}]{Freundlich2020b}
{Freundlich} J.,  et~al., 2020, \mn@doi [\mnras] {10.1093/mnras/staa2790},
  \href {https://ui.adsabs.harvard.edu/abs/2020MNRAS.499.2912F} {499, 2912}

\bibitem[\protect\citeauthoryear{{Fritz}, {Di Cintio}, {Battaglia}, {Brook}  \&
  {Taibi}}{{Fritz} et~al.}{2020}]{Fritz2020}
{Fritz} T.~K.,  {Di Cintio} A.,  {Battaglia} G.,  {Brook} C.,   {Taibi} S.,
  2020, \mn@doi [\mnras] {10.1093/mnras/staa1040}, \href
  {https://ui.adsabs.harvard.edu/abs/2020MNRAS.494.5178F} {494, 5178}

\bibitem[\protect\citeauthoryear{{Gaia Collaboration} et~al.,}{{Gaia
  Collaboration} et~al.}{2016}]{GaiaCollaboration2016}
{Gaia Collaboration} et~al., 2016, \mn@doi [\aap]
  {10.1051/0004-6361/201629272}, \href
  {https://ui.adsabs.harvard.edu/abs/2016A&A...595A...1G} {595, A1}

\bibitem[\protect\citeauthoryear{{Gaia Collaboration} et~al.,}{{Gaia
  Collaboration} et~al.}{2018}]{GaiaCollaboration2018d}
{Gaia Collaboration} et~al., 2018, \mn@doi [\aap]
  {10.1051/0004-6361/201833051}, \href
  {https://ui.adsabs.harvard.edu/abs/2018A&A...616A...1G} {616, A1}

\bibitem[\protect\citeauthoryear{{Gaia Collaboration} et~al.,}{{Gaia
  Collaboration} et~al.}{2023}]{GaiaCollaboration2022}
{Gaia Collaboration} et~al., 2023, \mn@doi [\aap]
  {10.1051/0004-6361/202243940}, \href
  {https://ui.adsabs.harvard.edu/abs/2023A&A...674A...1G} {674, A1}

\bibitem[\protect\citeauthoryear{Galassi et~al.,}{Galassi et~al.}{2001}]{gsl}
Galassi M.,  et~al., 2001, {{GNU Scientific Library Reference Manual}}.
{Network Theory Limited}, {Bristol}

\bibitem[\protect\citeauthoryear{{Gerhard}}{{Gerhard}}{1991}]{Gerhard1991}
{Gerhard} O.~E.,  1991, \mn@doi [\mnras] {10.1093/mnras/250.4.812}, \href
  {https://ui.adsabs.harvard.edu/abs/1991MNRAS.250..812G} {250, 812}

\bibitem[\protect\citeauthoryear{{Gibbons}, {Belokurov}  \& {Evans}}{{Gibbons}
  et~al.}{2014}]{Gibbons2014}
{Gibbons} S.~L.~J.,  {Belokurov} V.,   {Evans} N.~W.,  2014, \mn@doi [\mnras]
  {10.1093/mnras/stu1986}, \href
  {https://ui.adsabs.harvard.edu/abs/2014MNRAS.445.3788G} {445, 3788}

\bibitem[\protect\citeauthoryear{{Gieles} \& {Zocchi}}{{Gieles} \&
  {Zocchi}}{2015}]{Gieles2015}
{Gieles} M.,  {Zocchi} A.,  2015, \mn@doi [\mnras] {10.1093/mnras/stv1848},
  \href {https://ui.adsabs.harvard.edu/abs/2015MNRAS.454..576G} {454, 576}

\bibitem[\protect\citeauthoryear{{Grand}, {Deason}, {White}, {Simpson},
  {G{\'o}mez}, {Marinacci}  \& {Pakmor}}{{Grand} et~al.}{2019}]{Grand2019}
{Grand} R. J.~J.,  {Deason} A.~J.,  {White} S. D.~M.,  {Simpson} C.~M.,
  {G{\'o}mez} F.~A.,  {Marinacci} F.,   {Pakmor} R.,  2019, \mn@doi [\mnras]
  {10.1093/mnrasl/slz092}, \href
  {https://ui.adsabs.harvard.edu/abs/2019MNRAS.487L..72G} {487, L72}

\bibitem[\protect\citeauthoryear{{Granger} \& {Perez}}{{Granger} \&
  {Perez}}{2021}]{jupyter}
{Granger} B.~E.,  {Perez} F.,  2021, \mn@doi [Computing in Science and
  Engineering] {10.1109/MCSE.2021.3059263}, \href
  {https://ui.adsabs.harvard.edu/abs/2021CSE....23b...7G} {23, 7}

\bibitem[\protect\citeauthoryear{{Gratton}, {Bragaglia}, {Carretta}, {D'Orazi},
  {Lucatello}  \& {Sollima}}{{Gratton} et~al.}{2019}]{Gratton2019}
{Gratton} R.,  {Bragaglia} A.,  {Carretta} E.,  {D'Orazi} V.,  {Lucatello} S.,
   {Sollima} A.,  2019, \mn@doi [\aapr] {10.1007/s00159-019-0119-3}, \href
  {https://ui.adsabs.harvard.edu/abs/2019A&ARv..27....8G} {27, 8}

\bibitem[\protect\citeauthoryear{{Green} \& {Ting}}{{Green} \&
  {Ting}}{2020}]{Green2020}
{Green} G.~M.,  {Ting} Y.-S.,  2020, \mn@doi [arXiv e-prints]
  {10.48550/arXiv.2011.04673}, \href
  {https://ui.adsabs.harvard.edu/abs/2020arXiv201104673G} {p. arXiv:2011.04673}

\bibitem[\protect\citeauthoryear{{Green}, {Ting}  \& {Kamdar}}{{Green}
  et~al.}{2023}]{Green2022}
{Green} G.~M.,  {Ting} Y.-S.,   {Kamdar} H.,  2023, \mn@doi [\apj]
  {10.3847/1538-4357/aca3a7}, \href
  {https://ui.adsabs.harvard.edu/abs/2023ApJ...942...26G} {942, 26}

\bibitem[\protect\citeauthoryear{{Han}, {Wang}, {Cole}  \& {Frenk}}{{Han}
  et~al.}{2016a}]{Han2016b}
{Han} J.,  {Wang} W.,  {Cole} S.,   {Frenk} C.~S.,  2016a, \mn@doi [\mnras]
  {10.1093/mnras/stv2707}, \href
  {https://ui.adsabs.harvard.edu/abs/2016MNRAS.456.1003H} {456, 1003}

\bibitem[\protect\citeauthoryear{{Han}, {Wang}, {Cole}  \& {Frenk}}{{Han}
  et~al.}{2016b}]{Han2016a}
{Han} J.,  {Wang} W.,  {Cole} S.,   {Frenk} C.~S.,  2016b, \mn@doi [\mnras]
  {10.1093/mnras/stv2522}, \href
  {https://ui.adsabs.harvard.edu/abs/2016MNRAS.456.1017H} {456, 1017}

\bibitem[\protect\citeauthoryear{{Han}, {Wang}  \& {Li}}{{Han}
  et~al.}{2020}]{Han2019}
{Han} J.,  {Wang} W.,   {Li} Z.,  2020, in {Valluri} M.,  {Sellwood} J.~A.,
  eds,  IAU Symposium Vol. 353, Galactic Dynamics in the Era of Large Surveys.
  pp 109--112 (\mn@eprint {arXiv} {1909.02690}),
  \mn@doi{10.1017/S1743921319008020}

\bibitem[\protect\citeauthoryear{{Harris} et~al.,}{{Harris}
  et~al.}{2020}]{numpy}
{Harris} C.~R.,  et~al., 2020, \mn@doi [\nat] {10.1038/s41586-020-2649-2},
  \href {https://ui.adsabs.harvard.edu/abs/2020Natur.585..357H} {585, 357}

\bibitem[\protect\citeauthoryear{{Head}, {Kumar}, {Nahrstaedt}, {Louppe}  \&
  {Shcherbatyi}}{{Head} et~al.}{2021}]{skopt}
{Head} T.,  {Kumar} M.,  {Nahrstaedt} H.,  {Louppe} G.,   {Shcherbatyi} I.,
  2021, {scikit-optimize/scikit-optimize}, \mn@doi{10.5281/zenodo.5565057}

\bibitem[\protect\citeauthoryear{{Helmi} \& {White}}{{Helmi} \&
  {White}}{1999}]{Helmi1999}
{Helmi} A.,  {White} S. D.~M.,  1999, \mn@doi [\mnras]
  {10.1046/j.1365-8711.1999.02616.x}, \href
  {https://ui.adsabs.harvard.edu/abs/1999MNRAS.307..495H} {307, 495}

\bibitem[\protect\citeauthoryear{{Horta}, {Price-Whelan}, {Hogg}, {Johnston},
  {Widrow}, {Dalcanton}, {Ness}  \& {Hunt}}{{Horta} et~al.}{2024}]{Horta2023}
{Horta} D.,  {Price-Whelan} A.~M.,  {Hogg} D.~W.,  {Johnston} K.~V.,  {Widrow}
  L.,  {Dalcanton} J.~J.,  {Ness} M.~K.,   {Hunt} J. A.~S.,  2024, \mn@doi
  [\apj] {10.3847/1538-4357/ad16e8}, \href
  {https://ui.adsabs.harvard.edu/abs/2024ApJ...962..165H} {962, 165}

\bibitem[\protect\citeauthoryear{{Huang} et~al.,}{{Huang}
  et~al.}{2016}]{Huang2016}
{Huang} Y.,  et~al., 2016, \mn@doi [\mnras] {10.1093/mnras/stw2096}, \href
  {https://ui.adsabs.harvard.edu/abs/2016MNRAS.463.2623H} {463, 2623}

\bibitem[\protect\citeauthoryear{{Hunt} \& {Vasiliev}}{{Hunt} \&
  {Vasiliev}}{2025}]{Hunt2025}
{Hunt} J. A.~S.,  {Vasiliev} E.,  2025, \mn@doi [arXiv e-prints]
  {10.48550/arXiv.2501.04075}, \href
  {https://ui.adsabs.harvard.edu/abs/2025arXiv250104075H} {p. arXiv:2501.04075}

\bibitem[\protect\citeauthoryear{{Hunter}}{{Hunter}}{2007}]{matplotlib}
{Hunter} J.~D.,  2007, \mn@doi [Computing in Science and Engineering]
  {10.1109/MCSE.2007.55}, \href
  {https://ui.adsabs.harvard.edu/abs/2007CSE.....9...90H} {9, 90}

\bibitem[\protect\citeauthoryear{{Ibata} et~al.,}{{Ibata}
  et~al.}{2024}]{Ibata2023}
{Ibata} R.,  et~al., 2024, \mn@doi [\apj] {10.3847/1538-4357/ad382d}, \href
  {https://ui.adsabs.harvard.edu/abs/2024ApJ...967...89I} {967, 89}

\bibitem[\protect\citeauthoryear{{Jethwa}, {Erkal}  \& {Belokurov}}{{Jethwa}
  et~al.}{2016}]{Jethwa2016}
{Jethwa} P.,  {Erkal} D.,   {Belokurov} V.,  2016, \mn@doi [\mnras]
  {10.1093/mnras/stw1343}, \href
  {https://ui.adsabs.harvard.edu/abs/2016MNRAS.461.2212J} {461, 2212}

\bibitem[\protect\citeauthoryear{{Jiao}, {Hammer}, {Wang}, {Wang}, {Amram},
  {Chemin}  \& {Yang}}{{Jiao} et~al.}{2023}]{Jiao2023}
{Jiao} Y.,  {Hammer} F.,  {Wang} H.,  {Wang} J.,  {Amram} P.,  {Chemin} L.,
  {Yang} Y.,  2023, \mn@doi [\aap] {10.1051/0004-6361/202347513}, \href
  {https://ui.adsabs.harvard.edu/abs/2023A&A...678A.208J} {678, A208}

\bibitem[\protect\citeauthoryear{Jin, Zhang, Balakrishnan, Wainwright  \&
  Jordan}{Jin et~al.}{2016}]{Jin2016}
Jin C.,  Zhang Y.,  Balakrishnan S.,  Wainwright M.~J.,   Jordan M.~I.,  2016,
  in Lee D.~D.,  Sugiyama M.,  Luxburg U.~V.,  Guyon I.,   Garnett R.,  eds, ,
  Advances in {{Neural Information Processing Systems}} 29.
Curran Associates, Inc., pp 4116--4124

\bibitem[\protect\citeauthoryear{{Jing}}{{Jing}}{2000}]{Jing2000a}
{Jing} Y.~P.,  2000, \mn@doi [\apj] {10.1086/308809}, \href
  {https://ui.adsabs.harvard.edu/abs/2000ApJ...535...30J} {535, 30}

\bibitem[\protect\citeauthoryear{Kandasamy, Schneider  \& P{\'o}czos}{Kandasamy
  et~al.}{2015}]{Kandasamy2015}
Kandasamy K.,  Schneider J.,   P{\'o}czos B.,  2015, in Proceedings of the 24th
  {{International Conference}} on {{Artificial Intelligence}}. {{IJCAI}}'15.
AAAI Press, Buenos Aires, Argentina, pp 3605--3611

\bibitem[\protect\citeauthoryear{{Karukes}, {Benito}, {Iocco}, {Trotta}  \&
  {Geringer-Sameth}}{{Karukes} et~al.}{2019}]{Karukes2019}
{Karukes} E.~V.,  {Benito} M.,  {Iocco} F.,  {Trotta} R.,   {Geringer-Sameth}
  A.,  2019, \mn@doi [\jcap] {10.1088/1475-7516/2019/09/046}, \href
  {https://ui.adsabs.harvard.edu/abs/2019JCAP...09..046K} {2019, 046}

\bibitem[\protect\citeauthoryear{{Kent} \& {Gunn}}{{Kent} \&
  {Gunn}}{1982}]{Kent1982}
{Kent} S.~M.,  {Gunn} J.~E.,  1982, \mn@doi [\aj] {10.1086/113178}, \href
  {https://ui.adsabs.harvard.edu/abs/1982AJ.....87..945K} {87, 945}

\bibitem[\protect\citeauthoryear{{Kipper}, {Tenjes}, {Tempel}  \& {de
  Propris}}{{Kipper} et~al.}{2021}]{Kipper2021}
{Kipper} R.,  {Tenjes} P.,  {Tempel} E.,   {de Propris} R.,  2021, \mn@doi
  [\mnras] {10.1093/mnras/stab2104}, \href
  {https://ui.adsabs.harvard.edu/abs/2021MNRAS.506.5559K} {506, 5559}

\bibitem[\protect\citeauthoryear{{Koposov} et~al.,}{{Koposov}
  et~al.}{2008}]{Koposov2008}
{Koposov} S.,  et~al., 2008, \mn@doi [\apj] {10.1086/589911}, \href
  {https://ui.adsabs.harvard.edu/abs/2008ApJ...686..279K} {686, 279}

\bibitem[\protect\citeauthoryear{{Koposov} et~al.,}{{Koposov}
  et~al.}{2023}]{Koposov2023}
{Koposov} S.~E.,  et~al., 2023, \mn@doi [\mnras] {10.1093/mnras/stad551}, \href
  {https://ui.adsabs.harvard.edu/abs/2023MNRAS.521.4936K} {521, 4936}

\bibitem[\protect\citeauthoryear{{Kravtsov} \& {Winney}}{{Kravtsov} \&
  {Winney}}{2024}]{Kravtsov2024}
{Kravtsov} A.,  {Winney} S.,  2024, \mn@doi [The Open Journal of Astrophysics]
  {10.33232/001c.120316}, \href
  {https://ui.adsabs.harvard.edu/abs/2024OJAp....7E..50K} {7, 50}

\bibitem[\protect\citeauthoryear{{Leonard} \& {Tremaine}}{{Leonard} \&
  {Tremaine}}{1990}]{Leonard1990}
{Leonard} P. J.~T.,  {Tremaine} S.,  1990, \mn@doi [\apj] {10.1086/168638},
  \href {https://ui.adsabs.harvard.edu/abs/1990ApJ...353..486L} {353, 486}

\bibitem[\protect\citeauthoryear{{Li} \& {Han}}{{Li} \& {Han}}{2021}]{Li2021b}
{Li} Z.-Z.,  {Han} J.,  2021, \mn@doi [\apjl] {10.3847/2041-8213/ac0a7f}, \href
  {https://ui.adsabs.harvard.edu/abs/2021ApJ...915L..18L} {915, L18}

\bibitem[\protect\citeauthoryear{{Li} \& {White}}{{Li} \&
  {White}}{2008}]{Li2008}
{Li} Y.-S.,  {White} S. D.~M.,  2008, \mn@doi [\mnras]
  {10.1111/j.1365-2966.2007.12748.x}, \href
  {https://ui.adsabs.harvard.edu/abs/2008MNRAS.384.1459L} {384, 1459}

\bibitem[\protect\citeauthoryear{{Li}, {Jing}, {Qian}, {Yuan}  \& {Zhao}}{{Li}
  et~al.}{2017}]{Li2017}
{Li} Z.-Z.,  {Jing} Y.~P.,  {Qian} Y.-Z.,  {Yuan} Z.,   {Zhao} D.-H.,  2017,
  \mn@doi [\apj] {10.3847/1538-4357/aa94c0}, \href
  {https://ui.adsabs.harvard.edu/abs/2017ApJ...850..116L} {850, 116}

\bibitem[\protect\citeauthoryear{{Li}, {Qian}, {Han}, {Wang}  \& {Jing}}{{Li}
  et~al.}{2019}]{Li2019}
{Li} Z.-Z.,  {Qian} Y.-Z.,  {Han} J.,  {Wang} W.,   {Jing} Y.~P.,  2019,
  \mn@doi [\apj] {10.3847/1538-4357/ab4f6d}, \href
  {https://ui.adsabs.harvard.edu/abs/2019ApJ...886...69L} {886, 69}

\bibitem[\protect\citeauthoryear{{Li}, {Qian}, {Han}, {Li}, {Wang}  \&
  {Jing}}{{Li} et~al.}{2020a}]{Li2020a}
{Li} Z.-Z.,  {Qian} Y.-Z.,  {Han} J.,  {Li} T.~S.,  {Wang} W.,   {Jing} Y.~P.,
  2020a, \mn@doi [\apj] {10.3847/1538-4357/ab84f0}, \href
  {https://ui.adsabs.harvard.edu/abs/2020ApJ...894...10L} {894, 10}

\bibitem[\protect\citeauthoryear{{Li}, {Shao}, {Li}, {Yu}, {Zhong}  \&
  {Chen}}{{Li} et~al.}{2020b}]{Li2020b}
{Li} L.,  {Shao} Z.,  {Li} Z.-Z.,  {Yu} J.,  {Zhong} J.,   {Chen} L.,  2020b,
  \mn@doi [\apj] {10.3847/1538-4357/abaef3}, \href
  {https://ui.adsabs.harvard.edu/abs/2020ApJ...901...49L} {901, 49}

\bibitem[\protect\citeauthoryear{{Li}, {Han}, {Wang}, {Cui}, {Li}  \&
  {Yang}}{{Li} et~al.}{2021}]{Li2021d}
{Li} Q.,  {Han} J.,  {Wang} W.,  {Cui} W.,  {Li} Z.,   {Yang} X.,  2021,
  \mn@doi [\mnras] {10.1093/mnras/stab1633}, \href
  {https://ui.adsabs.harvard.edu/abs/2021MNRAS.505.3907L} {505, 3907}

\bibitem[\protect\citeauthoryear{{Li}, {Han}, {Wang}, {Cui}, {De Luca}, {Yang},
  {Zhou}  \& {Shi}}{{Li} et~al.}{2022}]{Li2022}
{Li} Q.,  {Han} J.,  {Wang} W.,  {Cui} W.,  {De Luca} F.,  {Yang} X.,  {Zhou}
  Y.,   {Shi} R.,  2022, \mn@doi [\mnras] {10.1093/mnras/stac1739}, \href
  {https://ui.adsabs.harvard.edu/abs/2022MNRAS.514.5890L} {514, 5890}

\bibitem[\protect\citeauthoryear{{Li}, {Dekel}, {Mandelker}, {Freundlich}  \&
  {Fran{\c{c}}ois}}{{Li} et~al.}{2023}]{Li2023}
{Li} Z.,  {Dekel} A.,  {Mandelker} N.,  {Freundlich} J.,   {Fran{\c{c}}ois}
  T.~L.,  2023, \mn@doi [\mnras] {10.1093/mnras/stac3233}, \href
  {https://ui.adsabs.harvard.edu/abs/2023MNRAS.518.5356L} {518, 5356}

\bibitem[\protect\citeauthoryear{{Lim}, {Raman}, {Buckley}  \& {Shih}}{{Lim}
  et~al.}{2024}]{Lim2022}
{Lim} S.~H.,  {Raman} K.~A.,  {Buckley} M.~R.,   {Shih} D.,  2024, \mn@doi
  [\mnras] {10.1093/mnras/stae1672}, \href
  {https://ui.adsabs.harvard.edu/abs/2024MNRAS.533..143L} {533, 143}

\bibitem[\protect\citeauthoryear{{Lim}, {Putney}, {Buckley}  \& {Shih}}{{Lim}
  et~al.}{2025}]{Lim2023}
{Lim} S.~H.,  {Putney} E.,  {Buckley} M.~R.,   {Shih} D.,  2025, \mn@doi
  [\jcap] {10.1088/1475-7516/2025/01/021}, \href
  {https://ui.adsabs.harvard.edu/abs/2025JCAP...01..021L} {2025, 021}

\bibitem[\protect\citeauthoryear{{{\L}okas} \& {Mamon}}{{{\L}okas} \&
  {Mamon}}{2003}]{Lokas2003}
{{\L}okas} E.~L.,  {Mamon} G.~A.,  2003, \mn@doi [\mnras]
  {10.1046/j.1365-8711.2003.06684.x}, \href
  {https://ui.adsabs.harvard.edu/abs/2003MNRAS.343..401L} {343, 401}

\bibitem[\protect\citeauthoryear{{Lucy}}{{Lucy}}{1974}]{Lucy1974}
{Lucy} L.~B.,  1974, \mn@doi [\aj] {10.1086/111605}, \href
  {https://ui.adsabs.harvard.edu/abs/1974AJ.....79..745L} {79, 745}

\bibitem[\protect\citeauthoryear{{Mackereth} \& {Bovy}}{{Mackereth} \&
  {Bovy}}{2020}]{Mackereth2020}
{Mackereth} J.~T.,  {Bovy} J.,  2020, \mn@doi [\mnras] {10.1093/mnras/staa047},
  \href {https://ui.adsabs.harvard.edu/abs/2020MNRAS.492.3631M} {492, 3631}

\bibitem[\protect\citeauthoryear{{Magorrian}}{{Magorrian}}{2014}]{Magorrian2014}
{Magorrian} J.,  2014, \mn@doi [\mnras] {10.1093/mnras/stt2031}, \href
  {https://ui.adsabs.harvard.edu/abs/2014MNRAS.437.2230M} {437, 2230}

\bibitem[\protect\citeauthoryear{{Magorrian}}{{Magorrian}}{2019}]{Magorrian2019}
{Magorrian} J.,  2019, \mn@doi [\mnras] {10.1093/mnras/stz037}, \href
  {https://ui.adsabs.harvard.edu/abs/2019MNRAS.484.1166M} {484, 1166}

\bibitem[\protect\citeauthoryear{{Malhan} \& {Ibata}}{{Malhan} \&
  {Ibata}}{2019}]{Malhan2019}
{Malhan} K.,  {Ibata} R.~A.,  2019, \mn@doi [\mnras] {10.1093/mnras/stz1035},
  \href {https://ui.adsabs.harvard.edu/abs/2019MNRAS.486.2995M} {486, 2995}

\bibitem[\protect\citeauthoryear{{Manwadkar} \& {Kravtsov}}{{Manwadkar} \&
  {Kravtsov}}{2022}]{Manwadkar2022}
{Manwadkar} V.,  {Kravtsov} A.~V.,  2022, \mn@doi [\mnras]
  {10.1093/mnras/stac2452}, \href
  {https://ui.adsabs.harvard.edu/abs/2022MNRAS.516.3944M} {516, 3944}

\bibitem[\protect\citeauthoryear{Maronna, Martin, Yohai  \&
  {Salibi{\'a}n-Barrera}}{Maronna et~al.}{2019}]{Maronna2019}
Maronna R.~A.,  Martin R.~D.,  Yohai V.~J.,   {Salibi{\'a}n-Barrera} M.,  2019,
  Robust {{Statistics}}: {{Theory}} and {{Methods}} (with {{R}}).
John Wiley \& Sons

\bibitem[\protect\citeauthoryear{{McMillan}}{{McMillan}}{2017}]{McMillan2017}
{McMillan} P.~J.,  2017, \mn@doi [\mnras] {10.1093/mnras/stw2759}, \href
  {https://ui.adsabs.harvard.edu/abs/2017MNRAS.465...76M} {465, 76}

\bibitem[\protect\citeauthoryear{{McMillan} \& {Binney}}{{McMillan} \&
  {Binney}}{2012}]{McMillan2012}
{McMillan} P.~J.,  {Binney} J.,  2012, \mn@doi [\mnras]
  {10.1111/j.1365-2966.2011.19879.x}, \href
  {https://ui.adsabs.harvard.edu/abs/2012MNRAS.419.2251M} {419, 2251}

\bibitem[\protect\citeauthoryear{{Merritt}}{{Merritt}}{1985}]{Merritt1985}
{Merritt} D.,  1985, \mn@doi [\aj] {10.1086/113810}, \href
  {https://ui.adsabs.harvard.edu/abs/1985AJ.....90.1027M} {90, 1027}

\bibitem[\protect\citeauthoryear{{Michie} \& {Bodenheimer}}{{Michie} \&
  {Bodenheimer}}{1963}]{Michie1963}
{Michie} R.~W.,  {Bodenheimer} P.~H.,  1963, \mn@doi [\mnras]
  {10.1093/mnras/126.3.269}, \href
  {https://ui.adsabs.harvard.edu/abs/1963MNRAS.126..269M} {126, 269}

\bibitem[\protect\citeauthoryear{{Nadler} et~al.,}{{Nadler}
  et~al.}{2020}]{Nadler2019}
{Nadler} E.~O.,  et~al., 2020, \mn@doi [\apj] {10.3847/1538-4357/ab846a}, \href
  {https://ui.adsabs.harvard.edu/abs/2020ApJ...893...48N} {893, 48}

\bibitem[\protect\citeauthoryear{{Naik}, {An}, {Burrage}  \& {Evans}}{{Naik}
  et~al.}{2022}]{Naik2022}
{Naik} A.~P.,  {An} J.,  {Burrage} C.,   {Evans} N.~W.,  2022, \mn@doi [\mnras]
  {10.1093/mnras/stac153}, \href
  {https://ui.adsabs.harvard.edu/abs/2022MNRAS.511.1609N} {511, 1609}

\bibitem[\protect\citeauthoryear{{Navarro}, {Frenk}  \& {White}}{{Navarro}
  et~al.}{1996}]{Navarro1996}
{Navarro} J.~F.,  {Frenk} C.~S.,   {White} S. D.~M.,  1996, \mn@doi [\apj]
  {10.1086/177173}, \href
  {https://ui.adsabs.harvard.edu/abs/1996ApJ...462..563N} {462, 563}

\bibitem[\protect\citeauthoryear{{Necib} \& {Lin}}{{Necib} \&
  {Lin}}{2022}]{Necib2022}
{Necib} L.,  {Lin} T.,  2022, \mn@doi [\apj] {10.3847/1538-4357/ac4244}, \href
  {https://ui.adsabs.harvard.edu/abs/2022ApJ...926..189N} {926, 189}

\bibitem[\protect\citeauthoryear{{Newton}, {Cautun}, {Jenkins}, {Frenk}  \&
  {Helly}}{{Newton} et~al.}{2018}]{Newton2018}
{Newton} O.,  {Cautun} M.,  {Jenkins} A.,  {Frenk} C.~S.,   {Helly} J.~C.,
  2018, \mn@doi [\mnras] {10.1093/mnras/sty1085}, \href
  {https://ui.adsabs.harvard.edu/abs/2018MNRAS.479.2853N} {479, 2853}

\bibitem[\protect\citeauthoryear{{Odland}}{{Odland}}{2018}]{kdepy}
{Odland} T.,  2018, {tommyod/KDEpy: Kernel Density Estimation in Python},
  \mn@doi{10.5281/zenodo.2392268}

\bibitem[\protect\citeauthoryear{{Osipkov}}{{Osipkov}}{1979}]{Osipkov1979}
{Osipkov} L.~P.,  1979, Soviet Astronomy Letters, \href
  {https://ui.adsabs.harvard.edu/abs/1979SvAL....5...42O} {5, 42}

\bibitem[\protect\citeauthoryear{{Ou}, {Eilers}, {Necib}  \& {Frebel}}{{Ou}
  et~al.}{2024}]{Ou2024}
{Ou} X.,  {Eilers} A.-C.,  {Necib} L.,   {Frebel} A.,  2024, \mn@doi [\mnras]
  {10.1093/mnras/stae034}, \href
  {https://ui.adsabs.harvard.edu/abs/2024MNRAS.528..693O} {528, 693}

\bibitem[\protect\citeauthoryear{{Pace}, {Erkal}  \& {Li}}{{Pace}
  et~al.}{2022}]{Pace2022}
{Pace} A.~B.,  {Erkal} D.,   {Li} T.~S.,  2022, \mn@doi [\apj]
  {10.3847/1538-4357/ac997b}, \href
  {https://ui.adsabs.harvard.edu/abs/2022ApJ...940..136P} {940, 136}

\bibitem[\protect\citeauthoryear{{Pardy} et~al.,}{{Pardy}
  et~al.}{2020}]{Pardy2019}
{Pardy} S.~A.,  et~al., 2020, \mn@doi [\mnras] {10.1093/mnras/stz3192}, \href
  {https://ui.adsabs.harvard.edu/abs/2020MNRAS.492.1543P} {492, 1543}

\bibitem[\protect\citeauthoryear{{Patel}, {Besla}, {Mandel}  \& {Sohn}}{{Patel}
  et~al.}{2018}]{Patel2018}
{Patel} E.,  {Besla} G.,  {Mandel} K.,   {Sohn} S.~T.,  2018, \mn@doi [\apj]
  {10.3847/1538-4357/aab78f}, \href
  {https://ui.adsabs.harvard.edu/abs/2018ApJ...857...78P} {857, 78}

\bibitem[\protect\citeauthoryear{{Pe{\~n}arrubia} \&
  {Fattahi}}{{Pe{\~n}arrubia} \& {Fattahi}}{2017}]{Penarrubia2017}
{Pe{\~n}arrubia} J.,  {Fattahi} A.,  2017, \mn@doi [\mnras]
  {10.1093/mnras/stx323}, \href
  {https://ui.adsabs.harvard.edu/abs/2017MNRAS.468.1300P} {468, 1300}

\bibitem[\protect\citeauthoryear{{Pe{\~n}arrubia}, {Koposov}  \&
  {Walker}}{{Pe{\~n}arrubia} et~al.}{2012}]{Penarrubia2012}
{Pe{\~n}arrubia} J.,  {Koposov} S.~E.,   {Walker} M.~G.,  2012, \mn@doi [\apj]
  {10.1088/0004-637X/760/1/2}, \href
  {https://ui.adsabs.harvard.edu/abs/2012ApJ...760....2P} {760, 2}

\bibitem[\protect\citeauthoryear{{Pe{\~n}arrubia}, {G{\'o}mez}, {Besla},
  {Erkal}  \& {Ma}}{{Pe{\~n}arrubia} et~al.}{2016}]{Penarrubia2016}
{Pe{\~n}arrubia} J.,  {G{\'o}mez} F.~A.,  {Besla} G.,  {Erkal} D.,   {Ma}
  Y.-Z.,  2016, \mn@doi [\mnras] {10.1093/mnrasl/slv160}, \href
  {https://ui.adsabs.harvard.edu/abs/2016MNRAS.456L..54P} {456, L54}

\bibitem[\protect\citeauthoryear{{Petersen} \& {Pe{\~n}arrubia}}{{Petersen} \&
  {Pe{\~n}arrubia}}{2020}]{Petersen2020}
{Petersen} M.~S.,  {Pe{\~n}arrubia} J.,  2020, \mn@doi [\mnras]
  {10.1093/mnrasl/slaa029}, \href
  {https://ui.adsabs.harvard.edu/abs/2020MNRAS.494L..11P} {494, L11}

\bibitem[\protect\citeauthoryear{{Petroff}}{{Petroff}}{2021}]{Petroff2021}
{Petroff} M.~A.,  2021, \mn@doi [arXiv e-prints] {10.48550/arXiv.2107.02270},
  \href {https://ui.adsabs.harvard.edu/abs/2021arXiv210702270P} {p.
  arXiv:2107.02270}

\bibitem[\protect\citeauthoryear{{Posti} \& {Helmi}}{{Posti} \&
  {Helmi}}{2019}]{Posti2018}
{Posti} L.,  {Helmi} A.,  2019, \mn@doi [\aap] {10.1051/0004-6361/201833355},
  \href {https://ui.adsabs.harvard.edu/abs/2019A&A...621A..56P} {621, A56}

\bibitem[\protect\citeauthoryear{{Posti}, {Binney}, {Nipoti}  \&
  {Ciotti}}{{Posti} et~al.}{2015}]{Posti2015}
{Posti} L.,  {Binney} J.,  {Nipoti} C.,   {Ciotti} L.,  2015, \mn@doi [\mnras]
  {10.1093/mnras/stu2608}, \href
  {https://ui.adsabs.harvard.edu/abs/2015MNRAS.447.3060P} {447, 3060}

\bibitem[\protect\citeauthoryear{{Price-Whelan} et~al.,}{{Price-Whelan}
  et~al.}{2021}]{Price-Whelan2021}
{Price-Whelan} A.~M.,  et~al., 2021, \mn@doi [\apj] {10.3847/1538-4357/abe1b7},
  \href {https://ui.adsabs.harvard.edu/abs/2021ApJ...910...17P} {910, 17}

\bibitem[\protect\citeauthoryear{{Read} et~al.,}{{Read}
  et~al.}{2021}]{Read2020}
{Read} J.~I.,  et~al., 2021, \mn@doi [\mnras] {10.1093/mnras/staa3663}, \href
  {https://ui.adsabs.harvard.edu/abs/2021MNRAS.501..978R} {501, 978}

\bibitem[\protect\citeauthoryear{{Rehemtulla}, {Valluri}  \&
  {Vasiliev}}{{Rehemtulla} et~al.}{2022}]{Rehemtulla2022}
{Rehemtulla} N.,  {Valluri} M.,   {Vasiliev} E.,  2022, \mn@doi [\mnras]
  {10.1093/mnras/stac400}, \href
  {https://ui.adsabs.harvard.edu/abs/2022MNRAS.511.5536R} {511, 5536}

\bibitem[\protect\citeauthoryear{{Reino}, {Rossi}, {Sanderson}, {Sellentin},
  {Helmi}, {Koppelman}  \& {Sharma}}{{Reino} et~al.}{2021}]{Reino2020}
{Reino} S.,  {Rossi} E.~M.,  {Sanderson} R.~E.,  {Sellentin} E.,  {Helmi} A.,
  {Koppelman} H.~H.,   {Sharma} S.,  2021, \mn@doi [\mnras]
  {10.1093/mnras/stab304}, \href
  {https://ui.adsabs.harvard.edu/abs/2021MNRAS.502.4170R} {502, 4170}

\bibitem[\protect\citeauthoryear{{Richardson}}{{Richardson}}{1972}]{Richardson1972}
{Richardson} W.~H.,  1972, Journal of the Optical Society of America
  (1917-1983), \href {https://ui.adsabs.harvard.edu/abs/1972JOSA...62...55R}
  {62, 55}

\bibitem[\protect\citeauthoryear{{Roche}, {Necib}, {Lin}, {Ou}  \&
  {Nguyen}}{{Roche} et~al.}{2024}]{Roche2024}
{Roche} C.,  {Necib} L.,  {Lin} T.,  {Ou} X.,   {Nguyen} T.,  2024, \mn@doi
  [\apj] {10.3847/1538-4357/ad58d7}, \href
  {https://ui.adsabs.harvard.edu/abs/2024ApJ...972...70R} {972, 70}

\bibitem[\protect\citeauthoryear{{Rothfuss}, {Ferreira}, {Walther}  \&
  {Ulrich}}{{Rothfuss} et~al.}{2019}]{Rothfuss2019}
{Rothfuss} J.,  {Ferreira} F.,  {Walther} S.,   {Ulrich} M.,  2019, \mn@doi
  [arXiv e-prints] {10.48550/arXiv.1903.00954}, \href
  {https://ui.adsabs.harvard.edu/abs/2019arXiv190300954R} {p. arXiv:1903.00954}

\bibitem[\protect\citeauthoryear{{Sanders} \& {Binney}}{{Sanders} \&
  {Binney}}{2015}]{Sanders2015b}
{Sanders} J.~L.,  {Binney} J.,  2015, \mn@doi [\mnras] {10.1093/mnras/stv578},
  \href {https://ui.adsabs.harvard.edu/abs/2015MNRAS.449.3479S} {449, 3479}

\bibitem[\protect\citeauthoryear{{Sanderson}, {Helmi}  \& {Hogg}}{{Sanderson}
  et~al.}{2015}]{Sanderson2015}
{Sanderson} R.~E.,  {Helmi} A.,   {Hogg} D.~W.,  2015, \mn@doi [\apj]
  {10.1088/0004-637X/801/2/98}, \href
  {https://ui.adsabs.harvard.edu/abs/2015ApJ...801...98S} {801, 98}

\bibitem[\protect\citeauthoryear{{Schaller} et~al.,}{{Schaller}
  et~al.}{2015}]{Schaller2015}
{Schaller} M.,  et~al., 2015, \mn@doi [\mnras] {10.1093/mnras/stv1067}, \href
  {https://ui.adsabs.harvard.edu/abs/2015MNRAS.451.1247S} {451, 1247}

\bibitem[\protect\citeauthoryear{{Schwarzschild}}{{Schwarzschild}}{1979}]{Schwarzschild1979}
{Schwarzschild} M.,  1979, \mn@doi [\apj] {10.1086/157282}, \href
  {https://ui.adsabs.harvard.edu/abs/1979ApJ...232..236S} {232, 236}

\bibitem[\protect\citeauthoryear{Scott}{Scott}{1979}]{Scott1979}
Scott D.~W.,  1979, \mn@doi [Biometrika] {10.1093/biomet/66.3.605}, 66, 605

\bibitem[\protect\citeauthoryear{Scott}{Scott}{2015}]{Scott2015}
Scott D.~W.,  2015, Multivariate {{Density Estimation}}: {{Theory}},
  {{Practice}}, and {{Visualization}}, 2nd edn.
Wiley, Hoboken, New Jersey

\bibitem[\protect\citeauthoryear{{Sharma} \& {Steinmetz}}{{Sharma} \&
  {Steinmetz}}{2006}]{Sharma2006}
{Sharma} S.,  {Steinmetz} M.,  2006, \mn@doi [\mnras]
  {10.1111/j.1365-2966.2006.11043.x}, \href
  {https://ui.adsabs.harvard.edu/abs/2006MNRAS.373.1293S} {373, 1293}

\bibitem[\protect\citeauthoryear{{Shen} et~al.,}{{Shen}
  et~al.}{2022}]{Shen2022}
{Shen} J.,  et~al., 2022, \mn@doi [\apj] {10.3847/1538-4357/ac3a7a}, \href
  {https://ui.adsabs.harvard.edu/abs/2022ApJ...925....1S} {925, 1}

\bibitem[\protect\citeauthoryear{Silverman}{Silverman}{1986}]{Silverman1986}
Silverman B.~W.,  1986, Density {{Estimation}} for {{Statistics}} and {{Data
  Analysis}}.
{Chapman and Hall}, Boca Raton

\bibitem[\protect\citeauthoryear{{Simon}}{{Simon}}{2019}]{Simon2019a}
{Simon} J.~D.,  2019, \mn@doi [\araa] {10.1146/annurev-astro-091918-104453},
  \href {https://ui.adsabs.harvard.edu/abs/2019ARA&A..57..375S} {57, 375}

\bibitem[\protect\citeauthoryear{{Slizewski}, {Dufresne}, {Murdock}, {Eadie},
  {Sanderson}, {Wetzel}  \& {Juri{\'c}}}{{Slizewski}
  et~al.}{2022}]{Slizewski2022}
{Slizewski} A.,  {Dufresne} X.,  {Murdock} K.,  {Eadie} G.,  {Sanderson} R.,
  {Wetzel} A.,   {Juri{\'c}} M.,  2022, \mn@doi [\apj]
  {10.3847/1538-4357/ac390b}, \href
  {https://ui.adsabs.harvard.edu/abs/2022ApJ...924..131S} {924, 131}

\bibitem[\protect\citeauthoryear{{Smith} et~al.,}{{Smith}
  et~al.}{2007}]{Smith2007}
{Smith} M.~C.,  et~al., 2007, \mn@doi [\mnras]
  {10.1111/j.1365-2966.2007.11964.x}, \href
  {https://ui.adsabs.harvard.edu/abs/2007MNRAS.379..755S} {379, 755}

\bibitem[\protect\citeauthoryear{{Sohn}, {Watkins}, {Fardal}, {van der Marel},
  {Deason}, {Besla}  \& {Bellini}}{{Sohn} et~al.}{2018}]{Sohn2018}
{Sohn} S.~T.,  {Watkins} L.~L.,  {Fardal} M.~A.,  {van der Marel} R.~P.,
  {Deason} A.~J.,  {Besla} G.,   {Bellini} A.,  2018, \mn@doi [\apj]
  {10.3847/1538-4357/aacd0b}, \href
  {https://ui.adsabs.harvard.edu/abs/2018ApJ...862...52S} {862, 52}

\bibitem[\protect\citeauthoryear{{Speagle}}{{Speagle}}{2020}]{Speagle2019}
{Speagle} J.~S.,  2020, \mn@doi [\mnras] {10.1093/mnras/staa278}, \href
  {https://ui.adsabs.harvard.edu/abs/2020MNRAS.493.3132S} {493, 3132}

\bibitem[\protect\citeauthoryear{{Ting}, {Rix}, {Bovy}  \& {van de Ven}}{{Ting}
  et~al.}{2013}]{Ting2013}
{Ting} Y.-S.,  {Rix} H.-W.,  {Bovy} J.,   {van de Ven} G.,  2013, \mn@doi
  [\mnras] {10.1093/mnras/stt1053}, \href
  {https://ui.adsabs.harvard.edu/abs/2013MNRAS.434..652T} {434, 652}

\bibitem[\protect\citeauthoryear{{Trick}, {Bovy}  \& {Rix}}{{Trick}
  et~al.}{2016}]{Trick2016}
{Trick} W.~H.,  {Bovy} J.,   {Rix} H.-W.,  2016, \mn@doi [\apj]
  {10.3847/0004-637X/830/2/97}, \href
  {https://ui.adsabs.harvard.edu/abs/2016ApJ...830...97T} {830, 97}

\bibitem[\protect\citeauthoryear{{Vasiliev}}{{Vasiliev}}{2018}]{Vasiliev2018}
{Vasiliev} E.,  2018, \mn@doi [arXiv e-prints] {10.48550/arXiv.1802.08255},
  \href {https://ui.adsabs.harvard.edu/abs/2018arXiv180208255V} {p.
  arXiv:1802.08255}

\bibitem[\protect\citeauthoryear{{Vasiliev}}{{Vasiliev}}{2019}]{Vasiliev2019}
{Vasiliev} E.,  2019, \mn@doi [\mnras] {10.1093/mnras/sty2672}, \href
  {https://ui.adsabs.harvard.edu/abs/2019MNRAS.482.1525V} {482, 1525}

\bibitem[\protect\citeauthoryear{{Vasiliev} \& {Baumgardt}}{{Vasiliev} \&
  {Baumgardt}}{2021}]{Vasiliev2021a}
{Vasiliev} E.,  {Baumgardt} H.,  2021, \mn@doi [\mnras]
  {10.1093/mnras/stab1475}, \href
  {https://ui.adsabs.harvard.edu/abs/2021MNRAS.505.5978V} {505, 5978}

\bibitem[\protect\citeauthoryear{{Vasiliev} \& {Valluri}}{{Vasiliev} \&
  {Valluri}}{2020}]{Vasiliev2020}
{Vasiliev} E.,  {Valluri} M.,  2020, \mn@doi [\apj] {10.3847/1538-4357/ab5fe0},
  \href {https://ui.adsabs.harvard.edu/abs/2020ApJ...889...39V} {889, 39}

\bibitem[\protect\citeauthoryear{{Vasiliev}, {Belokurov}  \&
  {Erkal}}{{Vasiliev} et~al.}{2021}]{Vasiliev2020a}
{Vasiliev} E.,  {Belokurov} V.,   {Erkal} D.,  2021, \mn@doi [\mnras]
  {10.1093/mnras/staa3673}, \href
  {https://ui.adsabs.harvard.edu/abs/2021MNRAS.501.2279V} {501, 2279}

\bibitem[\protect\citeauthoryear{{Virtanen} et~al.,}{{Virtanen}
  et~al.}{2020}]{scipy}
{Virtanen} P.,  et~al., 2020, \mn@doi [Nature Methods]
  {10.1038/s41592-019-0686-2}, \href
  {https://ui.adsabs.harvard.edu/abs/2020NatMe..17..261V} {17, 261}

\bibitem[\protect\citeauthoryear{{Walker} \& {Pe{\~n}arrubia}}{{Walker} \&
  {Pe{\~n}arrubia}}{2011}]{Walker2011}
{Walker} M.~G.,  {Pe{\~n}arrubia} J.,  2011, \mn@doi [\apj]
  {10.1088/0004-637X/742/1/20}, \href
  {https://ui.adsabs.harvard.edu/abs/2011ApJ...742...20W} {742, 20}

\bibitem[\protect\citeauthoryear{{Walker}, {Mateo}, {Olszewski},
  {Pe{\~n}arrubia}, {Evans}  \& {Gilmore}}{{Walker} et~al.}{2009}]{Walker2009a}
{Walker} M.~G.,  {Mateo} M.,  {Olszewski} E.~W.,  {Pe{\~n}arrubia} J.,  {Evans}
  N.~W.,   {Gilmore} G.,  2009, \mn@doi [\apj] {10.1088/0004-637X/704/2/1274},
  \href {https://ui.adsabs.harvard.edu/abs/2009ApJ...704.1274W} {704, 1274}

\bibitem[\protect\citeauthoryear{{Walsh}, {Willman}  \& {Jerjen}}{{Walsh}
  et~al.}{2009}]{Walsh2009}
{Walsh} S.~M.,  {Willman} B.,   {Jerjen} H.,  2009, \mn@doi [\aj]
  {10.1088/0004-6256/137/1/450}, \href
  {https://ui.adsabs.harvard.edu/abs/2009AJ....137..450W} {137, 450}

\bibitem[\protect\citeauthoryear{{Wang}, {Han}, {Cooper}, {Cole}, {Frenk}  \&
  {Lowing}}{{Wang} et~al.}{2015a}]{Wang2015b}
{Wang} W.,  {Han} J.,  {Cooper} A.~P.,  {Cole} S.,  {Frenk} C.,   {Lowing} B.,
  2015a, \mn@doi [\mnras] {10.1093/mnras/stv1647}, \href
  {https://ui.adsabs.harvard.edu/abs/2015MNRAS.453..377W} {453, 377}

\bibitem[\protect\citeauthoryear{{Wang}, {Lin}, {Pearce}, {Lux}, {Muldrew}  \&
  {Onions}}{{Wang} et~al.}{2015b}]{Wang2015}
{Wang} Y.,  {Lin} W.,  {Pearce} F.~R.,  {Lux} H.,  {Muldrew} S.~I.,   {Onions}
  J.,  2015b, \mn@doi [\apj] {10.1088/0004-637X/801/2/93}, \href
  {https://ui.adsabs.harvard.edu/abs/2015ApJ...801...93W} {801, 93}

\bibitem[\protect\citeauthoryear{{Wang}, {Han}, {Cole}, {Frenk}  \&
  {Sawala}}{{Wang} et~al.}{2017}]{Wang2017d}
{Wang} W.,  {Han} J.,  {Cole} S.,  {Frenk} C.,   {Sawala} T.,  2017, \mn@doi
  [\mnras] {10.1093/mnras/stx1334}, \href
  {https://ui.adsabs.harvard.edu/abs/2017MNRAS.470.2351W} {470, 2351}

\bibitem[\protect\citeauthoryear{{Wang}, {Han}, {Cole}, {More}, {Frenk}  \&
  {Schaller}}{{Wang} et~al.}{2018}]{Wang2018a}
{Wang} W.,  {Han} J.,  {Cole} S.,  {More} S.,  {Frenk} C.,   {Schaller} M.,
  2018, \mn@doi [\mnras] {10.1093/mnras/sty706}, \href
  {https://ui.adsabs.harvard.edu/abs/2018MNRAS.476.5669W} {476, 5669}

\bibitem[\protect\citeauthoryear{{Wang}, {Han}, {Cautun}, {Li}  \&
  {Ishigaki}}{{Wang} et~al.}{2020}]{Wang2019b}
{Wang} W.,  {Han} J.,  {Cautun} M.,  {Li} Z.,   {Ishigaki} M.~N.,  2020,
  \mn@doi [Science China Physics, Mechanics, and Astronomy]
  {10.1007/s11433-019-1541-6}, \href
  {https://ui.adsabs.harvard.edu/abs/2020SCPMA..6309801W} {63, 109801}

\bibitem[\protect\citeauthoryear{{Wang}, {Hammer}  \& {Yang}}{{Wang}
  et~al.}{2022a}]{Wang2022c}
{Wang} J.,  {Hammer} F.,   {Yang} Y.,  2022a, \mn@doi [\mnras]
  {10.1093/mnras/stab3258}, \href
  {https://ui.adsabs.harvard.edu/abs/2022MNRAS.510.2242W} {510, 2242}

\bibitem[\protect\citeauthoryear{{Wang} et~al.,}{{Wang}
  et~al.}{2022b}]{Wang2022f}
{Wang} W.,  et~al., 2022b, \mn@doi [\apj] {10.3847/1538-4357/ac9b19}, \href
  {https://ui.adsabs.harvard.edu/abs/2022ApJ...941..108W} {941, 108}

\bibitem[\protect\citeauthoryear{{Wang}, {Chrob{\'a}kov{\'a}},
  {L{\'o}pez-Corredoira}  \& {Sylos Labini}}{{Wang} et~al.}{2023}]{Wang2023f}
{Wang} H.-F.,  {Chrob{\'a}kov{\'a}} {\v{Z}}.,  {L{\'o}pez-Corredoira} M.,
  {Sylos Labini} F.,  2023, \mn@doi [\apj] {10.3847/1538-4357/aca27c}, \href
  {https://ui.adsabs.harvard.edu/abs/2023ApJ...942...12W} {942, 12}

\bibitem[\protect\citeauthoryear{{Watkins}, {Evans}  \& {An}}{{Watkins}
  et~al.}{2010}]{Watkins2010a}
{Watkins} L.~L.,  {Evans} N.~W.,   {An} J.~H.,  2010, \mn@doi [\mnras]
  {10.1111/j.1365-2966.2010.16708.x}, \href
  {https://ui.adsabs.harvard.edu/abs/2010MNRAS.406..264W} {406, 264}

\bibitem[\protect\citeauthoryear{{Watkins}, {van der Marel}, {Sohn}  \&
  {Evans}}{{Watkins} et~al.}{2019}]{Watkins2018}
{Watkins} L.~L.,  {van der Marel} R.~P.,  {Sohn} S.~T.,   {Evans} N.~W.,  2019,
  \mn@doi [\apj] {10.3847/1538-4357/ab089f}, \href
  {https://ui.adsabs.harvard.edu/abs/2019ApJ...873..118W} {873, 118}

\bibitem[\protect\citeauthoryear{{Webb} \& {Carlberg}}{{Webb} \&
  {Carlberg}}{2021}]{Webb2020}
{Webb} J.~J.,  {Carlberg} R.~G.,  2021, \mn@doi [\mnras]
  {10.1093/mnras/stab353}, \href
  {https://ui.adsabs.harvard.edu/abs/2021MNRAS.502.4547W} {502, 4547}

\bibitem[\protect\citeauthoryear{{Wegg}, {Gerhard}  \& {Bieth}}{{Wegg}
  et~al.}{2019}]{Wegg2018}
{Wegg} C.,  {Gerhard} O.,   {Bieth} M.,  2019, \mn@doi [\mnras]
  {10.1093/mnras/stz572}, \href
  {https://ui.adsabs.harvard.edu/abs/2019MNRAS.485.3296W} {485, 3296}

\bibitem[\protect\citeauthoryear{{Wilkinson} \& {Evans}}{{Wilkinson} \&
  {Evans}}{1999}]{Wilkinson1999}
{Wilkinson} M.~I.,  {Evans} N.~W.,  1999, \mn@doi [\mnras]
  {10.1046/j.1365-8711.1999.02964.x}, \href
  {https://ui.adsabs.harvard.edu/abs/1999MNRAS.310..645W} {310, 645}

\bibitem[\protect\citeauthoryear{{Wojtak}, {{\L}okas}, {Mamon},
  {Gottl{\"o}ber}, {Klypin}  \& {Hoffman}}{{Wojtak} et~al.}{2008}]{Wojtak2008}
{Wojtak} R.,  {{\L}okas} E.~L.,  {Mamon} G.~A.,  {Gottl{\"o}ber} S.,  {Klypin}
  A.,   {Hoffman} Y.,  2008, \mn@doi [\mnras]
  {10.1111/j.1365-2966.2008.13441.x}, \href
  {https://ui.adsabs.harvard.edu/abs/2008MNRAS.388..815W} {388, 815}

\bibitem[\protect\citeauthoryear{{Wolf}, {Martinez}, {Bullock}, {Kaplinghat},
  {Geha}, {Mu{\~n}oz}, {Simon}  \& {Avedo}}{{Wolf} et~al.}{2010}]{Wolf2010a}
{Wolf} J.,  {Martinez} G.~D.,  {Bullock} J.~S.,  {Kaplinghat} M.,  {Geha} M.,
  {Mu{\~n}oz} R.~R.,  {Simon} J.~D.,   {Avedo} F.~F.,  2010, \mn@doi [\mnras]
  {10.1111/j.1365-2966.2010.16753.x}, \href
  {https://ui.adsabs.harvard.edu/abs/2010MNRAS.406.1220W} {406, 1220}

\bibitem[\protect\citeauthoryear{{Xue} et~al.,}{{Xue} et~al.}{2008}]{Xue2008}
{Xue} X.~X.,  et~al., 2008, \mn@doi [\apj] {10.1086/589500}, \href
  {https://ui.adsabs.harvard.edu/abs/2008ApJ...684.1143X} {684, 1143}

\bibitem[\protect\citeauthoryear{{Yang}, {Lai}  \& {Lin}}{{Yang}
  et~al.}{2012}]{Yang2012}
{Yang} M.-S.,  {Lai} C.-Y.,   {Lin} C.-Y.,  2012, \mn@doi [Pattern Recognition]
  {10.1016/j.patcog.2012.04.031}, \href
  {https://ui.adsabs.harvard.edu/abs/2012PatRe..45.3950Y} {45, 3950}

\bibitem[\protect\citeauthoryear{{Yang}, {Boruah}  \& {Afshordi}}{{Yang}
  et~al.}{2020}]{Yang2020}
{Yang} T.,  {Boruah} S.~S.,   {Afshordi} N.,  2020, \mn@doi [\mnras]
  {10.1093/mnras/staa441}, \href
  {https://ui.adsabs.harvard.edu/abs/2020MNRAS.493.3061Y} {493, 3061}

\bibitem[\protect\citeauthoryear{{Yang}, {Zhu}, {Tahmasebzadeh}, {Xue}  \&
  {Liu}}{{Yang} et~al.}{2022}]{Yang2022}
{Yang} C.,  {Zhu} L.,  {Tahmasebzadeh} B.,  {Xue} X.-X.,   {Liu} C.,  2022,
  \mn@doi [\aj] {10.3847/1538-3881/ac9900}, \href
  {https://ui.adsabs.harvard.edu/abs/2022AJ....164..241Y} {164, 241}

\bibitem[\protect\citeauthoryear{Yi, Delaigle  \& Gustafson}{Yi
  et~al.}{2021}]{Yi2021}
Yi G.~Y.,  Delaigle A.,   Gustafson P.,  eds, 2021, Handbook of {{Measurement
  Error Models}}.
{Chapman and Hall/CRC}, New York, \mn@doi{10.1201/9781315101279}

\bibitem[\protect\citeauthoryear{{Zhai}, {Xue}, {Zhang}, {Li}, {Zhao}  \&
  {Yang}}{{Zhai} et~al.}{2018}]{Zhai2018}
{Zhai} M.,  {Xue} X.-X.,  {Zhang} L.,  {Li} C.-D.,  {Zhao} G.,   {Yang} C.-Q.,
  2018, \mn@doi [Research in Astronomy and Astrophysics]
  {10.1088/1674-4527/18/9/113}, \href
  {https://ui.adsabs.harvard.edu/abs/2018RAA....18..113Z} {18, 113}

\bibitem[\protect\citeauthoryear{{Zhou}, {Li}, {Huang}  \& {Zhang}}{{Zhou}
  et~al.}{2023}]{Zhou2023a}
{Zhou} Y.,  {Li} X.,  {Huang} Y.,   {Zhang} H.,  2023, \mn@doi [\apj]
  {10.3847/1538-4357/acadd9}, \href
  {https://ui.adsabs.harvard.edu/abs/2023ApJ...946...73Z} {946, 73}

\bibitem[\protect\citeauthoryear{{Zhu} et~al.,}{{Zhu} et~al.}{2018}]{Zhu2018b}
{Zhu} L.,  et~al., 2018, \mn@doi [\mnras] {10.1093/mnras/stx2409}, \href
  {https://ui.adsabs.harvard.edu/abs/2018MNRAS.473.3000Z} {473, 3000}

\bibitem[\protect\citeauthoryear{{van den Bosch}, {van de Ven}, {Verolme},
  {Cappellari}  \& {de Zeeuw}}{{van den Bosch} et~al.}{2008}]{vandenBosch2008a}
{van den Bosch} R.~C.~E.,  {van de Ven} G.,  {Verolme} E.~K.,  {Cappellari} M.,
    {de Zeeuw} P.~T.,  2008, \mn@doi [\mnras]
  {10.1111/j.1365-2966.2008.12874.x}, \href
  {https://ui.adsabs.harvard.edu/abs/2008MNRAS.385..647V} {385, 647}

\makeatother
\end{thebibliography}

%\begin{thebibliography}{99}
%\end{thebibliography}

%%%%%%%%%%%%%%%%%%%%%%%%%%%%%%%%%%%%%%%%%%%%%%%%%%

%%%%%%%%%%%%%%%%% APPENDICES %%%%%%%%%%%%%%%%%%%%%

\appendix

\section{Construction of empirical DF}
\label{sec:kde_odf}

We construct the smooth orbital distribution $p(E,L)$ from a given sample,
largely following the procedure of \citet{Li2019},
originally used to build the stacking orbital distribution of satellite galaxies in simulations.

The $(E, L)$ are distributed in a sharp triangular shape unevenly,
making it non-trivial to handle boundaries during density estimation.
It is more convenient to perform this task in the parameter space of $(E, \varepsilon^2)$,
where $\varepsilon = L / L_{\max} (E) \in [0, 1]$ and $L_{\max} (E)$ is the maximum angular momentum allowed for an orbit with energy $E$.
Without observational limits, $L_{\max}$ corresponds to the angular momentum of the circular orbit with given energy,
and $\varepsilon$ is known as \emph{{{orbital circularity}}}. 
When considering tracers restricted within radius range $[r_{\min}, r_{\max}]$, 
$L_{\max} (E)$ becomes
\[ L_{\max} (E) = \left\{ \begin{array}{ll}
     r_{\min}  \sqrt{2 [E - \Phi (r_{\min})]},
        & r_{\mathrm{cir}} (E) < r_{\min},\\
     \sqrt{G M (< r_{{\mathrm{cir}}}) r_{{\mathrm{cir}}}}, 
        & r_{\min} \leq r_{\mathrm{cir}} (E) \leq r_{\max},\\
     r_{\max}  \sqrt{2 [E - \Phi (r_{\max})]}, 
        & r_{\max} < r_{\mathrm{cir}} (E),
   \end{array} \right. \]
where $r_{{\mathrm{cir}}}$ is the radius of the circular orbit with energy $E$.

Given the trial potential, we calculate $\{E_i, \varepsilon^2_i \}_{i = 1}^N$ 
and their weights $\{\omega_i\}$ (equal to 1 unless a selection function is involved).
We then construct the smooth density distribution in $(E, \varepsilon^2)$ space via KDE,
\begin{gather}
p (E, \varepsilon^2) = \frac{1}{\sum_{i=1}^N \omega_i} {\textstyle\sum_{i = 1}^N} \omega_i \mathcal{K} \left( \frac{E - E_i}{h\sigma_E} \right) \mathcal{K}\left(\frac{\varepsilon^2 - \varepsilon_i^2}{h\sigma_{\!\varepsilon^2}} \right),
  \label{eq:kde}
\end{gather}
where $\sigma_E$ and $\sigma_{\!\varepsilon^2}$ are the standard deviations of
$\{ E_i \}$ and $\{ \varepsilon^2_i \}$ respectively, and $h$ is the smoothing bandwidth.
$\mathcal{K}$ denotes the Gaussian smoothing kernel, $\mathcal{K} ({x/\sigma}) =
({\sqrt{2 \pi}\sigma})^{-1} \exp[{- ({{x}/\sigma})^2 / 2}]$. We adopt \citet{Scott1979}'s rule of thumb,
\begin{gather}
  h = n_{{\mathrm{eff}}}^{- 1 / (n_\mathrm{dim}+ 4)},
\label{eq:scott_rule}
\end{gather}
where $n_\mathrm{dim}=2$ for 2D smoothing,
and $n_{{\mathrm{eff}}} = ( \sum \omega_i )^2 / ( \sum \omega_i^2 )$ is the effective sample size. 
The smoothing size is thus larger for a small sample or unequal weights.
Reflecting boundaries are applied at $\varepsilon^2 = 0$ and 1 and $E = \Phi (r_{\min})$.
By construction, $p (E, \varepsilon^2)$ satisfies $\int_{\Phi
(r_{\min})}^{\infty} \int_0^1 p (E, \varepsilon^2) \dif E \dif \varepsilon^2 = 1$. 

The orbital distribution and DF (\refeqnalt{eq:f2}) are then
\begin{align}
  p (E, L) &= \frac{2L}{L_{\max}^2 (E)} p (E, \varepsilon^2), \\
  f_{\Phi} ({\boldsymbol{r}}, {\boldsymbol{v}}) &= \frac{p (E, \varepsilon^2)}{4 \pi^2 L_{\max}^2 (E) T_r (E, L)} .
\end{align}
The DF $f_\Phi$ is guaranteed to be a smooth function, since $p(E,\varepsilon^2)$ is smoothed.
Direct smoothing in $f (E, L)$ is meaningless, because it does not conserve the mass [recall that $f$ is a distribution in $(\bm{x},\bm{v})$ rather than $(E,L)$].

\subsection{Additional remarks on KDE}
\label{sec:kde_cmp}

KDE is a well-established technique with wide applications \citep{Silverman1986, Scott2015}.
Theory and practice suggest that the bandwidth is more important than the specific choice of kernel.
We find that the Scott rule is indeed close to optimal for our test cases.
When the sample is very extended with long tails, standard deviation $\sigma$ may overestimate the smoothing scale.
A robust alternative is the normalized median absolute deviation (MADN, \citealt[p36]{Maronna2019}),
$\hat{\sigma}=1.4826\,{\mathrm{median}} \{ | x - {\mathrm{median}} \{ x \} | \}$.
Additional attention is needed to determine the best kernel size (e.g., using cross-validation) for multimodal distributions, where Scott's rule is perhaps not optimal.
We also tested adaptive smoothing size $h$ based on $k$-nearest neighbor \citep[see e.g.,][]{Li2019}, but it yields worse performance.
% We do not consider weight-dependent kernel sizes here.
% In principle, a larger kernel size could be used for points with higher weights,
% which might be helpful when dealing with non-equal weights.
% Such technical improvements are left for future exploration.

Besides KDE, several other techniques for density estimation have potential or existing applications in dynamical modeling methods (not limited to \empdf).
Examples include Voronoi tessellation, entropy-based binary decomposition \citep[EnBid]{Sharma2006}, $k$-th nearest neighbor \citep{Silva2024},
Gaussian mixtures \citep{Magorrian2014},
penalized B-splines \citep{Vasiliev2018},
neural networks \citep{Rothfuss2019},
%\footnote{\url{https://github.com/freelunchtheorem/Conditional_Density_Estimation}}
and normalizing flows \citep{Dinh2016,Lim2022}.
However, it is crucial to remember that dynamical modeling requires more than just smoothing for orbits within a specific potential. The chosen technique must also provide consistent and continuous density estimation (and thus likelihood) when continuously changing the trial potential.
The advantage of KDE on this point becomes more evident in the next subsection.

\subsection{Comments on Magorrian 2014 and Gaussian mixtures}
\label{sec:other_de}

\citet{Magorrian2014} proposed a very elegant DF method based on Infinite Gaussian Mixture Models (GMMs) regulated by a Dirichlet process.
It models the distribution of actions as the sum of Gaussian blobs whose number, weights, locations, and shapes are to be determined during fitting.
There is a clear technical resemblance with \empdf, as both construct orbital distribution (through GMMs vs KDE) and use likelihood in 6D phase space.

While mathematically elegant, Infinite GMMs face several practical challenges that hinder their general adoption for dynamical modeling.
GMMs lack a closed-form solution and require iterative optimization \citep{Bishop2006}, which is sensitive to initialization and can converge to undesired local optima \citep{Archambeau2003a,Yang2012,Jin2016}.
Moreover, GMMs may exhibit unpredictable variations in the number, location, and size of Gaussian blobs when the underlying potential changes, possibly leading to discontinuous likelihood and unreliable parameter inference \citep[Appendix C3]{Magorrian2014}.
In contrast, KDE has no free parameters, and the resultant likelihood responds continuously to changes in potential.
KDE also outperforms GMMs in computational efficiency, particularly for large datasets (e.g., by using fast Fourier transform).

More importantly, the underlying philosophy differs. \citet{Magorrian2014} can be regarded as a highly flexible analytical model with many free parameters, whereas \empdf considers the time-averaged DF of the tracers themselves deterministically.
% In GMMs, each blob represents many points, whereas in KDE, each point represents a blob.

\begin{figure*}
  \includegraphics[width=0.95\linewidth]{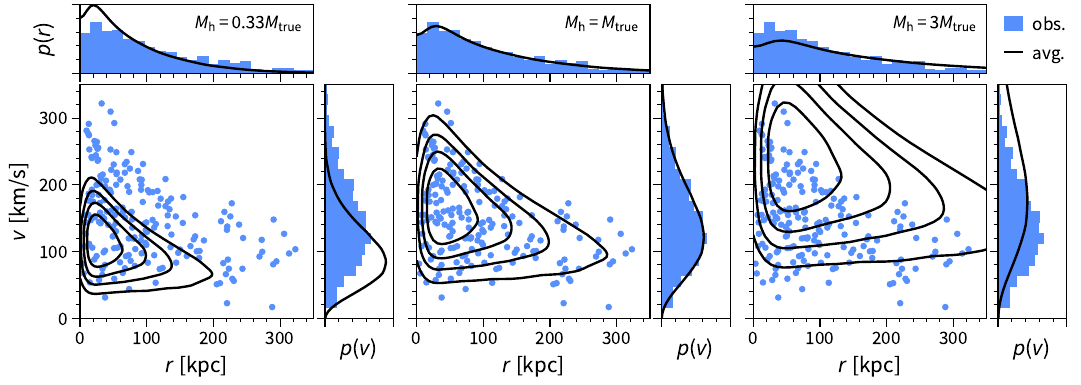}
  \renewcommand{\thefigure}{E\arabic{figure}}
  \setcounter{figure}{0}
  \vspace{-0.5em}
  \caption{
  Similar to \reffig{fig:illustrate},
  but illustrating how the simulation-informed DF scales with the potential.
  The blue dots and shades show the observation data.
  The black contour and distributions represent the model predicted by the scaling relation.
  }
  \label{fig:illustrate_scaling}
\end{figure*}

\section{Likelihood and Kullback-Leibler divergence}
\label{sec:lik_KLdiv}

To quantify the difference between the time-averaged DF and the instantaneous DF, a natural choice based on information theory is the Kullback-Leibler (KL) divergence (aka. relative entropy), a measure of how one probability distribution differs from another (\citealt{Bishop2006}, \S1.6.1). % also cf. cross entropy.
For two distributions $p(x)$ and $q(x)$, 
the KL divergence from $p$ to $q$ is defined as
\begin{equation}
    {\mathrm{KL}} (q \| p) = \int q(x)\ln \frac{q(x)}{p(x)} \dif x.
\end{equation}
Suppose that the data sample $\{x_i\}$ was generated from an unknown distribution $q(x)$ that we wish to approximate using some parametric distribution $p(x\mid\Theta)$, controlled by a set of parameters $\Theta$.
Specifically, in our problem 
$q$ is the underlying DF of the tracers represented by observation $\{ {\bm{w}}_i \}$,
and $p$ is the empirical DF which depends on the parametrized potential.

One way to determine $\Theta$ is to minimize the KL divergence between $q(x)$ and $p(x\mid\Theta)$. Although we do not know $q(x)$ explicitly, the integration over $q(x)$ can be approximated by summing over ${x_i}$,
${\mathrm{KL}} (q \| p)_\Theta = \frac{1}{N} \sum_i [\ln q(x_i) - \ln p(x_i \mid \Theta)]=-\frac{1}{N} \sum_i \ln p(x_i\mid \Theta)+\mathrm{const}$, where `const' arises because $q(x)$ does not depend on $\Theta$.
Therefore, minimizing KL divergence is equivalent to maximizing the log likelihood for $\Theta$, $\sum_i \ln p(x_i \mid \Theta)$.

We should note that while the mathematical form of the KL divergence (or log likelihood) resembles entropy, their interpretations differ. 
Entropy measures the information of a single distribution, whereas KL divergence quantifies the difference between two distributions.

\section{Self-correction for observational errors}
\label{sec:error_corr}
The observational errors tend to smear and bias the DF constructed empirically from observed kinematics using KDE.
For example, the observed tangential velocity and thus angular momentum 
are biased high on average, especially for distant MW tracers (see e.g., Fig. 5 and 6 in \citealt{Li2020a}).
Our goal is to derive the underlying DF unbiasedly by correcting the observational noises.
A standard technique is the deconvolution KDE (\citealt{Delaigle2008,Yi2021}; also cf. deconvolution Gaussian mixtures by \citealt{Bovy2011}).
However, computing this deconvolution KDE is nontrivial as it requires a Fourier transform of the noise function, which is often non-analytic and varies for each tracer (heteroscedasticity).

Here we propose a novel method to correct heteroscedastic errors in KDE using \emph{iterative reweighted importance sampling} (IRIS). This method can be regarded as an implementation of the Richardson--Lucy deconvolution algorithm \citep{Richardson1972, Lucy1974} but for discrete samples rather than images.
While IRIS is applicable to general problems, such as correcting the Eddington bias in galaxy luminosity functions, we specifically focus on its application in emPDF below.

If the underlying DF $f ({\bm{w}})$ is known in prior,
for an observed tracer $\bm{w}_i$,
its true kinematics $\bm{w}_{\mathrm{tr}}$ 
is given by the posterior distribution,
\begin{gather}
  p ({\bm{w}}_{\mathrm{tr}} \mid {\bm{w}_i})
  \propto p_{\mathrm{err}}
  ({\bm{w}_i}\mid {\bm{w}}_{\mathrm{tr}}) f
  ({\bm{w}}_{\mathrm{tr}}) .
\label{eq:post_err}
\end{gather}
The key idea of the self-correction is that
the sum of the posterior distributions of a tracer sample provides a much closer 
representation of the true DF than the direct observation $\{\bm{w}_i\}$.

The above posterior distribution can be achieved by importance sampling.
For each tracer $\bm{w}_i$,
we sample $m$ subparticles $\{{\bm{w}}_{ij} \}_{j = 1}^m$
according to $p ({\bm{w}}_{ij}) \propto p_{\mathrm{err}}
({\bm{w}}_i \mid {\bm{w}}_{ij})$,
and assign weights as $\omega_{ij} \propto f ({\bm{w}}_{ij})$,
with normalization $\sum_j \omega_{ij} = 1$. 
In this way, $\{{\bm{w}}_{ij}, \omega_{ij} \}_j$ represents a
weighted realization of \refeqn{eq:post_err},
and the ensemble $\{{\bm{w}}_{ij}, \omega_{ij} \}_{i, j}$ provides a 
better representation of the underlying DF than
the original $\{{\bm{w}}_i \}_i$.

Though we do not know the true DF $f ({\bm{w}})$ in the beginning, we can resort to an iterative procedure to approach a self-consistent DF and largely remove the observation scatter.
Starting with the empricial DF $f(\bm{w})$ constructed from observed kinematics $\{\bm{w}_i\}$ (\refeqnalt{eq:nel})
as a rough guess, we initialize $\omega_{ij} \propto f ({\bm{w}}_{ij})$ (alternatively, initializing with $\omega_{ij} = 1 / m$ also works).
We then rebuild $f(\bm{w})$ from 
$\{{\bm{w}}_{ij}, \omega_{ij} \}_{i, j}$ with weighted KDE,
and recompute $\omega_{ij}$ by $f(\bm{w})$ in iteration.
For the KDE smoothing bandwidth, we use the Scott's rule (\refeqnalt{eq:scott_rule})
with the effective sample size,
$n_{{\mathrm{eff}}} = ( \sum_{ij} \omega_{ij} )^2 / ( \sum_{ij} \omega_{ij}^2 )$.
Clearly, the computational cost of this scheme is significantly higher than in cases without observational error due to the large number of subparticles involved.%
\footnote{
  One may wonder if we can reduce the computation of KDE 
  by using the mean posterior values
  $\avg{\bm{w}_\mathrm{tr}}_i=\sum_j \omega_{ij} \bm{w}_{ij}$ ($N$ particles) instead of the ensemble subparticles $\bm{w}_{ij}$ ($N\times m$ subparticles) to construct the empirical DF. 
  Unfortunately, this does not converge to the correct underlying distribution.
}
% In principle, it might be better to exclude the contribution of the other sub-particles sharing the same targets tracer,
% which is however not practical considering the computation costs.

Preliminary numerical experiments with 1D and 2D distributions confirm the viability of the above procedure.
The underlying distribution can be closely recovered from a sample with large observational errors comparable to its intrinsic spread.
Taking $m=1000$ with 10 iterations usually yields sufficiently accurate results, achieving approximate convergence, 
though these parameters may vary depending on the specific problem and dimension.
Validation for actual dynamical modeling problems is left for future work.

When the selection function is also involved,
we should replace the $\omega_{ij}$ with 
$\omega_{ij}\cdot\omega_\mathrm{sel}(\bm{w}_{ij})$ 
to construct the empirical DF (\refeqnalt{eq:nel}),
where $\omega_\mathrm{sel}(\bm{w}_{ij})$ is
the correction factor for selection function (\refeqnalt{eq:w1}).%
\footnote{
A subtle note: a selection in observation might perform on either the 
true distance or the observed distance, depending on how the sample is selected.
The two cases have slightly different likelihood functions.
The actual difference is usually negligible, 
so here we treat it as if the sample is selected by 
${\bm{w}}_{\mathrm{tr}}$ for simplicity.
}

Besides the empirical DF, the observational errors also affect the likelihood (see \refeqnalt{eq:mc_int_err}),
$ \textstyle p_{\mathrm{ob}} ({\bm{w}_i}\mid {\bm{\Theta}}_{\Phi}) = 
  \int p_{\mathrm{err}} ({\bm{w}_i}\mid {\bm{w}}_{\mathrm{tr}}) \allowbreak f_\Phi ({\bm{w}})
   \dif^6{\bm{w}_\mathrm{tr}}$.
It is easy to compute the integral by Monte Carlo integration with importance sampling \citep[\S3.2]{Li2020a},%
\footnote{
  Strictly speaking, \refeqn{eq:p_oberr2} differs from \refeqn{eq:mc_int_err} by a constant factor
  (see footnote 5 in \citealt{Li2020a}), but this does not affect the parameter inference.
}
\begin{gather}
   p_{\mathrm{ob}} ({\bm{w}_i}\mid {\bm{\Theta}}_{\Phi}) = 
  \frac{1}{m}\textstyle\sum_j f_\Phi ({\bm{w}_{ij}}).
\label{eq:p_oberr2}
\end{gather}

\section{Spherical Jeans equation} 
\label{sec:SJE}

The spherical Jeans equation (\citetalias{Binney2008a}, eq. 4.214) writes,
\begin{gather}
  \frac{d (\nu \sigma_r^2)}{\nu dr} + \frac{2 \sigma_r^2}{r}  \left( 1 -
  \frac{\sigma_t^2}{2 \sigma_r^2} \right) = - \frac{d \Phi}{dr}
\end{gather}
where the tracer number density $\nu$,
velocity dispersions $\sigma_r^2 = \avg{v_r^2}$ and $\sigma_t^2 = \avg{v_t^2}$,
and velocity anisotropy $\beta = 1 - {\sigma_t^2}/{2 \sigma_r^2}$ are functions of radius.
These quantities are measured from observation.
The mass within a given radius is thus
\begin{gather}
  M(<r) = - \left[ \frac{d \ln (\nu \sigma_r^2)}{d \ln r} + 2 \beta \right] 
  \frac{r\sigma_r^2}{G}.
\label{eq:sje_mass}
\end{gather}
We adopt the ``backward'' implementation by \citealt{Li2021d}.
We divide the tracers into $m$ radial bins ($m=5$ for $N_\mathrm{tracers}=160$, and 20 for larger sample size).
In each bin, we measure $\sigma_r^2, \sigma_t^2, \beta$ and $\nu \sigma_r^2$.
The gradient of $\nu \sigma_r^2$ is estimated by finite difference.
We compute $M_i$ for $i$-th bin using \refeqn{eq:sje_mass} and determine the covariance matrix $\Sigma_{ij}=\avg{\delta M_i \delta M_j}$ using bootstrap resampling of the tracers.
For a given potential model, we calculate $\hat M_i=M_\mathrm{model}(<r_i)$.
We then search for parameters of the best-fit potential by minimizing $\chi^2=(\hat M_i-M_i)\Sigma_{ij}^{-1}(\hat M_j-M_j)$ with the Python package \textsc{iminuit}.%
\footnote{\url{https://github.com/scikit-hep/iminuit}}

This backward method offers more direct estimates of the mass profile but also suffers from observational errors directly and requires relatively large sample sizes.
In contrast, the forward approach assumes some parametric form for the potential and tracer distribution, then computes expected velocity distributions.
It can apply to small samples and handle observational errors and selection functions more effectively.
Yet similar to analytical DF methods, improper model assumptions may lead to biases.
% Forward approaches commonly approximate velocity distributions at a given radius as Gaussians, which however proves inaccurate.
% Moreover, given these assumptions about tracer density profiles and velocity anisotropy profiles,
% it is more efficient to directly use analytical DFs (if available) based on the same assumptions but with higher efficiency.

\section{Illustration of simulation-informed DF}

The simulation-informed DF method \citep{Li2020a} assumes that the kinematics of tracers scale with the potential well.
It constructs a template DF by rescaling and stacking halo tracers in simulations.
The DF model for any virial mass and concentration is obtained by rescaling this template DF using the characteristic scales of the NFW potential, $\rs$ and $\vs$.
This scaling procedure is illustrated in \reffig{fig:illustrate_scaling}, clarifying its difference from \empdf.
When applied to different potentials,
the predicted distribution of simulation-informed DF changes more dramatically than \empdf in response to variations in potential.
This explains why \empdf provides tighter constraints and exhibits a different parameter degeneracy direction.
% The shape of the simulation-informed DF is known a priori but scales with the trial potential.

\section{Mocking observational effects}
\label{sec:more_tests}

\begin{figure}
  \centering
  \includegraphics[width=1\columnwidth]{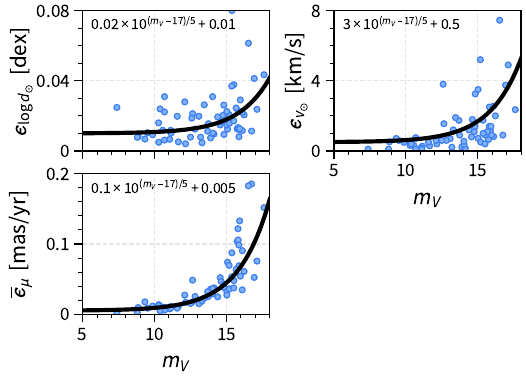}
  \vspace{-3ex}
    \caption{
  Observational errors of MW satellites and globular clusters
  with \gaia DR3 measurements.
  Blue symbols show the errors in Heliocentric distance $d_\odot$, proper motion $\mu$,
  and line-of-sight velocity ${v_\mathrm{los}}$
  as functions of apparent magnitude for MW tracers within $[20, 300] \kpc$.
  Black curves show the fitting relations according to the formulas in each panel.
  }
  \label{fig:obserr_appmag}
\end{figure}

Following \refsec{sec:mock}, we generate mock tracers within $[20, 300]\kpc$
as the parent sample, on which we add the observational effects
to mimic the MW satellite galaxies used in this work.
For simplicity, we place the mock observer at the center of the halo,
and thus the observables, Heliocentric distance $d_\odot$, proper motion $\mu$,
and line-of-sight velocity ${v_\mathrm{los}}$,
become $d_\odot=r$, $\mu=v_t/r$, and $v_\mathrm{los}=v_r$, respectively.

\textbf{Selection function.}
We mimic the MW satellites as a flux-limited sample,
where tracers fainter than $m_{V, \max}=17$ are unobservable (cf. \reffig{fig:mw_tracers}).
The maximal observable radius of a tracer with absolute magnitude $M_V$ is then
$\robsmax (M_V) = 10^{- 0.2 (M_V - 17 + 10)} \kpc$. 
The complete satellite luminosity function at faint end is
$N (<\!M_V) = 10^{0.156 M_V + 2.21}$ \citep{Newton2018}.
Using the mapping between $M_V$ and $\robsmax$, we obtain
$N (> \robsmax) \propto \robsmax^{-0.78}$.
Thus, the distribution of $\robsmax$ for MW satellites follows
\begin{gather}
  p (\robsmax) \propto \robsmax^{- 1.78}.
  \label{eq:p_robs}
\end{gather}
We assign each mock tracer with an $\robsmax$, randomly sampled from $p (\robsmax)$.
Tracers with $r>\robsmax$ are considered unobservable and removed from the mock sample.
% This magnitude cut simplifies generating mock samples; however, when applying to MW observations, a more delicate selection function is used.

\textbf{Observational error.}
As shown in \reffig{fig:obserr_appmag},
typical observational errors in $d_\odot$, $\mu$,
and ${v_\mathrm{los}}$ of MW satellites (and globular clusters)
with \gaia DR3 measurements can be approximated by
\begin{gather}
  \epsilon_{X} = c_1 10^{0.2 (m_V - 17)} + c_0 = c_1 r/\robsmax + c_0,
  \label{eq:gaia_obs_err}
\end{gather}
where $X=\log d_\odot, \mu$, or ${v_\mathrm{los}}$,
and the parameters $c_0$ and $c_1$ are provided in the figure.
The observational errors are assumed to be Gaussian and mutually independent.
Random observational errors are added according to $r/\robsmax$ of each mock tracer following \refeqn{eq:gaia_obs_err}.

% This relation arises because the number of detectable stars above $m_V=17$ is roughly $10^{-0.4 (m_V - 17)}$.
% We did not consider the expected number of tracers in this test, but rather focused on the influence of the selection function's form.
% In a sense, we assume the faintest measurable satellites $r\sim\robsmax$ always have similar observational errors.

% The observational error level is determined by two factors, the measurement
% precision for one single star, and the number of measurable stars.
% Combining both, \gaia DR3 has an improvement of factor of 2 for the proper motions of satellite galaxies and globular clusters.
% In the future, we will go to deeper surveys (e.g., \gaia DR4) to detect more DGs and GCs.
% Therefore, $c_0$ will decrease as the number of tracers increases in principle.

%%%%%%%%%%%%%%%%%%%%%%%%%%%%%%%%%%%%%%%%%%%%%%%%%%
% \balance

% Don't change these lines
\bsp	% typesetting comment
\label{lastpage}
\end{document}